\batchmode
\makeatletter
\def\input@path{{"/home/jacob/Documents/Work/My Papers/Magnetic Forces Can Do Work (2020)/"}}
\makeatother
\documentclass[twoside,twocolumn,english,superscriptaddress,nobibnotes,nofootinbib]{revtex4-2}
\usepackage[latin9]{inputenc}
\usepackage{color}
\usepackage{babel}
\usepackage{mathrsfs}
\usepackage{mathtools}
\usepackage{amsmath}
\usepackage{amssymb}
\usepackage[pdftex,unicode=true,
 bookmarks=true,bookmarksnumbered=true,bookmarksopen=false,
 breaklinks=false,pdfborder={0 0 0},pdfborderstyle={},backref=false,colorlinks=true]
 {hyperref}
\hypersetup{pdftitle={Can Magnetic Forces Do Work?},
 pdfauthor={Jacob A. Barandes},
 pdfsubject={Physics},
 pdfkeywords={Classical physics; particle physics; gauge theories; electromagnetism},
 linkcolor=blue, urlcolor=blue, citecolor=blue}

\makeatletter

\usepackage{url}

\let\originalleft\left
\let\originalright\right
\renewcommand{\left}{\mathopen{}\mathclose\bgroup\originalleft}
\renewcommand{\right}{\aftergroup\egroup\originalright}

\def\smalloverbrace#1{\mathop{\vbox{\m@th\ialign{##\crcr%
      \noalign{\kern3\p@}%
      \tiny\downbracefill\crcr\noalign{\kern3\p@\nointerlineskip}%
      $\hfil\displaystyle{#1}\hfil$\crcr}}}\limits}

\def\smallunderbrace#1{\mathop{\vtop{\m@th\ialign{##\crcr
   $\hfil\displaystyle{#1}\hfil$\crcr
   \noalign{\kern3\p@\nointerlineskip}%
   \tiny\upbracefill\crcr\noalign{\kern3\p@}}}}\limits}

\makeatother

\begin{document}
\title{Can Magnetic Forces Do Work?}
\author{Jacob A. Barandes}
\email{jacob\_barandes@harvard.edu}

\affiliation{Jefferson Physical Laboratory, Harvard University, Cambridge, MA 02138}
\date{\today}
\begin{abstract}
Standard lore holds that magnetic forces are incapable of doing mechanical
work. More precisely, the claim is that whenever it appears that a
magnetic force is doing work, the work is actually being done by another
force, with the magnetic force serving only as an indirect mediator.
However, the most familiar instances of magnetic forces acting in
everyday life, such as when bar magnets lift other bar magnets, appear
to present manifest evidence of magnetic forces doing work. These
sorts of counterexamples are often dismissed as arising from quantum
effects that lie outside the classical regime. In this paper, we show
that quantum theory is not needed to account for these phenomena,
and that classical electromagnetism admits a model of elementary magnetic
dipoles on which magnetic forces can indeed do work. In order to develop
this model, we revisit the foundational principles of the classical
theory of electromagnetism, showcase the importance of constraints
from relativity, examine the structure of the multipole expansion,
and study the connection between the Lorentz force law and conservation
of energy and momentum.
\end{abstract}
\maketitle

\global\long\def\vec#1{{\bf #1}}%
\global\long\def\vecgreek#1{\boldsymbol{#1}}%
\global\long\def\dotprod{\cdot}%
\global\long\def\crossprod{\times}%
\global\long\def\tud#1#2#3{#1^{#2}{}_{#3}}%
\global\long\def\tdu#1#2#3{#1_{#2}{}^{#3}}%
\global\long\def\defeq{\equiv}%
\global\long\def\Trace{\mathrm{Tr}}%
\global\long\def\transp{\mathrm{T}}%
\global\long\def\refvalue{0}%
\global\long\def\parens#1{(#1)}%
\global\long\def\bigparens#1{\big(#1\big)}%
\global\long\def\Bigparens#1{\Big(#1\Big)}%
\global\long\def\biggparens#1{\bigg(#1\bigg)}%
\global\long\def\Biggparens#1{\Bigg(#1\Bigg)}%
\global\long\def\bracks#1{[#1]}%
\global\long\def\bigbracks#1{\big[#1\big]}%
\global\long\def\Bigbracks#1{\Big[#1\Big]}%
\global\long\def\biggbracks#1{\bigg[#1\bigg]}%
\global\long\def\Biggbracks#1{\Bigg[#1\Bigg]}%
\global\long\def\curlies#1{\{#1\}}%
\global\long\def\bigcurlies#1{\big\{#1\big\}}%
\global\long\def\Bigcurlies#1{\Big\{#1\Big\}}%
\global\long\def\biggcurlies#1{\bigg\{#1\bigg\}}%
\global\long\def\Biggcurlies#1{\Bigg\{#1\Bigg\}}%
\global\long\def\verts#1{\vert#1\vert}%
\global\long\def\bigverts#1{\big\vert#1\big\vert}%
\global\long\def\Bigverts#1{\Big\vert#1\Big\vert}%
\global\long\def\biggverts#1{\bigg\vert#1\bigg\vert}%
\global\long\def\Biggverts#1{\Bigg\vert#1\Bigg\vert}%

\section{Introduction}

The question of whether magnetic forces can do mechanical work presents
a marvelous opportunity for exploring basic definitions in analytical
mechanics and the fundamental structure of classical electromagnetism.
In this paper, which builds on \citep{Barandes:2019mcl}, we show
that classically extending Maxwell's theory of electromagnetism to
include \emph{elementary} dipoles\textemdash meaning dipole moments
that are permanent and intrinsic\textemdash allows magnetic forces
to do work.\footnote{For a synopsis of the results obtained in this paper, see \citep{Barandes:2021omfaw}.}

We start by carefully reviewing the relevant ingredients of classical
mechanics, including the precise definition of mechanical work, as
well as the Lagrangian formulation and its generalizations. We then
turn to a detailed study of relativistic classical particles with
intrinsic spin and electric and magnetic multipole moments. Along
the way, we provide a new, classical argument for why a particle\textquoteright s
elementary dipole moments must be collinear with its spin axis. Next,
extending the work of \citep{BargmannMichelTelegdi:1959pppmhef,HansonRegge:1974rst,BalachandranMarmoSkagerstamStern:1983gsfb,Souriau:1997sds,Rivas:2002ktsp},
we couple the electromagnetic field to a classical relativistic particle
with intrinsic spin and elementary electric and magnetic dipole moments.
We derive the particle's equations of motion together with the overall
system's energy-momentum tensor and its angular-momentum flux tensor,
and then show both from the equations of motion and from local conservation
of energy and momentum that magnetic forces can do work on the particle
if its elementary magnetic dipole moment is nonzero. We conclude by
computing the system\textquoteright s Belinfante-Rosenfeld energy-momentum
tensor, which is another new result.

\subsection{Mechanical Preliminaries}

Recall that the net force $\vec F$ on a mechanical object is equal
to the instantaneous rate at which the object's momentum $\vec p$
changes with time $t$: 
\begin{equation}
\vec F=\frac{d\vec p}{dt}.\label{eq:DefForceFromMomentum}
\end{equation}
 Let $m$ be the object's inertial mass, let $\vec X$ be its position
vector, and let $\vec v\defeq d\vec X/dt$ be its velocity. In the
Newtonian case, the object's momentum is related to its velocity according
to 
\begin{equation}
\vec p\defeq m\vec v\quad\bracks{\textrm{Newtonian}},\label{eq:DefMomentumNewtonian}
\end{equation}
 meaning that under the assumption that $m$ is constant, \eqref{eq:DefForceFromMomentum}
becomes Newton's second law, 
\begin{equation}
\vec F=m\vec a,\label{eq:NewtonsSecondLaw}
\end{equation}
 with $\vec a\defeq d\vec v/dt$ the object's acceleration.

Still assuming the Newtonian case, the object's kinetic energy is
\begin{equation}
T\defeq\frac{1}{2}m\vec v^{2}=\frac{\vec p^{2}}{2m}\quad\bracks{\textrm{Newtonian}}.\label{eq:DefKineticEnergyNewtonian}
\end{equation}
A simple calculation then shows that the rate of change in the kinetic
energy of an object of constant mass $m$ is given by the dot product
of the object's velocity $\vec v$ and the force $\vec F$: 
\begin{equation}
\frac{dT}{dt}=\vec v\dotprod\frac{d\vec p}{dt}=\vec v\dotprod\vec F=\frac{d\vec X}{dt}\dotprod\vec F.\label{eq:DifferentialWorkEnergyTheorem}
\end{equation}

\subsection{The Definition of Mechanical Work}

\emph{By definition}, we say that a given force does mechanical work
on a classical object if the object moves through space and the vector
representing the force has a nonzero component along the object's
path.

More precisely, the work $W$ done by the force on the object is
the dot product of the force vector $\vec F$ and the object's incremental
displacement vector $d\vec X$, integrated over the total displacement
from the object's initial location $A$ to its final location $B$:
\begin{equation}
W\defeq\int_{A}^{B}\negthickspace d\vec X\dotprod\vec F.\label{eq:DefWork}
\end{equation}
Assuming for simplicity that $\vec F$ is the only force doing work
on the object, and integrating the relation \eqref{eq:DifferentialWorkEnergyTheorem}
over the time duration of the object's trajectory, we can use the
fundamental theorem of calculus to obtain the work-energy theorem,
\begin{equation}
W=\Delta T,\label{eq:WorkEnergyTheorem}
\end{equation}
 which establishes that the work $W$ done by the force $\vec F$
on the object translates into an overall change $\Delta T$ in the
object's kinetic energy $T$.

As a different question, one may ask whether a given force $\vec F$
acting on an object arises from some other source of energy, and,
if so, what that energy is and where it comes from. The simplest example
is provided by a conservative force, which is a force on an object
that is a function $\vec F\parens{\vec X}$ only of the object's instantaneous
position $\vec X$ and with the additional property that any work
\eqref{eq:DefWork} done by the force only ever depends on the endpoints
$A$ and $B$ of whatever arbitrary path the object takes.

On the one hand, forces that do work need not be conservative, as
dissipative forces like friction make clear. On the other hand, conservative
forces need not do work, such as a conservative force that acts centripetally
on an object and is therefore always perpendicular to the object's
motion, meaning that it has an always-vanishing dot product $d\vec X\dotprod\vec F\parens{\vec X}=0$
with the object's incremental displacements $d\vec X$.

Given a conservative force $\vec F\parens{\vec X}$, if we replace
the upper limit of integration in the definition \eqref{eq:DefWork}
of $W$ with a variable position $\vec X$, then the result is a well-defined
function of $\vec X$ that, together with an overall minus sign, defines
the object's potential energy $V\parens{\vec X}$ due to that force,
\begin{equation}
V\parens{\vec X}\defeq-\int^{\vec X}d\vec X^{\prime}\dotprod\vec F\parens{\vec X^{\prime}},\label{eq:DefPotentialEnergy}
\end{equation}
 where we neglect the lower limit of integration because it merely
determines an irrelevant additive constant. Taking the gradient of
both sides of this definition \eqref{eq:DefPotentialEnergy} of $V\parens{\vec X}$,
we see that we can express a conservative force as the negative gradient
of its corresponding potential energy: 
\begin{equation}
\vec F\parens{\vec X}=-\nabla V\parens{\vec X}.\label{eq:DefConservativeForceFromPotentialEnergy}
\end{equation}

Once again assuming for simplicity that $\vec F\parens{\vec X}$ is
the only force acting on the object, and combining the integral definition
\eqref{eq:DefWork} of the work done together with the relationship
\eqref{eq:DefConservativeForceFromPotentialEnergy} between the force
and its potential energy, we see that the work $W$ done by the force
on the object is equal to the overall change $\Delta V$ in the object's
potential energy: 
\begin{equation}
W=-\Delta V.\label{eq:WorkEqualsMinusChangePotential}
\end{equation}
 It follows from the work-energy theorem \eqref{eq:WorkEnergyTheorem},
$W=\Delta T$, that the sum of the change $\Delta T$ in the object's
kinetic energy and the change $\Delta V$ in the object's potential
energy is zero: 
\begin{equation}
\Delta T+\Delta V=\Delta\parens{T+V}=0.\label{eq:TotalChangeKineticPlusPotentialZero}
\end{equation}
 We therefore conclude that there exists an associated conserved total energy
$E$: 
\begin{equation}
E=T+V=\textrm{constant}.\label{eq:TotalMechanicalEnergyConserved}
\end{equation}
 Indeed, taking the time derivative of $E$, and using \eqref{eq:DifferentialWorkEnergyTheorem}
to calculate $dT/dt$ together with the chain rule to calculate $dV/dt$,
we have 
\begin{align}
\frac{dE}{dt} & =\frac{dT}{dt}+\frac{dV}{dt}\nonumber \\
 & =\vec v\dotprod\vec F+\frac{d\vec X}{dt}\dotprod\nabla V\nonumber \\
 & =\vec v\dotprod\vec F+\vec v\dotprod\parens{-\vec F}=0.\label{eq:ExplicitCalculationMechanicalEnergyConservation}
\end{align}

\subsection{The Maxwell Equations and the Lorentz Force Law}

We next review the fundamentals of the classical theory of electromagnetism,
taking this opportunity to establish the various conventions that
we will be using in this paper.\footnote{For more comprehensive pedagogical treatments, see \citep{Griffiths:2017ie,Jackson:1998ce,Vanderlinde:2005cet,Zangwill:2012me}.}

Working in SI units, we let $\epsilon_{0}$ and $\mu_{0}$ respectively
denote the permittivity of free space and the permeability of free
space. We use $\vec E=\parens{E_{x},E_{y},E_{z}}$ for the electric
field, $\vec B=\parens{B_{x},B_{y},B_{z}}$ for the magnetic field,
$\rho$ for the volume density of electric charge, and $\vec J=\parens{J_{x},J_{y},J_{z}}$
for the current density or charge flux density, meaning the rate
of charge flow per unit time, per unit cross-sectional area. We can
then write down the four Maxwell equations in their standard form:
\begin{align}
\nabla\dotprod\vec E & =\frac{\rho}{\epsilon_{0}},\label{eq:GaussEq}\\
\nabla\dotprod\vec B & =0,\label{eq:NoNameEq}\\
\nabla\!\crossprod\!\vec E & =-\frac{\partial\vec B}{\partial t},\label{eq:FaradayEq}\\
\nabla\!\crossprod\!\vec B & =\mu_{0}\vec J+\epsilon_{0}\mu_{0}\frac{\partial\vec E}{\partial t}.\label{eq:AmpereEq}
\end{align}
 We will respectively call these the electric Gauss equation, the
magnetic Gauss equation, the Faraday equation, and the Ampère equation.
The first and fourth equations contain the source functions $\rho$
and $\vec J$, and are called the inhomogeneous Maxwell equations.
The second and third equations do not involve source functions, and
are called the homogeneous Maxwell equations. Note that $\epsilon_{0}$,
$\mu_{0}$, and the speed of light $c$ are related by 
\begin{equation}
\frac{1}{\sqrt{\epsilon_{0}\mu_{0}}}=c.\label{eq:PermittivityPermeabilitySpeedOfLight}
\end{equation}

The Maxwell equations tell us how charged sources generate electric
and magnetic fields. The fields, in turn, cause changes to the motion
of those charged sources. To provide a precise formulation of this
latter statement, one traditionally supplements the Maxwell equations
with an additional axiom called the Lorentz force law, whose textbook
form expresses the force $\vec F$ on a particle of charge $q$ and
velocity $\vec v$ due to an external electric field $\vec E_{\textrm{ext}}$
and an external magnetic field $\vec B_{\textrm{ext}}$ as 
\begin{equation}
\vec F=q\vec E_{\textrm{ext}}+q\vec v\crossprod\vec B_{\textrm{ext}}.\label{eq:LorentzForceLaw}
\end{equation}
 The electric and magnetic forces on the particle are therefore given
individually by 
\begin{align}
\vec F_{\textrm{el}}\ \ \, & =q\vec E_{\textrm{ext}},\label{eq:LorentzForceLawElectric}\\
\vec F_{\textrm{mag}} & =q\vec v\crossprod\vec B_{\textrm{ext}}.\label{eq:LorentzForceLawMagnetic}
\end{align}
 Note that the particle's velocity $\vec v$ is assumed to be constant
here to avoid complications involving radiation and backreactive self-forces.

\subsection{Models of Magnetic Dipoles}

We will eventually show that magnetic forces can do work on certain
kinds of magnetic dipoles. First, however, we should take a moment
to explain why this claim has historically been questioned.

According to the usual Ampère model, classical magnetic dipoles are
\emph{composite} entities consisting of charged particles\textemdash that
is, electric monopoles\textemdash moving around in current loops.
For such a composite magnetic dipole, the textbook Lorentz force law
\eqref{eq:LorentzForceLaw} makes clear that magnetic forces cannot
do work. The simple reason is that the magnetic force $\vec F_{\textrm{mag}}$
on each electric monopole in a given current loop is proportional
to the cross product $\vec v\crossprod\vec B_{\textrm{ext}}$ of the
particle's velocity $\vec v\defeq d\vec X/dt$ and the external magnetic
field $\vec B_{\textrm{ext}}$, so the magnetic force $\vec F_{\textrm{mag}}$
is always perpendicular to the particle's incremental displacements
$d\vec X$. By its definition \eqref{eq:DefWork}, $W=\int d\vec X\dotprod\vec F$,
work is equal to the dot product of force and incremental displacement,
integrated over the full displacement. Because $d\vec X\dotprod\vec F_{\textrm{mag}}=0$,
the work done by the magnetic force in this context always vanishes.\footnote{For more detailed examples, see Section~8.3 of \citep{Griffiths:2017ie}.}

Notice also that the magnetic force $\vec F_{\textrm{mag}}=q\vec v\crossprod\vec B_{\textrm{ext}}$
on electric monopoles is explicitly velocity-dependent, and so cannot
represent a conventionally conservative force. By contrast, the electric
force $\vec F_{\textrm{el}}\parens{\vec X}=q\vec E_{\textrm{ext}}\parens{\vec X}$
due to a time-independent electric field $\vec E_{\textrm{ext}}\parens{\vec X}$
depends only on the electric monopole's position $\vec X$. Moreover,
the static version of the Faraday equation \eqref{eq:FaradayEq},
$\nabla\crossprod\vec E=0$, ensures that the electric force $\vec F_{\textrm{el}}$
is expressible in terms of a potential energy $V$ as $\vec F=-\nabla V$,
in keeping with \eqref{eq:DefConservativeForceFromPotentialEnergy},
so the static electric force on an electric monopole is conservative.

One could, in principle, evade the preceding conclusions about magnetic
forces by considering composite magnetic dipoles according to the
Gilbert model, in which the magnetic dipoles instead consist of pairs
of fundamental magnetic \emph{monopoles}. However, employing the
Gilbert model would require generalizing Maxwell's theory of electromagnetism
to include magnetic monopoles, as well as generalizing the Lorentz
force law accordingly to describe forces acting on them.

Experiments indicate that many kinds of particles, including electrons,
possess permanent, elementary magnetic \emph{dipole} moments that
do not seem to arise from underlying classical loops of current or
as pairs of magnetic monopoles. At a truly fundamental level, these
elementary magnetic dipole moments are quantum-mechanical in nature,
but, then, so is electric charge, and we obviously still include electric
charges as basic sources in Maxwell's classical theory of electromagnetism.

It is therefore worth studying how we might similarly include elementary
dipoles as basic sources in a classical extension of Maxwell's theory
of electromagnetism, as well as determine from first principles how
they should interact with electric and magnetic fields\textemdash without
assuming the textbook Lorentz force law \eqref{eq:LorentzForceLaw}
as one of our starting ingredients. Such an investigation could then
be expected to shed light on the specific issues of magnetic forces
and work done on elementary dipoles. 

Ultimately, we will show that if we are given an external electric
field $\vec E_{\textrm{ext}}$ and an external magnetic field $\vec B_{\textrm{ext}}$,
then the following generalization of the Lorentz force law describes
the corresponding electromagnetic force $\vec F$ that acts on a particle
with charge $q$, elementary electric dipole moment $\vecgreek{\pi}$,
and elementary magnetic dipole moment $\vecgreek{\mu}$ traveling
at a constant velocity $\vec v$ that is slow compared with the speed
of light $c$: 
\begin{equation}
\vec F=q\vec E_{\textrm{ext}}+q\vec v\crossprod\vec B_{\textrm{ext}}+\nabla\parens{\vecgreek{\pi}\dotprod\vec E_{\textrm{ext}}}+\nabla\parens{\vecgreek{\mu}\dotprod\vec B_{\textrm{ext}}}.\label{eq:LorentzForceLawNonRelApproxIntro}
\end{equation}
 On the one hand, this formula again implies that magnetic forces
on electric monopoles are proportional to $\vec v\crossprod\vec B_{\textrm{ext}}$
and are therefore incapable of doing work on them. On the other hand,
this argument does not hold for the term $\nabla\parens{\vecgreek{\mu}\dotprod\vec B_{\textrm{ext}}}$
describing the magnetic force on an elementary magnetic dipole, thereby
allowing magnetic forces to do work in that case. We will confirm
this last statement explicitly by deriving the force law \eqref{eq:LorentzForceLawNonRelApproxIntro}
in detail, first from the equations of motion for a particle with
elementary electric and magnetic dipole moments coupled to the electromagnetic
field, and then again from fundamental principles of local energy
and momentum conservation.

\subsection{The Lorentz-Covariant Formulation of Electromagnetism}

In order to establish the claimed expression \eqref{eq:LorentzForceLawNonRelApproxIntro}
for the appropriate generalization of the Lorentz force law without
assuming a composite model for dipoles, we will need to develop a
formulation of elementary dipoles within the classical theory of electromagnetism.
More broadly, we will see that the Lorentz force law, rather than
being a separate postulate of the theory, emerges naturally from constraints
provided by relativity as well as by local conservation of energy
and momentum.

For these purposes, we will need to review the Lorentz-covariant formulation
of classical electromagnetism, once again taking the opportunity to
establish our notational conventions.\footnote{See Chapter 12 of \citep{Jackson:1998ce} and Chapter 11 of \citep{Vanderlinde:2005cet}
for more extensive treatments.} Working always in Cartesian coordinates, we will use Latin indices
$i,j,k,l,\dotsc$ that each run through the three values $x,y,z$
for three-dimensional vectors and tensors, and we will use Greek indices
$\mu,\nu,\rho,\sigma,\dotsc$ that each run through the four values
$t,x,y,z$ for four-dimensional Lorentz vectors and Lorentz tensors.
We have four-dimensional spacetime coordinates 
\begin{align}
x^{\mu} & =\parens{x^{t},x^{x},x^{y},x^{z}}^{\mu}\defeq\parens{c\,t,x,y,z}^{\mu}\nonumber \\
 & =\parens{c\,t,\vec x}^{\mu}\label{eq:Def4DCartesianCoordinates}
\end{align}
 and four-dimensional spacetime derivatives 
\begin{align}
\partial_{\mu} & \defeq\frac{\partial}{\partial x^{\mu}}=\parens{\partial_{t},\partial_{x},\partial_{y},\partial_{z}}_{\mu}\nonumber \\
 & =\biggparens{\frac{1}{c}\frac{\partial}{\partial t},\frac{\partial}{\partial x},\frac{\partial}{\partial y},\frac{\partial}{\partial z}}_{\mu}\nonumber \\
 & =\biggparens{\frac{1}{c}\frac{\partial}{\partial t},\nabla}_{\mu},\label{eq:Def4DSpacetimeDerivative}
\end{align}
 and we will follow the standard Einstein summation convention in
which we implicitly sum all repeated upper-lower index pairs over
their full range of values. We will employ the ``mostly positive''
Minkowski metric tensor, 
\begin{equation}
\eta_{\mu\nu}\defeq\eta^{\mu\nu}\defeq\begin{pmatrix}-1 & 0 & 0 & 0\\
0 & +1 & 0 & 0\\
0 & 0 & +1 & 0\\
0 & 0 & 0 & +1
\end{pmatrix}_{\mathclap{\mu\nu}},\label{eq:MinkMetric}
\end{equation}
 which means that if we raise or lower a Lorentz index on a Lorentz
four-vector $v^{\mu}$ (or, more generally, on a Lorentz tensor $\tud T{\mu\nu\cdots}{\rho\sigma\cdots}$)
according to 
\begin{equation}
\left.\begin{aligned}v_{\mu} & =\eta_{\mu\nu}v^{\nu},\\
v^{\mu} & =\eta^{\mu\nu}v_{\nu},
\end{aligned}
\quad\right\} \label{eq:DefRaisingLowerIndices}
\end{equation}
 then raising or lowering a $t$ index entails a change in overall
sign, whereas raising or lowering an $x$, $y$, or $z$ index has
no effect: 
\begin{equation}
\left.\begin{aligned}v^{t} & =-v_{t},\\
v^{x} & =\ \ v_{x},\\
v^{y} & =\ \ v_{y},\\
v^{z} & =\ \ v_{z}.
\end{aligned}
\quad\right\} \label{eq:DefRaisingLowerIndicesExplicit}
\end{equation}

As in \citep{Barandes:2019mcl}, we introduce a set of matrices $\tud{\bracks{\sigma_{\mu\nu}}}{\alpha}{\beta}$
called the Lorentz generators, 
\begin{equation}
\tud{\bracks{\sigma_{\mu\nu}}}{\alpha}{\beta}=-i\delta_{\mu}^{\alpha}\eta_{\nu\beta}+i\eta_{\mu\beta}\delta_{\nu}^{\alpha},\label{eq:LorentzGeneratorsMixedIndices}
\end{equation}
 which have the commutation relations 
\begin{align}
 & \bracks{\sigma_{\mu\nu},\sigma_{\rho\sigma}}\defeq\sigma_{\mu\nu}\sigma_{\rho\sigma}-\sigma_{\rho\sigma}\sigma_{\mu\nu}\nonumber \\
 & \qquad=i\eta_{\mu\rho}\sigma_{\nu\sigma}-i\eta_{\mu\sigma}\sigma_{\nu\rho}-i\eta_{\nu\rho}\sigma_{\mu\sigma}+i\eta_{\nu\sigma}\sigma_{\mu\rho},\label{eq:LorentzGeneratorsCommutator}
\end{align}
 form a basis for all antisymmetric Lorentz tensors with two indices,
\begin{equation}
A^{\alpha\beta}=-A^{\beta\alpha}=\frac{i}{2}A^{\mu\nu}\bracks{\sigma_{\mu\nu}}^{\alpha\beta},\label{eq:AntisymmTensorFromLorentzGeneratorsBasis}
\end{equation}
 and satisfy the key identities 
\begin{equation}
\frac{1}{2}\Trace\bracks{\sigma^{\mu\nu}\sigma_{\rho\sigma}}=i\bracks{\sigma_{\rho\sigma}}^{\mu\nu}\label{eq:LorentzGeneratorsTraceProd}
\end{equation}
 and 
\begin{equation}
\frac{1}{2}\Trace\bracks{\sigma^{\mu\nu}A}=iA^{\mu\nu}.\label{eq:LorentzGeneratorsTracePeelsOffAntisymmTensor}
\end{equation}
 We can express any Lorentz-transformation matrix $\Lambda_{\textrm{inf}}$
that differs infinitesimally from the identity as 
\begin{equation}
\Lambda_{\textrm{inf}}=1-\frac{i}{2}d\theta^{\mu\nu}\sigma_{\mu\nu}.\label{eq:InfinitesimalLorentzTransfFromGenerators}
\end{equation}
 Here $d\theta^{\mu\nu}=-d\theta^{\nu\mu}$ is an antisymmetric array
of small parameters given by 
\begin{equation}
d\theta^{\mu\nu}=\begin{pmatrix}0 & d\eta_{x} & d\eta_{y} & d\eta_{z}\\
-d\eta_{x} & 0 & d\theta_{z} & -d\theta_{y}\\
-d\eta_{y} & -d\theta_{z} & 0 & d\theta_{x}\\
-d\eta_{z} & d\theta_{y} & -d\theta_{x} & 0
\end{pmatrix}^{\mathclap{\mu\nu}},\label{eq:InfinitesimalBoostAngleTensor}
\end{equation}
 and describes a passive boost in the direction of the three-vector
$d\vecgreek{\eta}\defeq\parens{d\theta^{tx},d\theta^{ty},d\theta^{tz}}$
with magnitude $\verts{d\vecgreek{\eta}}$, together with a passive
rotation around the direction of the three-vector $d\vecgreek{\theta}\defeq\parens{d\theta^{yz},d\theta^{zx},d\theta^{xy}}$
by an angle $\verts{d\vecgreek{\theta}}$.

The electric field $\vec E=\parens{E_{x},E_{y},E_{z}}$ and magnetic
field $\vec B=\parens{B_{x},B_{y},B_{z}}$ transform as three-vectors
under rotations, but they mix together in a complicated manner under
Lorentz boosts. We can correctly capture this transformation behavior
by packaging the electric and magnetic fields into an antisymmetric,
Lorentz-covariant tensor $F^{\mu\nu}$, called the Faraday tensor,
that is defined by 
\begin{equation}
F^{\mu\nu}\defeq\begin{pmatrix}0 & E_{x}/c & E_{y}/c & E_{z}/c\\
-E_{x}/c & 0 & B_{z} & -B_{y}\\
-E_{y}/c & -B_{z} & 0 & B_{x}\\
-E_{z}/c & B_{y} & -B_{x} & 0
\end{pmatrix}^{\mathclap{\mu\nu}}=-F^{\nu\mu}.\label{eq:FaradayTensor}
\end{equation}
 Introducing the totally antisymmetric, four-index Levi-Civita symbol,
\begin{align}
\epsilon_{\mu\nu\rho\sigma} & \defeq\begin{cases}
+1 & \textrm{for \ensuremath{\mu\nu\rho\sigma} an even permutation of \ensuremath{txyz}},\\
-1 & \textrm{for \ensuremath{\mu\nu\rho\sigma} an odd permutation of \ensuremath{txyz}},\\
0 & \textrm{otherwise}
\end{cases}\nonumber \\
 & =-\epsilon^{\mu\nu\rho\sigma},\label{eq:4DLeviCivita}
\end{align}
  the dual Faraday tensor $\tilde{F}_{\mu\nu}$ is defined according
to 
\begin{align}
\tilde{F}_{\mu\nu} & \defeq\frac{1}{2}\epsilon_{\mu\nu\rho\sigma}F^{\rho\sigma}=\begin{pmatrix}0 & B_{x} & B_{y} & B_{z}\\
-B_{x} & 0 & E_{z}/c & -E_{y}/c\\
-B_{y} & -E_{z}/c & 0 & E_{x}/c\\
-B_{z} & E_{y}/c & -E_{x}/c & 0
\end{pmatrix}_{\mathclap{\mu\nu}}\nonumber \\
 & =-\tilde{F}_{\nu\mu}.\label{eq:DualFaradayTensor}
\end{align}
 We collect the charge density $\rho$ and the current density (or
charge flux density) $\vec J$ into the Lorentz-covariant current density
defined by 
\begin{equation}
j^{\mu}\defeq\parens{\rho c,J_{x},J_{y},J_{z}}^{\mu},\label{eq:Def4DCurrentDensity}
\end{equation}
 meaning that 
\begin{equation}
j^{\mu}=\begin{cases}
\textrm{density of charge} & \textrm{for }\mu=t,\\
\textrm{flux density of charge} & \textrm{for }\mu=x,y,z.
\end{cases}\label{eq:InterpretationCurrentDensity}
\end{equation}

The Maxwell equations \eqref{eq:GaussEq}\textendash \eqref{eq:AmpereEq}
are then expressible in Lorentz-covariant form as the pair of tensor
equations 
\begin{align}
\partial_{\mu}F^{\mu\nu} & =-\mu_{0}j^{\nu},\label{eq:4DInhomogMaxwellEq}\\
\partial_{\mu}\tilde{F}^{\mu\nu} & =0,\label{eq:4DHomogMaxwellEq}
\end{align}
 the first of which encompasses the inhomogeneous Maxwell equations
\eqref{eq:GaussEq} and \eqref{eq:AmpereEq}, and the second of which
encompasses the homogeneous Maxwell equations \eqref{eq:NoNameEq}
and \eqref{eq:FaradayEq}. In addition, the second Lorentz-covariant
equation \eqref{eq:4DHomogMaxwellEq} is equivalent to the electromagnetic Bianchi identity:
\begin{equation}
\partial^{\mu}F^{\nu\rho}+\partial^{\rho}F^{\mu\nu}+\partial^{\nu}F^{\rho\mu}=0.\label{eq:EMBianchiIdentity}
\end{equation}

Taking the spacetime divergence of the inhomogeneous Maxwell equation
\eqref{eq:4DInhomogMaxwellEq} yields the equation of local current conservation,
\begin{equation}
\partial_{\mu}j^{\mu}=0,\label{eq:4DLocalCurrentConservationEq}
\end{equation}
 which, in three-vector notation, becomes the continuity equation
for electric charge: 
\begin{equation}
\frac{\partial\rho}{\partial t}=-\nabla\dotprod\vec J.\label{eq:ChargeContinuityEq}
\end{equation}
 This continuity equation also follows from taking the divergence
of the Ampère equation \eqref{eq:AmpereEq}, using the vector-calculus
identity $\nabla\dotprod\parens{\nabla\crossprod\vec B}=0$, and then
invoking the electric Gauss equation \eqref{eq:GaussEq}.

Meanwhile, by the Helmholtz theorem from vector calculus, the homogeneous
Maxwell equation \eqref{eq:4DHomogMaxwellEq}, $\partial_{\mu}\tilde{F}^{\mu\nu}=0$,
implies the existence of a four-vector field $A_{\mu}$, called the
electromagnetic gauge potential, in terms of which we can express
the Faraday tensor $F_{\mu\nu}$ as the following antisymmetric pair
of spacetime derivatives: 
\begin{equation}
F_{\mu\nu}=\partial_{\mu}A_{\nu}-\partial_{\nu}A_{\mu}.\label{eq:FaradayTensorFromDerivsGaugePot}
\end{equation}
 We give conventional names to the components of the gauge potential
$A_{\mu}$ according to 
\begin{equation}
A_{\mu}=\parens{-\Phi/c,\vec A}_{\mu},\label{eq:4DGaugePotAsComponents}
\end{equation}
 where $\Phi$ is called the scalar potential and $\vec A$ is called
the vector potential. A comparison between \eqref{eq:FaradayTensorFromDerivsGaugePot}
and the definition \eqref{eq:FaradayTensor} of the Faraday tensor
$F_{\mu\nu}$ then yields the following relationships between the
potentials $\Phi$ and $\vec A$ and the electromagnetic fields $\vec E$
and $\vec B$: 
\begin{align}
\vec E & =-\nabla\Phi-\frac{\partial\vec A}{\partial t},\label{eq:ElectricFieldFromPot}\\
\vec B & =\nabla\crossprod\vec A.\label{eq:MagneticFieldFromPot}
\end{align}

The Faraday tensor $F_{\mu\nu}$ is unchanged under gauge transformations,
meaning any redefinition of the gauge potential $A_{\mu}$ by the
addition of the total spacetime derivative of an arbitrary scalar
function $f$: 
\begin{equation}
A_{\mu}\mapsto A_{\mu}+\partial_{\mu}f.\label{eq:4DGaugeTransformation}
\end{equation}
 Translating this gauge transformation into three-vector language,
the electromagnetic fields $\vec E$ and $\vec B$ are correspondingly
invariant under the combined transformation 
\begin{align}
\Phi & \mapsto\Phi-\frac{\partial f}{\partial t},\label{eq:3DGaugeTransformationScalarPot}\\
\vec A & \mapsto\vec A+\nabla f,\label{eq:3DGaugeTransformationVectorPot}
\end{align}
 where the minus sign in the first of these two formulas comes from
the minus sign in the definition \eqref{eq:4DGaugePotAsComponents}
relating $A_{t}$ and $\Phi$.

Because the electromagnetic fields $\vec E$ and $\vec B$ are unmodified
by simultaneously carrying out \eqref{eq:3DGaugeTransformationScalarPot}
and \eqref{eq:3DGaugeTransformationVectorPot}, gauge transformations
have no physical significance for observable quantities. Gauge transformations
therefore express a redundancy in the description of electromagnetism
when we formulate the theory in terms of potentials.

\section{The Lagrangian Formulation and its Generalizations}

In order to talk fundamentally about momentum, energy, force, and
work for systems that go beyond classical particles, such as the electromagnetic
field and our model for elementary dipoles, we will find it necessary
to employ the Lagrangian formulation of classical dynamics, which
we will review here.\footnote{We present a much more detailed survey in \citep{Barandes:2019mcl}.}

\subsection{The Lagrangian Formulation for a Classical System}

Consider a general classical system with degrees of freedom $q_{\alpha}$
and rates of change $\dot{q}_{\alpha}$, with an action functional
$S\bracks q$ given as the integral of the system's Lagrangian $L\parens{q,\dot{q},t}$
from an arbitrary initial time $t_{A}$ to an arbitrary final time
$t_{B}$: 
\begin{equation}
S\bracks q\defeq\int_{t_{A}}^{t_{B}}\negthickspace\negthickspace dt\,L.\label{eq:ClassicalActionFromLagrangian}
\end{equation}
 To say that this action functional encodes the system's dynamics
is to say that if we extremize $S\bracks q$ over all candidate trajectories
that share the same initial and final conditions, 
\begin{align}
 & \delta S\bracks q=0,\nonumber \\
 & \quad\textrm{with \ensuremath{q_{\alpha}\parens{t_{A}}} and \ensuremath{q_{\alpha}\parens{t_{B}}} held fixed for all \ensuremath{\alpha},}\label{eq:VariationActionFunctional}
\end{align}
 then the resulting Euler-Lagrange equations 
\begin{equation}
\frac{\partial L}{\partial q_{\alpha}}-\frac{d}{dt}\biggparens{\frac{\partial L}{\partial\dot{q}_{\alpha}}}=0\label{eq:EulerLagrangeEquations}
\end{equation}
 fully capture the system's equations of motion.

We define the system's canonical momentum $p_{\alpha}$ conjugate
to $q_{\alpha}$ in terms of the system's Lagrangian $L$ as the partial
derivative of $L$ with respect to the corresponding rate of change
$\dot{q}_{\alpha}$: 
\begin{equation}
p_{\alpha}\defeq\frac{\partial L}{\partial\dot{q}_{\alpha}}.\label{eq:DefCanonicalMomenta}
\end{equation}
 Assuming that we can solve these definitions for the rates of change
$\dot{q}_{\alpha}$ as functions of the canonical coordinates $q_{\alpha}$
and canonical momenta $p_{\alpha}$, the system's Hamiltonian $H\parens{q,p,t}$,
which roughly describes the system's energy, is then defined as a
function of the variables $q_{\alpha}$, $p_{\alpha}$, and $t$ as
the Legendre transformation 
\begin{align}
H & \defeq\sum_{\alpha}\frac{\partial L}{\partial\dot{q}_{\alpha}}\dot{q}_{\alpha}-L,\nonumber \\
 & =\sum_{\alpha}p_{\alpha}\dot{q}_{\alpha}-L.\label{eq:DefHamiltonian}
\end{align}
Employing the chain rule together with the Euler-Lagrange equations,
it follows from a straightforward calculation that the time derivative
of the Hamiltonian \eqref{eq:DefHamiltonian} is given by 
\begin{equation}
\frac{dH}{dt}=-\frac{\partial L}{\partial t}.\label{eq:TimeDerivHamiltonianFromLagr}
\end{equation}
 An important implication of this result is that if the system's Lagrangian
has no explicit dependence on the time $t$, meaning no dependence
on $t$ except arising through the degrees of freedom $q_{\alpha}\parens t$
for a given candidate trajectory, then the Hamiltonian is constant
in time, $dH/dt=0$. 

The Euler-Lagrange equations \eqref{eq:EulerLagrangeEquations} are
equivalent to the canonical equations of motion: 
\begin{equation}
\left.\begin{aligned}\dot{q}_{\alpha} & =\frac{\partial H}{\partial p_{\alpha}},\\
\dot{p}_{\alpha} & =-\frac{\partial H}{\partial q_{\alpha}}.
\end{aligned}
\qquad\right\} \label{eq:CanonicalEquationsOfMotion}
\end{equation}
 The canonical equations of motion therefore provide an alternative
way of encoding the system's dynamics, known as the Hamiltonian formulation.

\subsection{A Pair of Interacting Systems}

We will now study a simple example that will turn out to be highly
relevant to our work ahead.

In this example, which we will call the $xy$ system, we consider
a pair of subsystems, the first of which has a single degree of freedom
$x$, and the second of which has a single degree of freedom $y$.
We define the dynamics of the overall $xy$ system by choosing an
action functional 
\begin{equation}
S\bracks{x,y}\defeq\int dt\,L\label{eq:2DOFDerivExampleAction}
\end{equation}
 based on a Lagrangian defined by 
\begin{equation}
L\defeq\frac{1}{2}m\dot{x}^{2}+\frac{1}{2}M\dot{y}^{2}-ay^{2}-bxy+c\dot{x}y.\label{eq:2DOFDerivExampleLagrangian}
\end{equation}
 Here $m$, $M$, $a$, $b$, and $c$ are constants and, as usual,
dots denote time derivatives.  The constants $m$ and $M$ play the
role of inertial masses, and $a$, $b$, and $c$ can be interpreted
as coupling constants.

The Euler-Lagrange equations \eqref{eq:EulerLagrangeEquations} for
$x$ and $y$ then respectively yield the equations of motion 
\begin{align}
m\ddot{x} & =-by-c\dot{y},\label{eq:2DOFDerivExampleEOMFirst}\\
M\ddot{y} & =-2ay-bx+c\dot{x}.\label{eq:2DOFDerivExampleEOMSecond}
\end{align}
  The intuitive interpretation of these coupled differential equations
is that the right-hand sides describe interaction forces between the
two subsystems. 

On the one hand, notice that the force terms involving the constants
$a$ and $b$ are conservative, in the sense that they can be derived
from a potential energy 
\begin{equation}
V\parens{x,y}\defeq ay^{2}+bxy\label{eq:2DOFDerivExamplePotentialEnergy}
\end{equation}
 according to \eqref{eq:DefConservativeForceFromPotentialEnergy}:
\begin{align}
F_{x} & \defeq-\frac{\partial V}{\partial x}=-by,\label{eq:2DOFDerivExampleForceFirst}\\
F_{y} & \defeq-\frac{\partial V}{\partial y}=-2ay-bx.\label{eq:2DOFDerivExampleForceSecond}
\end{align}
 On the other hand, the force terms involving the constant $c$ depend
on the rates of change $\dot{x}$ and $\dot{y}$, and so are manifestly
not conservative.

The $xy$ system's canonical momenta are, from the general definition
\eqref{eq:DefCanonicalMomenta}, given by 
\begin{align}
p_{x} & \defeq\frac{\partial L}{\partial\dot{x}}=m\dot{x}+cy,\label{eq:2DOFDerivExampleCanonicalMomentumFirst}\\
p_{y} & \defeq\frac{\partial L}{\partial\dot{y}}=M\dot{y}.\label{eq:2DOFDerivExampleCanonicalMomentumSecond}
\end{align}
 Solving these equations to obtain $\dot{x}$ and $\dot{y}$ in terms
of the canonical variables $x$, $y$, $p_{x}$, and $p_{y}$, we
obtain 
\begin{align}
\dot{x} & =\frac{p_{x}-cy}{m},\label{eq:2DOFDerivExampleVelFromMomFirst}\\
\dot{y} & =\frac{p_{y}}{M}.\label{eq:2DOFDerivExampleVelFromMomSecond}
\end{align}
 Then a short calculation of the $xy$ system's Hamiltonian \eqref{eq:DefHamiltonian}
yields the result 
\begin{align}
H & \defeq p_{x}\dot{x}+p_{y}\dot{y}-L\nonumber \\
 & =\frac{\parens{p_{x}-cy}^{2}}{2m}+\frac{p_{y}^{2}}{2M}+ay^{2}+bxy.\label{eq:2DOFDerivExampleHamiltonian}
\end{align}
 One can verify that the canonical equations of motion \eqref{eq:CanonicalEquationsOfMotion}
derived from this Hamiltonian give back the original equations of
motion \eqref{eq:2DOFDerivExampleEOMFirst} and \eqref{eq:2DOFDerivExampleEOMSecond}.
Moreover, because the Lagrangian \eqref{eq:2DOFDerivExampleLagrangian}
has no explicit time dependence, $\partial L/\partial t=0$, our formula
\eqref{eq:TimeDerivHamiltonianFromLagr} guarantees that $H$ is constant
in time, 
\begin{equation}
\frac{dH}{dt}=0,\label{eq:2DOFDerivExampleGaugeExampleHamiltonianConserved}
\end{equation}
 as one can check explicitly. 

Substituting the formulas \eqref{eq:2DOFDerivExampleVelFromMomFirst}
for $\dot{x}$ and \eqref{eq:2DOFDerivExampleVelFromMomSecond} for
$\dot{y}$ into the Hamiltonian \eqref{eq:2DOFDerivExampleHamiltonian},
we can rewrite the Hamiltonian of the $xy$ system as 
\[
\frac{1}{2}m\dot{x}^{2}+\frac{1}{2}M\dot{y}^{2}+ay^{2}+bxy.
\]
 The first two terms look like Newtonian kinetic energies \eqref{eq:DefKineticEnergyNewtonian}
for the $x$ and $y$ subsystems individually, 
\begin{align}
T_{x} & \defeq\frac{1}{2}m\dot{x}^{2},\label{eq:2DOFDerivExampleKineticEnergyFirst}\\
T_{y} & \defeq\frac{1}{2}M\dot{y}^{2},\label{eq:2DOFDerivExampleKineticEnergySecond}
\end{align}
 and we recognize the final two terms as making up the potential energy
defined in \eqref{eq:2DOFDerivExamplePotentialEnergy}: 
\[
V\parens{x,y}=ay^{2}+bxy.
\]
 It is therefore natural to interpret $H$ as the total energy $E$
of the overall $xy$ system, 
\begin{align}
E\defeq H & =T_{x}+T_{y}+V\parens{x,y}\nonumber \\
 & =\frac{1}{2}m\dot{x}^{2}+\frac{1}{2}M\dot{y}^{2}+ay^{2}+bxy,\label{eq:2DOFDerivExampleTotalEnergyAsHamiltonian}
\end{align}
 where, from \eqref{eq:2DOFDerivExampleGaugeExampleHamiltonianConserved},
this energy is conserved: 
\begin{equation}
\frac{dE}{dt}=0.\label{eq:2DOFDerivExampleTotalEnergyConserved}
\end{equation}

Observe that we are always free to modify the definition \eqref{eq:2DOFDerivExampleTotalEnergyAsHamiltonian}
of the total energy $E$ by adding on terms with vanishing time derivative,
$d\parens{\cdots}/dt=0$, as such terms do not alter the conservation
equation \eqref{eq:2DOFDerivExampleTotalEnergyConserved}. Notice
also that the velocity-dependent interaction term $cx\dot{y}$ in
the Lagrangian \eqref{eq:2DOFDerivExampleLagrangian} does not appear
in the $xy$ system's conserved energy. 

Crucially, neither the $x$ subsystem nor the $y$ subsystem has a
separately conserved energy on its own. Furthermore, although we can
derive each of the two equations of motion \eqref{eq:2DOFDerivExampleEOMFirst}
and \eqref{eq:2DOFDerivExampleEOMSecond} individually as the canonical
equations of motion \eqref{eq:CanonicalEquationsOfMotion} for the
two individual Hamiltonians defined by 
\begin{align}
H_{x} & \defeq\frac{\parens{p_{x}-cy}^{2}}{2m}+bxy,\label{eq:2DOFDerivExampleHamiltonianFirst}\\
H_{y} & \defeq\frac{p_{y}^{2}}{2M}+ay^{2}+bxy-c\dot{x}y,\label{eq:2DOFDerivExampleHamiltonianSecond}
\end{align}
 the \emph{overall} $xy$ system's Hamiltonian \eqref{eq:2DOFDerivExampleHamiltonian}
is not equal to the sum of the two individual Hamiltonians $H_{x}$
and $H_{y}$, due to a double-counting of the interaction term $bxy$,
as well as due to the appearance of the velocity-dependent interaction
term $-c\dot{x}y$: 
\begin{equation}
H\ne H_{x}+H_{y}.\label{eq:2DOFDerivExampleTotalHamiltonianNotSumIndividuals}
\end{equation}

It is therefore up to us to decide whether to interpret the interaction
terms $ay^{2}$ and $bxy$ as belonging to one of the two individual
subsystems or the other. If, for example, we choose to regard the
$y$ subsystem as a ``force field'' acting on the $x$ subsystem,
then it would be natural to regard the interaction terms as part of
the energy of the $y$ subsystem, and we would correspondingly define
\emph{non-conserved} energies for the two subsystems individually
as 
\begin{align}
E_{x} & \defeq\frac{1}{2}m\dot{x}^{2},\label{eq:2DOFDerivExampleEnergyFirst}\\
E_{y} & \defeq\frac{1}{2}M\dot{y}^{2}+ay^{2}+bxy.\label{eq:2DOFDerivExampleEnergySecond}
\end{align}
 In this case, the conserved total energy \eqref{eq:2DOFDerivExampleTotalEnergyAsHamiltonian}
of the overall $xy$ system is the sum of these two energies: 
\begin{equation}
E=E_{x}+E_{y}.\label{eq:2DOFDerivExampleEnergyAsSumOfIndividuals}
\end{equation}

Notice that in splitting up $E$ as in \eqref{eq:2DOFDerivExampleEnergyAsSumOfIndividuals},
we have effectively taken the energy $E_{x}$ of the $x$ subsystem
to be solely its kinetic energy $T_{x}\defeq\parens{1/2}m\dot{x}^{2}$.
Additionally, the conservation law \eqref{eq:2DOFDerivExampleTotalEnergyConserved}
for the total energy $E$ immediately implies that the time derivative
of either $E_{x}$ or $E_{y}$ yields the \emph{opposite} of the rate
at which the other subsystem's energy is changing: 
\begin{equation}
\frac{dE_{x}}{dt}=-\frac{dE_{y}}{dt}.\label{eq:2DOFDerivExampleEnergyTimeDerivGivesOtherRate}
\end{equation}
 Observe that the left-hand side is given explicitly by 
\[
\frac{dE_{x}}{dt}=m\ddot{x}\dot{x}=\parens{\textrm{force}}\parens{\textrm{speed}},
\]
 so it precisely represents the rate at which work is being done on
the $x$ subsystem. 

Looking back at the velocity-dependent interaction term $c\dot{x}y$,
notice that we can use the product rule in reverse (that is, ``integration
by parts'' without an actual integration) to replace it with $-cx\dot{y}$,
up to a total time derivative: 
\begin{equation}
c\dot{x}y=-cx\dot{y}+\frac{d}{dt}\parens{cxy}.\label{eq:2DOFDerivExampleVelDepIntIBP}
\end{equation}
 By the fundamental theorem of calculus, a total time derivative in
a Lagrangian leads to terms in the action functional \eqref{eq:ClassicalActionFromLagrangian},
$S\defeq\int dt\,L$, that depend only on the fixed initial and final
conditions, and that are therefore constants that do not affect the
variation condition \eqref{eq:VariationActionFunctional} or the Euler-Lagrange
equations \eqref{eq:EulerLagrangeEquations}.

Indeed, one can verify explicitly that the alternative Lagrangian
defined by 
\begin{equation}
L^{\prime}\defeq\frac{1}{2}m\dot{x}^{2}+\frac{1}{2}M\dot{y}^{2}-ay^{2}-bxy-cx\dot{y},\label{eq:2DOFDerivExampleLagrangianAlt}
\end{equation}
 which differs from our original Lagrangian \eqref{eq:2DOFDerivExampleLagrangian}
by only the total time derivative of $cxy$, 
\begin{equation}
L=L^{\prime}+\frac{d}{dt}\parens{cxy},\label{eq:2DOFDerivExampleCompareLagrangians}
\end{equation}
 leads to precisely the same equations of motion \eqref{eq:2DOFDerivExampleEOMFirst}
and \eqref{eq:2DOFDerivExampleEOMSecond} for the $xy$ system as
before. The new Lagrangian $L^{\prime}$ yields respective canonical
momenta 
\begin{align}
p_{x}^{\prime} & \defeq\frac{\partial L^{\prime}}{\partial\dot{x}}=m\dot{x},\label{eq:2DOFDerivExampleCanonicalMomentumFirstAlt}\\
p_{y}^{\prime} & \defeq\frac{\partial L^{\prime}}{\partial\dot{y}}=M\dot{y}-cx,\label{eq:2DOFDerivExampleCanonicalMomentumSecondAlt}
\end{align}
 and Hamiltonian 
\begin{equation}
H^{\prime}=\frac{p_{x}^{\prime2}}{2m}+\frac{\parens{p_{y}^{\prime}+cx}^{2}}{2M}+ay^{2}+bxy,\label{eq:2DOFDerivExampleHamiltonianAlt}
\end{equation}
 which formally disagree with the canonical momenta \eqref{eq:2DOFDerivExampleCanonicalMomentumFirst}
and \eqref{eq:2DOFDerivExampleCanonicalMomentumSecond}, as well as
with the Hamiltonian \eqref{eq:2DOFDerivExampleHamiltonian} derived
from our original Lagrangian $L$. However, if we write the Hamiltonians
$H$ and $H^{\prime}$ in terms of $\dot{x}$ and $\dot{y}$, then
we see that they actually describe precisely the same conserved total
energy \eqref{eq:2DOFDerivExampleTotalEnergyAsHamiltonian} for the
$xy$ system, 
\[
E=\frac{1}{2}m\dot{x}^{2}+\frac{1}{2}M\dot{y}^{2}+ay^{2}+bxy,
\]
 thereby confirming that it does not physically matter whether we
use $L$ or $L^{\prime}$ as the $xy$ system's Lagrangian. In essence,
by switching from $L$ to $L^{\prime}$, we have merely carried out
a canonical transformation of the form 
\begin{equation}
\begin{pmatrix}x\\
p_{x}\\
y\\
p_{y}
\end{pmatrix}\mapsto\begin{pmatrix}x^{\prime}\\
p_{x}^{\prime}\\
y^{\prime}\\
p_{y}^{\prime}
\end{pmatrix}=\begin{pmatrix}x\\
p_{x}-cy\\
y\\
p_{y}-cx
\end{pmatrix},\label{eq:2DOFDerivExampleCanonicalTransformation}
\end{equation}
 but we obviously have not changed the underlying physics. 

\subsection{The Lagrangian Formulation for a Relativistic Massive Particle with
Spin}

As reviewed in \citep{Barandes:2019mcl}, one can reformulate the
Lagrangian description of a generic classical system in a manifestly
covariant language by introducing an arbitrary smooth, strictly monotonic
parametrization $t\mapsto t\parens{\lambda}$ in place of the time
$t$, in which case one arrives at the following alternative formula
for the system's Lagrangian: 
\begin{equation}
S\bracks{q,t}=\int d\lambda\,\mathscr{L}\parens{q,\dot{q},t,\dot{t}}.\label{eq:ReparamInvActionFunctional}
\end{equation}
 Here dots now denote derivatives with respect to the parameter $\lambda$,
and we have introduced a manifestly covariant Lagrangian according
to 
\begin{equation}
\mathscr{L}\parens{q,\dot{q},t,\dot{t}}\defeq\dot{t}\,L\parens{q,\dot{q}/\dot{t},t}.\label{eq:ReparamInvLagrangian}
\end{equation}
 This formalism puts the system's degrees of freedom $q_{\alpha}$
and the time $t$ on a similar footing, with the system's original
Hamiltonian $H$ expressible as the ``canonical momentum'' conjugate
to $-t$.

We are now ready to turn to the Lagrangian formulation for a relativistic
point particle with spin. We will need to be careful to distinguish
between the coordinates $x^{\mu}$ of arbitrary points in spacetime\textemdash such
as in the arguments of field variables\textemdash and the specific
coordinates $X^{\mu}$ of our particle's location in spacetime. We
will therefore continue to use capital letters for the particle's
spacetime coordinates, 
\begin{equation}
X^{\mu}\parens{\lambda}=\parens{c\,T\parens{\lambda},\vec X\parens{\lambda}}^{\mu},\label{eq:ElemMultipoleCoordDOF}
\end{equation}
 where $\lambda$ is a smooth, strictly monotonic parameter for the
particle's four-dimensional worldline.\footnote{For maximum generality and to avoid introducing any unnecessary constraints
into the particle's Lagrangian formulation, it is convenient to wait
until after deriving the particle's equations of motion before imposing
the simplifying condition that $\lambda$ is the particle's proper
time $\tau$.} 

We will assume that the particle has a positive mass $m>0$, a future-directed
four-momentum $p^{\mu}$ whose temporal component $p^{t}>0$ encodes
the particle's relativistic-kinetic energy $E$ and whose spatial
components $\vec p=\parens{p_{x},p_{y},p_{z}}$ encode the particle's
relativistic three-momentum, 
\begin{equation}
p^{\mu}\defeq\parens{E/c,\vec p}^{\mu},\label{eq:DefFreeParticleFourMom}
\end{equation}
and an intrinsic spin that is encoded in an antisymmetric spin tensor
$S^{\mu\nu}=-S^{\nu\mu}$ whose independent components define a pair
of three-vectors 
\begin{align}
\vec S & \defeq\parens{S^{yz},S^{zx},S^{xy}},\label{eq:DefSpin3Vec}\\
\tilde{\vec S} & \defeq\parens{S^{tx},S^{ty},S^{tz}}.\label{eq:DefDualSpin3Vec}
\end{align}
 The particle's Pauli-Lubanski pseudovector is then given by 
\begin{equation}
W^{\mu}=-\frac{1}{2}\epsilon^{\mu\nu\rho\sigma}p_{\nu}S_{\rho\sigma}.\label{eq:PauliLubanski4Vec}
\end{equation}

As explained in \citep{Barandes:2019mcl}, a massive particle with
positive energy $E=p^{t}c>0$ is a classical system whose phase space
provides an irreducible representation (or, more precisely, a transitive
group action or homogeneous space) of the Poincaré group. In somewhat
more detail, we start from the unique reference state\footnote{We will eventually show that the fixed reference values $p_{0}^{\mu}$
and $S_{0}^{\mu\nu}$ can be taken to correspond to the particle's
rest frame. Keep in mind that up to this point in our discussion,
we have not yet provided a precise relationship between the particle's
four-momentum $p^{\mu}$ and its four-velocity $dX^{\mu}/d\lambda$.} 
\begin{equation}
\parens{X_{0},p_{0},S_{\refvalue}}\defeq\parens{0,\parens{mc,\vec 0},S_{\refvalue}}.\label{eq:ReferenceState}
\end{equation}
 Then all the other states $\parens{X,p,S}$ in the particle's phase
space are, by construction, related to this reference state by an
appropriate Poincaré transformation $\parens{a,\Lambda}\in\mathbb{R}^{4}\ltimes O\parens{1,3}$:
\begin{equation}
\parens{X,p,S}=\parens{a,\Lambda\parens{mc,\vec 0},\Lambda S_{\refvalue}\Lambda^{\transp}}.\label{eq:PhaseSpacePointFromPoincTransfFromRef}
\end{equation}
 In this formalism, the coordinates $X^{\mu}=a^{\mu}$ and the Lorentz-transformation
matrix $\tud{\Lambda}{\mu}{\nu}$, which all vary along the particle's
worldline, are treated as the particle's fundamental phase-space variables,
with the constraint that $\tud{\Lambda}{\mu}{\nu}$, as a Lorentz
transformation, must leave the Minkowski metric tensor \eqref{eq:MinkMetric}
invariant: 
\begin{equation}
\Lambda^{\transp}\eta\Lambda=\eta=\mathrm{diag}\parens{-1,+1,+1,+1}.\label{eq:LorentzTransfsPreserveMinkMetric}
\end{equation}

This irreducible representation or transitive group action of the
Poincaré group is characterized by the fixed scalar quantities 
\begin{align}
p^{2} & \defeq p_{\mu}p^{\mu}\defeq-m^{2}c^{2},\label{eq:Def4DMassSquaredAsInvariant}\\
W^{2} & \defeq W_{\mu}W^{\mu}\defeq w^{2},\label{eq:SquarePauliLubanskiAsInvariant}\\
\frac{1}{2}S^{2} & \defeq\frac{1}{2}S_{\mu\nu}S^{\mu\nu}\defeq s^{2}=\vec S^{2}-\tilde{\vec S}^{2},\label{eq:Def4DSpinSquaredAsInvariant}
\end{align}
 as well as the fixed pseudoscalar quantity 
\begin{equation}
\frac{1}{8}\epsilon_{\mu\nu\rho\sigma}S^{\mu\nu}S^{\rho\sigma}\defeq\tilde{s}^{2}=\vec S\dotprod\tilde{\vec S}.\label{eq:Def4DDualSpinSquaredAsInvariant}
\end{equation}
 The constancy of the quantities \eqref{eq:Def4DMassSquaredAsInvariant}\textendash \eqref{eq:Def4DDualSpinSquaredAsInvariant}
is a fundamental feature of the particle's phase space whether or
not interactions are present, and leads to several self-consistency
conditions, the most important of which is that the contraction of
the particle's four-momentum with its spin tensor must vanish:\footnote{Like the analogous Lorenz equation $\partial_{\mu}A^{\mu}=0$ in the
Proca field theory, as well as in Lorenz gauge in electromagnetism,
the condition \eqref{eq:FourMomSpinTensorZeroPhysicalCondition} ends
up eliminating unphysical spin states.} 
\begin{equation}
p_{\mu}S^{\mu\nu}=0.\label{eq:FourMomSpinTensorZeroPhysicalCondition}
\end{equation}

As shown in \citep{Barandes:2019mcl}, we can use the following manifestly
covariant action functional of the form \eqref{eq:ReparamInvActionFunctional}
for the case in which the particle is free from interactions: 
\begin{align}
 & S_{\textrm{particle}}\bracks{X,\Lambda}=\int d\lambda\,\mathscr{L}_{\textrm{particle}}\nonumber \\
 & \qquad\qquad=\int d\lambda\,\biggparens{p_{\mu}\dot{X}^{\mu}+\frac{1}{2}\Trace\bracks{S\dot{\Lambda}\Lambda^{-1}}}\nonumber \\
 & \qquad\qquad=\int d\lambda\,\biggparens{p_{\mu}\dot{X}^{\mu}+\frac{1}{2}S_{\mu\nu}\dot{\theta}^{\mu\nu}}.\label{eq:FreeParticleActionFunctionalWithSpin}
\end{align}
 The degrees of freedom in this description are the particle's spacetime
coordinates $X^{\mu}\parens{\lambda}$ and a variable Lorentz-transformation
matrix $\tud{\Lambda}{\mu}{\nu}\parens{\lambda}$. The particle's
four-momentum $p^{\mu}\parens{\lambda}$ and its spin tensor $S^{\mu\nu}\parens{\lambda}$
are given respectively in terms of their fixed reference values $p_{\refvalue}^{\mu}$
and $S_{\refvalue}^{\mu\nu}$ in the reference state \eqref{eq:ReferenceState}
according to 
\begin{align}
p^{\mu}\parens{\lambda} & \defeq\tud{\Lambda}{\mu}{\nu}\parens{\lambda}p_{\refvalue}^{\nu},\label{eq:4MomFromRefParam}\\
S^{\mu\nu}\parens{\lambda} & \defeq\tud{\Lambda}{\mu}{\rho}\parens{\lambda}S_{\refvalue}^{\rho\sigma}\tdu{\parens{\Lambda^{\transp}}}{\sigma}{\nu}\parens{\lambda}\nonumber \\
 & =-\frac{i}{2}\Trace\bracks{\sigma^{\mu\nu}\Lambda\parens{\lambda}S_{\refvalue}\Lambda^{-1}\parens{\lambda}}.\label{eq:SpinTensorFromRefParam}
\end{align}
 Note that neither $p^{\mu}\parens{\lambda}$ nor $S^{\mu\nu}\parens{\lambda}$
depends on the particle's spacetime degrees of freedom $X^{\mu}\parens{\lambda}$
before the equations of motion are imposed. Here, again, $\tud{\bracks{\sigma_{\mu\nu}}}{\alpha}{\beta}$
are the Lorentz generators \eqref{eq:LorentzGeneratorsMixedIndices},
and we can use our formula \eqref{eq:InfinitesimalLorentzTransfFromGenerators}
for an infinitesimal Lorentz transformation to express the derivative
of $\Lambda\parens{\lambda}$ with respect to the worldline parameter
$\lambda$ in terms of the rates of change $\dot{\theta}^{\mu\nu}$
in the corresponding boost and angular parameters as 
\begin{equation}
\dot{\Lambda}\parens{\lambda}=-\frac{i}{2}\dot{\theta}^{\mu\nu}\parens{\lambda}\sigma_{\mu\nu}\Lambda\parens{\lambda}.\label{eq:ParamDerivLorentzTransf}
\end{equation}

From the definition \eqref{eq:ReferenceState} of our reference state,
we see that the reference value of the particle's four-momentum is
\begin{equation}
p_{\refvalue}^{\mu}\defeq\parens{mc,\vec 0}^{\mu}=mc\,\delta_{t}^{\mu}.\label{eq:4MomRef}
\end{equation}
 It follows that the particle's four-momentum \eqref{eq:4MomFromRefParam}
is given for general states in the particle's phase space by 
\begin{equation}
p^{\mu}\parens{\lambda}=mc\,\tud{\Lambda}{\mu}t\parens{\lambda}.\label{eq:FourMomentumFromRest}
\end{equation}
 The self-consistency condition \eqref{eq:FourMomSpinTensorZeroPhysicalCondition}
then tells us that the reference value $S_{\refvalue}^{\mu\nu}$ of
the particle's spin tensor satisfies 
\begin{equation}
mc\,S_{\refvalue}^{t\nu}=0,\label{eq:MassivePosEnergyRefCondSpinTensor}
\end{equation}
 and therefore has the general form 
\begin{equation}
S_{\refvalue}^{\mu\nu}=\begin{pmatrix}0 & 0 & 0 & 0\\
0 & 0 & S_{\refvalue,z} & -S_{\refvalue,y}\\
0 & -S_{\refvalue,z} & 0 & S_{\refvalue,x}\\
0 & S_{\refvalue,y} & -S_{\refvalue,x} & 0
\end{pmatrix}^{\mathclap{\mu\nu}}.\label{eq:MassivePosEnergyRefSpinTensor}
\end{equation}

Notice that the reference value \eqref{eq:MassivePosEnergyRefSpinTensor}
of the particle's spin tensor therefore determines a spin three-vector
$\vec S_{0}\defeq\parens{S_{0,x},S_{0,y},S_{0,z}}$ that picks out
some specific direction in three-dimensional space. Hence, the particle's
reference state \eqref{eq:ReferenceState} spontaneously breaks the
full three-dimensional rotation group down to just the subgroup of
rotations around the axis defined by $\vec S_{0}$.

\subsection{The Limit of Vanishing Spin}

We now specialize momentarily to the case of a free particle without
spin, $S^{\mu\nu}=0$. In principle, we can then solve the condition
$p^{2}\defeq-m^{2}c^{2}$ from \eqref{eq:Def4DMassSquaredAsInvariant}
for $p^{t}\defeq E/c$ to obtain the mass-shell relation 
\begin{equation}
E^{2}=\vec p^{2}c^{2}+m^{2}c^{4}.\label{eq:MassShellRelation}
\end{equation}
 Setting our parameter $\lambda\defeq t$ to be the background time
coordinate and switching back to the traditional, non-covariant Lagrangian
formulation, we end up with the Hamiltonian 
\begin{equation}
H=E=\sqrt{\vec p^{2}c^{2}+m^{2}c^{4}}.\label{eq:ReparamExampleHamiltonian}
\end{equation}
 The canonical equations of motion \eqref{eq:CanonicalEquationsOfMotion}
derived from this Hamiltonian then imply that the individual components
of the particle's three-velocity, 
\begin{equation}
\vec v\defeq\frac{d\vec X}{dt}=\parens{v_{x},v_{y},v_{z}},\label{eq:FreeParticleDef3Vel}
\end{equation}
 are given by 
\begin{equation}
v_{i}\defeq\frac{dX^{i}}{dt}=\frac{\partial H}{\partial p^{i}}.\label{eq:FreeParticle3VelFromCanEOM}
\end{equation}
 These equations yield the following relationship between the particle's
three-velocity $\vec v$, its three-momentum $\vec p$, and its energy
$E$: 
\begin{equation}
\vec v=\frac{\vec pc^{2}}{E}=\frac{\vec pc^{2}}{\sqrt{\vec p^{2}c^{2}+m^{2}c^{4}}}.\label{eq:RelativisticVelocityMomentumEnergyRelationship}
\end{equation}
  Solving for $\vec p$ in terms of $\vec v$ gives the formula 
\begin{equation}
\vec p=\gamma m\vec v,\label{eq:FreeParticle3MomentumRelativistic}
\end{equation}
 where the Lorentz factor $\gamma$ is defined by 
\begin{equation}
\gamma\defeq\frac{1}{\sqrt{1-\vec v^{2}/c^{2}}}.\label{eq:DefLorentzFactor}
\end{equation}
 Using $\gamma$, we can also express the particle's relativistic
energy $E$ as 
\begin{equation}
E=\gamma mc^{2},\label{eq:FreeParticleRelativisticEnergy}
\end{equation}
  and so we find that the four-momentum \eqref{eq:DefFreeParticleFourMom}
takes the form 
\begin{equation}
p^{\mu}=\parens{E/c,\vec p}^{\mu}=\parens{\gamma mc,\gamma m\vec v}^{\mu}=mu^{\mu}.\label{eq:FreeParticleFourMomentumFrom4Vel}
\end{equation}
 Here $u^{\mu}$ is the particle's normalized four-velocity, 
\begin{equation}
u^{\mu}\defeq\parens{\gamma c,\gamma\vec v}^{\mu}=\gamma\frac{dX^{\mu}}{dt},\label{eq:ParticleFourVelocity}
\end{equation}
 where by ``normalized,'' we mean that $u^{\mu}$ satisfies the
normalization condition 
\begin{equation}
u^{2}\defeq u_{\mu}u^{\mu}=-c^{2}.\label{eq:ParticleFourVelocityNormalization}
\end{equation}

It then follows from a straightforward calculation that the particle's
action functional \eqref{eq:FreeParticleActionFunctionalWithSpin}
reduces to the non-covariant form 
\begin{align}
S_{\textrm{particle}}\bracks{\vec X} & =\int dt\,p_{\mu}\frac{dX^{\mu}}{dt}=-mc^{2}\int dt/\gamma\nonumber \\
 & =-mc^{2}\int dt\,\sqrt{1-\vec v^{2}/c^{2}}.\label{eq:FreeRelativisticParticleActionFunctionalCoordTime}
\end{align}
By another calculation, one can also show that in the non-relativistic
limit, $\vec v^{2}\ll c^{2}$, \eqref{eq:FreeRelativisticParticleActionFunctionalCoordTime}
reduces to the action functional for a Newtonian particle with Lagrangian
$\parens{1/2}m\vec v^{2}-mc^{2}$, describing a particle with kinetic
energy $\parens{1/2}m\vec v^{2}$ and ``intrinsic potential energy''
$mc^{2}$. 

By definition, the squared proper-time interval $d\tau^{2}$ is the
infinitesimal squared arc length of the particle's worldline, up to
a factor of $-c^{2}$, so 
\begin{align*}
-c^{2}d\tau^{2} & =\eta_{\mu\nu}dX^{\mu}dX^{\nu}\\
 & =\eta_{\mu\nu}\parens{c\,dT,d\vec X}^{\mu}\parens{c\,dT,d\vec X}^{\nu}\\
 & =-c^{2}dT^{2}+d\vec X^{2}\\
 & =-c^{2}dt^{2}\parens{1-\vec v^{2}/c^{2}}.
\end{align*}
 We therefore obtain the familiar time-dilation formula 
\begin{equation}
d\tau=\frac{dt}{\gamma},\label{eq:TimeDilation}
\end{equation}
 so we can write the particle's normalized four-velocity \eqref{eq:ParticleFourVelocity}
as 
\begin{equation}
u^{\mu}=\frac{dX^{\mu}}{d\tau},\label{eq:ParticleFourVelocityFromProperTime}
\end{equation}
and we can compactly express the formula \eqref{eq:FreeRelativisticParticleActionFunctionalCoordTime}
for the particle's action functional as the particle's Lorentz-invariant,
integrated proper time $\int d\tau$, up to a proportionality factor
of $-mc^{2}$: 
\begin{equation}
S_{\textrm{particle}}\bracks{\vec X}\defeq-mc^{2}\int d\tau.\label{eq:FreeRelativisticParticleActionFunctionalProperTime}
\end{equation}

It is important to note that if a particle with intrinsic spin $S^{\mu\nu}\ne0$
and elementary dipole moments is interacting with a nonvanishing electromagnetic
field, then the particle's four-momentum will not necessarily take
the familiar form \eqref{eq:FreeParticleFourMomentumFrom4Vel}, $p^{\mu}=mu^{\mu}$,
that holds for a free particle, as we will show explicitly later.

\subsection{The Dynamics of a Relativistic Massive Particle with Spin}

Once again allowing the particle to have a nonzero spin tensor, $S^{\mu\nu}\ne0$,
we can vary the particle's action functional \eqref{eq:FreeParticleActionFunctionalWithSpin},
\[
S_{\textrm{particle}}\bracks{X,\Lambda}=\int d\lambda\,\biggparens{p_{\mu}\dot{X}^{\mu}+\frac{1}{2}\Trace\bracks{S\dot{\Lambda}\Lambda^{-1}}},
\]
 to obtain the particle's equations of motion, in accordance with
the extremization condition \eqref{eq:VariationActionFunctional}.

Extremizing the particle's action functional with respect to its spacetime
coordinates $X^{\mu}$ yields 
\begin{equation}
\dot{p}^{\mu}=0.\label{eq:FreeRelativisticParticleFourMomEOM}
\end{equation}
 This equation of motion implies that the particle's energy and momentum
are constant in time, as would be expected for an isolated particle
that is not subject to external forces.

As shown in \citep{Barandes:2019mcl}, extremizing the particle's
action functional \eqref{eq:FreeParticleActionFunctionalWithSpin}
with respect to the variable Lorentz-transformation matrix $\tud{\Lambda}{\mu}{\nu}\parens{\lambda}$
yields 
\begin{equation}
\dot{J}^{\mu\nu}=\dot{L}^{\mu\nu}+\dot{S}^{\mu\nu}=0,\label{eq:FreeRelativisticParticleAngMomEOM}
\end{equation}
 where $J^{\mu\nu}=-J^{\nu\mu}$ is the particle's antisymmetric total
angular-momentum tensor, 
\begin{equation}
J^{\mu\nu}\defeq L^{\mu\nu}+S^{\mu\nu},\label{eq:FreeParticleTotAngMomTensor}
\end{equation}
 and $L^{\mu\nu}=-L^{\nu\mu}$ is the particle's antisymmetric orbital
angular-momentum tensor, 
\begin{equation}
L^{\mu\nu}\defeq X^{\mu}p^{\nu}-X^{\nu}p^{\mu}.\label{eq:FreeParticleOrbAngMomTensor}
\end{equation}
 The equation of motion \eqref{eq:FreeRelativisticParticleAngMomEOM}
tells us that the particle's total angular-momentum tensor is conserved,
as would be expected in the absence of external torques.

As derived in \citep{Barandes:2019mcl}, the four-momentum $p^{\mu}$
of a massive free particle with nonzero spin is related to its four-velocity
$u^{\mu}$ according to $p^{\mu}=mu^{\mu}\propto\dot{X}^{\mu}$, as
in \eqref{eq:FreeParticleFourMomentumFrom4Vel} for the case of vanishing
spin. It follows that the particle's orbital angular-momentum tensor
$L^{\mu\nu}$ is constant by itself, 
\begin{equation}
\dot{L}^{\mu\nu}=0,\label{eq:FreeRelativisticParticleOrbAngMomConst}
\end{equation}
 so the particle's spin tensor is likewise separately conserved, 
\begin{equation}
\dot{S}^{\mu\nu}=0.\label{eq:FreeRelativisticParticleSpinTensorConst}
\end{equation}

\subsection{The Lagrangian Formulation for Classical Field Theories and Electromagnetism}

The Lagrangian formulation naturally accommodates the case of a classical
field theory with local field degrees of freedom $\varphi_{\alpha}\parens x$
and an action functional $S\bracks{\varphi}$ defined in terms of
a Lagrangian density $\mathcal{L}\parens{\varphi,\partial\varphi,x}$
as 
\begin{equation}
S\bracks{\varphi}=\int dt\int d^{3}x\,\mathcal{L},\label{eq:ClassicalFieldActionFromLagrangianDensity}
\end{equation}
 where $d^{3}x$ denotes the usual three-dimensional volume element:
 
\begin{equation}
d^{3}x\defeq dx\,dy\,dz.\label{eq:Def3DVolumeElement}
\end{equation}
 The extremization condition \eqref{eq:VariationActionFunctional}
on the action functional $S\bracks{\varphi}$ yields a field-theoretic
generalization of the Euler-Lagrange equations \eqref{eq:EulerLagrangeEquations}
given by: 
\begin{equation}
\frac{\partial\mathcal{L}}{\partial\varphi_{\alpha}}-\partial_{\mu}\biggparens{\frac{\partial\mathcal{L}}{\partial\parens{\partial_{\mu}\varphi_{\alpha}}}}=0.\label{eq:EulerLagrangeEquationsField}
\end{equation}

We now turn to the specific case of electromagnetism. If we temporarily
assume the absence of electromagnetic sources, meaning that we take
the four-dimensional current density \eqref{eq:Def4DCurrentDensity}
to be zero, 
\begin{equation}
j^{\mu}\defeq\parens{\rho c,\vec J}^{\mu}=0,\label{eq:4DCurrentDensityNoSources}
\end{equation}
 then we can encode the Maxwell equations \eqref{eq:GaussEq}\textendash \eqref{eq:AmpereEq}
in a Lagrangian formulation using the Lorentz-invariant, translation-invariant,
gauge-invariant Lagrangian density 
\begin{equation}
\mathcal{L}_{\textrm{field}}\defeq-\frac{1}{4\mu_{0}}F^{\mu\nu}F_{\mu\nu}.\label{eq:MaxwellLagrangianDensityVacuum}
\end{equation}
 The corresponding action functional is then defined by 
\begin{align}
S_{\textrm{field}}\bracks A & \defeq\int dt\int d^{3}x\,\mathcal{L}_{\textrm{field}}\nonumber \\
 & =\int dt\int d^{3}x\,\biggparens{-\frac{1}{4\mu_{0}}F^{\mu\nu}F_{\mu\nu}},\label{eq:MaxwellActionInVacuum}
\end{align}
 where $F_{\mu\nu}=\partial_{\mu}A_{\nu}-\partial_{\nu}A_{\mu}$ from
\eqref{eq:FaradayTensorFromDerivsGaugePot} is the Faraday tensor,
and where we regard the gauge potential $A_{\mu}$ as constituting
the Maxwell theory's set of underlying degrees of freedom. Indeed,
the field-theoretic Euler-Lagrange equations \eqref{eq:EulerLagrangeEquationsField}
yield 
\begin{align*}
 & \frac{\partial\mathcal{L}_{\textrm{field}}}{\partial A_{\nu}}-\partial_{\mu}\biggparens{\frac{\partial\mathcal{L}_{\textrm{field}}}{\partial\parens{\partial_{\mu}A_{\nu}}}}\\
 & \qquad=0-\partial_{\mu}\biggparens{-\frac{1}{\mu_{0}}F^{\mu\nu}}=0,
\end{align*}
 which immediately gives us the inhomogeneous Maxwell equation \eqref{eq:4DInhomogMaxwellEq}
with vanishing current density $j^{\nu}=0$: 
\begin{equation}
\partial_{\mu}F^{\mu\nu}=0.\label{eq:4DInhomogMaxwellEqInVacuum}
\end{equation}
 The homogeneous Maxwell equation \eqref{eq:4DHomogMaxwellEq}, meanwhile,
follows immediately from the relation $F_{\mu\nu}=\partial_{\mu}A_{\nu}-\partial_{\nu}A_{\mu}$
together with the definition \eqref{eq:DualFaradayTensor} of the
dual Faraday tensor $\tilde{F}_{\mu\nu}\defeq\parens{1/2}\epsilon_{\mu\nu\rho\sigma}F^{\rho\sigma}$:
\begin{equation}
\partial_{\mu}\tilde{F}^{\mu\nu}=0.\label{eq:4DHomogMaxwellEqInVacuum}
\end{equation}

\section{Elementary Multipoles}

\subsection{The Multipole Expansion of the Current Density}

For our first step toward modeling electromagnetic multipoles\textemdash meaning
not just electric monopoles, but also electric and magnetic dipoles,
electric and magnetic quadrupoles, and higher multipoles\textemdash we
express the Lorentz-covariant current density $j^{\nu}$ from \eqref{eq:Def4DCurrentDensity}
as a series expansion of local terms with increasingly many spacetime
derivatives $\partial_{\mu}$, where the requirements of Lorentz covariance
dictate the schematic structure 
\begin{equation}
j^{\nu}=\parens{\cdots}^{\nu}+\partial_{\mu}\parens{\cdots}^{\mu\nu}+\partial_{\mu}\partial_{\rho}\parens{\cdots}^{\mu\rho\nu}+\dotsb.\label{eq:4DCurrentDensityDerivativeExpansionSchematic}
\end{equation}
 As we will see, the series \eqref{eq:4DCurrentDensityDerivativeExpansionSchematic}
represents a multipole expansion 
\begin{equation}
j^{\nu}=j_{\textrm{e}}^{\nu}+j_{\textrm{d}}^{\nu}+j_{\textrm{q}}^{\nu}+\dotsb,\label{eq:4DCurrentDensityMultipoleExpansion}
\end{equation}
 where each term $j_{\textrm{e}}^{\nu},j_{\textrm{d}}^{\nu},j_{\textrm{q}}^{\nu}$
has a specific physical interpretation.
\begin{itemize}
\item The four-vector $j_{\textrm{e}}^{\nu}$ represents the total contribution
to $j^{\nu}$ from charged sources whose spatial densities involve
no derivatives. We will show that $j_{\textrm{e}}^{\nu}$ describes
the distribution of electric monopoles throughout physical space.
\item The four-vector $j_{\textrm{d}}^{\nu}$ represents the net contribution
from all charged sources whose spatial densities involve a single
spacetime divergence. Reading off the expression for $j_{\textrm{d}}^{\nu}$
from the expansion \eqref{eq:4DCurrentDensityDerivativeExpansionSchematic},
we see that Lorentz covariance implies that $j_{\textrm{d}}^{\nu}$
is expressible in terms of a tensor field $M^{\mu\nu}$ according
to 
\begin{equation}
j_{\textrm{d}}^{\nu}=\partial_{\mu}M^{\mu\nu}.\label{eq:4DDipoleCurrentDensity}
\end{equation}
 We will show later that $j_{\textrm{d}}^{\nu}$ represents the spatial
distribution of electric and magnetic dipoles, so we will call $M^{\mu\nu}$
the dipole-density tensor. 
\item Similarly, the four-vector $j_{\textrm{q}}^{\nu}$ is given in terms
of a pair of spacetime divergences of a tensor field $N^{\mu\rho\nu}$,
\begin{equation}
j_{\textrm{q}}^{\nu}=\partial_{\mu}\partial_{\rho}N^{\mu\rho\nu},\label{eq:4DQuadrupoleCurrentDensity}
\end{equation}
 and represents the spatial distribution of electric and magnetic
quadrupoles.
\item Subsequent terms in the series \eqref{eq:4DCurrentDensityMultipoleExpansion}
represent still-higher multipoles and involve incrementally more spacetime
divergences.
\end{itemize}
We can now write the schematic multipole expansion \eqref{eq:4DCurrentDensityDerivativeExpansionSchematic}
in the more concrete form 
\begin{align}
j^{\nu} & =j_{\textrm{e}}^{\nu}+j_{\textrm{d}}^{\nu}+j_{\textrm{q}}^{\nu}+\dotsb\nonumber \\
 & =j_{\textrm{e}}^{\nu}+\partial_{\mu}M^{\mu\nu}+\partial_{\mu}\partial_{\rho}N^{\mu\rho\nu}+\dotsb.\label{eq:4DCurrentDensityDerivativeExpansionExplicit}
\end{align}
 To ensure individual local conservation laws for each category of
elementary multipole, we take the tensors $M^{\mu\nu},N^{\mu\sigma\nu},\dots$
to obey the antisymmetry conditions 
\begin{align}
M^{\mu\nu} & =-M^{\nu\mu},\label{eq:4DDipoleTensorAntisymmetry}\\
N^{\mu\rho\nu} & =-N^{\nu\rho\mu}=-N^{\mu\nu\rho},\label{eq:4DQuadrupoleTensorAntisymmetry}
\end{align}
 and so on. It then follows immediately from the symmetry $\partial_{\mu}\partial_{\nu}=\partial_{\nu}\partial_{\mu}$
of mixed partial derivatives that the current density for each kind
of multipole separately obeys its own local conservation equation.
That is, we have 
\begin{align}
\partial_{\nu}j_{\textrm{e}}^{\nu} & =0,\label{eq:4DLocalElectricMonopoleConservation}\\
\partial_{\nu}j_{\textrm{d}}^{\nu} & =0,\label{eq:4DLocalDipoleConservation}\\
\partial_{\nu}j_{\textrm{q}}^{\nu} & =0,\label{eq:4DLocalQuadrupoleConservation}
\end{align}
 and so forth.

Note that the local conservation law \eqref{eq:4DLocalDipoleConservation}
for the dipole current density $j_{\textrm{d}}^{\nu}$ is not related
to the fact that the elementary dipole moments of our particles are
permanent. Nor does one need to invoke quantum mechanics and quantization
of angular momentum to explain their permanence, either. In our model,
the intrinsic spin and the associated elementary dipole moments of
a classical particle are permanent, invariant features of the particle,
in the same sense that the rest mass of the particle is a permanent,
invariant feature. As detailed in \citep{Barandes:2019mcl}, the invariance
of a classical particle's rest mass and the invariance of its intrinsic
spin follow from group-theoretic considerations in constructing the
particle's phase space (or, in the analogous quantum case, the particle's
Hilbert space). That is, the particle's phase space simply lacks the
degrees of freedom that would be necessary to allow the rest mass
or the intrinsic spin of the particle to be able to change.

Notice that we can recast the multipole expansion \eqref{eq:4DCurrentDensityDerivativeExpansionExplicit}
as 
\begin{equation}
j^{\nu}=j_{\textrm{e}}^{\nu}+\partial_{\mu}\parens{M^{\mu\nu}+\partial_{\rho}N^{\mu\rho\nu}+\cdots}.\label{eq:4DCurrentDensityMultipoleExpansionOrganized}
\end{equation}
 Introducing the multipole-density tensor, 
\begin{equation}
Q^{\mu\nu}\defeq M^{\mu\nu}+\partial_{\rho}N^{\mu\rho\nu}+\dotsb,\label{eq:4DAbbrevMultipoleTensor}
\end{equation}
 which is antisymmetric on its two indices, 
\begin{equation}
Q^{\mu\nu}=-Q^{\nu\mu},\label{eq:MultipoleTensorAntisymmetric}
\end{equation}
 we can therefore write the multipole expansion for $j^{\nu}$ more
compactly as 
\begin{equation}
j^{\nu}=j_{\textrm{e}}^{\nu}+\partial_{\mu}Q^{\mu\nu}.\label{eq:4DAbbrevMultipoleExpansion}
\end{equation}

\subsection{The Auxiliary Faraday Tensor}

Correspondingly, we define the antisymmetric auxiliary Faraday tensor
$H^{\mu\nu}=-H^{\nu\mu}$ to absorb all source contributions from
dipoles and higher multipoles, together with the constant $\mu_{0}$:
\begin{align}
H^{\mu\nu} & \defeq\frac{1}{\mu_{0}}F^{\mu\nu}+Q^{\mu\nu}\nonumber \\
 & =\frac{1}{\mu_{0}}F^{\mu\nu}+M^{\mu\nu}+\partial_{\rho}N^{\mu\rho\nu}+\dotsb.\label{eq:Def4DAuxiliaryFaradayTensor}
\end{align}
 We can then re-cast the inhomogeneous Maxwell equation \eqref{eq:4DInhomogMaxwellEq},
$\partial_{\mu}F^{\mu\nu}=-\mu_{0}j^{\nu}$, in the alternative form
\begin{equation}
\partial_{\mu}H^{\mu\nu}=-j_{\textrm{e}}^{\nu},\label{eq:4DInhomogMaxwellEqFromAuxiliaryTensor}
\end{equation}
 where again $j_{\textrm{e}}^{\nu}$ represents contributions to the
current density that arise solely from electric monopoles: 
\begin{equation}
j_{\textrm{e}}^{\nu}=\parens{\rho_{\textrm{e}}c,\vec J_{\textrm{e}}}^{\nu}.\label{eq:4DCurrentDensityElectricMonopoles}
\end{equation}

The auxiliary Faraday tensor $H^{\mu\nu}$ can be expressed in terms
of the electric displacement field $\vec D\defeq\parens{H^{tx}/c,H^{ty}/c,H^{tz}/c}$
and the auxiliary magnetic field $\vec H\defeq\parens{H^{yz},H^{zx},H^{xy}}$
according to 
\begin{equation}
H^{\mu\nu}\defeq\begin{pmatrix}0 & cD_{x} & cD_{y} & cD_{z}\\
-cD_{x} & 0 & H_{z} & -H_{y}\\
-cD_{y} & -H_{z} & 0 & H_{x}\\
-cD_{z} & H_{y} & -H_{x} & 0
\end{pmatrix}^{\mathclap{\mu\nu}}.\label{eq:4DAuxiliaryFaradayTensorFrom3VecFields}
\end{equation}
 These definitions permit us to write the auxiliary version \eqref{eq:4DInhomogMaxwellEqFromAuxiliaryTensor}
of the inhomogeneous Maxwell equation in three-vector form as the
pair of equations 
\begin{align}
\nabla\dotprod\vec D & =\rho_{\textrm{e}},\label{eq:GaussEqFromElectricDisp}\\
\nabla\!\crossprod\!\vec H & =\vec J_{\textrm{e}}+\frac{\partial\vec D}{\partial t}.\label{eq:AmpereEqFromMagneticAuxiliary}
\end{align}
 These two equations can be used in place of the three-vector inhomogeneous
Maxwell equations \eqref{eq:GaussEq} (the electric Gauss equation)
and \eqref{eq:AmpereEq} (the Ampère equation).

The formulation of electromagnetism in terms of this pair of alternative
three-vector equations is particularly suited to the study of ``macroscopic''
electromagnetic fields in charged matter. In that case, the total
current density $j^{\nu}$ is regarded as a coarse-grained spatial
average over appropriately large regions of the physical material
in question, with the result that electromagnetic multipoles arise,
in part, \emph{emergently} from the averaging process. Indeed, in
textbooks, the equations \eqref{eq:GaussEqFromElectricDisp} and \eqref{eq:AmpereEqFromMagneticAuxiliary}
are conventionally derived from this sort of averaging.

In this paper, by contrast, we have obtained these equations by expanding
our fundamental charged sources as a series \eqref{eq:4DCurrentDensityDerivativeExpansionSchematic}
in spacetime derivatives and imposing Lorentz covariance. In this
way, we are expressly allowing for the possibility of \emph{elementary}
electromagnetic multipoles.

\subsection{The Lorentz-Invariant Pointlike Volume Density}

If we wish, we can regard our elementary electric monopoles as providing
a classical model of electrons and other elementary particles, and
our elementary magnetic dipoles as providing a classical model of
their magnetic dipole moments. In order to study the behavior of pointlike
electric monopoles and elementary multipoles in detail, we will need
to review the formalism of Dirac delta functions in three and four
dimensions.

Consider a product of three delta functions describing an abstract
volume density sharply localized at a spatial point $\vec x^{\prime}=\parens{x^{\prime},y^{\prime},z^{\prime}}$:
\begin{equation}
\delta^{3}\parens{\vec x-\vec x^{\prime}}\defeq\delta\parens{x-x^{\prime}}\,\delta\parens{y-y^{\prime}}\,\delta\parens{z-z^{\prime}}.\label{eq:3DProdDiracDeltas}
\end{equation}
 The defining feature of this three-dimensional delta function is
that its integral $\int d^{3}x\,\parens{\cdots}\defeq\int dx\,dy\,dz\,\parens{\cdots}$
over any spatial volume $\mathcal{V}$ containing the point $\vec x^{\prime}=\parens{x^{\prime},y^{\prime},z^{\prime}}$
yields the number $1$, whereas its integral over any spatial volume
not containing the point $\vec x^{\prime}$ yields $0$: 
\begin{equation}
\int_{\mathcal{V}}\!d^{3}x\,\delta^{3}\parens{\vec x-\vec x^{\prime}}=\begin{cases}
1 & \textrm{if \ensuremath{\mathcal{V}} contains \ensuremath{\vec x^{\prime}}},\\
0 & \textrm{if \ensuremath{\mathcal{V}} does not contain \ensuremath{\ensuremath{\vec x^{\prime}}}}.
\end{cases}\label{eq:3DDeltaFuncIntegralCases}
\end{equation}

We can extend this construction to four-dimensional spacetime. An
isolated spacetime event with coordinates 
\begin{equation}
x^{\prime\mu}=\parens{c\,t^{\prime},x^{\prime},y^{\prime},z^{\prime}}^{\mu}\label{eq:4DSourceSpacetimePoint}
\end{equation}
 corresponds to a product of four delta functions, 
\begin{align}
 & \delta^{4}\parens{x-x^{\prime}}\nonumber \\
 & \qquad\defeq\delta\parens{c\,t-c\,t^{\prime}}\,\delta\parens{x-x^{\prime}}\,\delta\parens{y-y^{\prime}}\,\delta\parens{z-z^{\prime}},\label{eq:Def4DDeltaFuncEvent}
\end{align}
 with the defining feature that its integral $\int d^{4}x\,\parens{\cdots}\defeq\int c\,dt\,dx\,dy\,dz\,\parens{\cdots}$
over any four-dimensional region $\mathcal{M}$ of spacetime yields
the number $1$ or $0$ depending on whether that region contains
the spacetime point labeled by $x^{\prime\mu}$: 
\begin{align}
 & \int_{\mathcal{M}}\negthickspace d^{4}x\,\delta^{4}\parens{x-x^{\prime}}\nonumber \\
 & \qquad=\begin{cases}
1 & \textrm{if \ensuremath{\mathcal{M}} contains \ensuremath{x^{\prime\mu}}},\\
0 & \textrm{if \ensuremath{\mathcal{M}} does not contain \ensuremath{x^{\prime\mu}}}.
\end{cases}\label{eq:4DDeltaFuncIntegralCases}
\end{align}
 Under an arbitrary Lorentz transformation 
\begin{equation}
x^{\mu}\mapsto\tud{\Lambda}{\mu}{\nu}x^{\nu},\label{eq:SpacetimeCoordsLorentzTransf}
\end{equation}
 the four-dimensional integration measure 
\begin{equation}
d^{4}x\defeq c\,dt\,dx\,dy\,dz\label{eq:4DLorentzInVolumeMeasure}
\end{equation}
 incurs a trivial Jacobian factor of 
\begin{equation}
\verts{\det\Lambda}=1,\label{eq:AbsValDetLorentzTransfEq1}
\end{equation}
 and is therefore invariant. The defining condition \eqref{eq:4DDeltaFuncIntegralCases}
then implies that the four-dimensional delta function $\delta^{4}\parens{x-x^{\prime}}$
is likewise invariant under Lorentz transformations.

Generalizing from an isolated spacetime event to the worldline trajectory
of a particle, we replace $x^{\prime\mu}=\parens{c\,t^{\prime},x^{\prime},y^{\prime},z^{\prime}}^{\mu}$
with appropriate coordinate functions 
\begin{equation}
X^{\mu}\parens{\lambda}=\parens{c\,T\parens{\lambda},X\parens{\lambda},Y\parens{\lambda},Z\parens{\lambda}}^{\mu}\label{eq:4DWorldlineSpacetimeCoordFuncs}
\end{equation}
 of a smooth, strictly monotonic parameter $\lambda$. Our Lorentz-invariant
four-dimensional delta function \eqref{eq:Def4DDeltaFuncEvent} becomes
\begin{align}
 & \delta^{4}\parens{x-X}\nonumber \\
 & \qquad\defeq\delta\parens{c\,t-c\,T}\,\delta\parens{x-X}\,\delta\parens{y-Y}\,\delta\parens{z-Z}.\label{eq:4DDeltaFunc}
\end{align}

Infinitesimal durations of the particle's Lorentz-invariant proper
time $\tau$ are related to corresponding intervals of the coordinate
time $t$ according to the usual formula \eqref{eq:TimeDilation}
for time dilation, 
\[
d\tau=\frac{dt}{\gamma},
\]
 where again $\gamma$ is the particle's Lorentz factor, defined as
in \eqref{eq:DefLorentzFactor} according to 
\[
\gamma\defeq\frac{1}{\sqrt{1-\vec v^{2}/c^{2}}}.
\]
 The integral of the product of the Lorentz-invariant quantity $dt/\gamma$
and the Lorentz-invariant delta function $\delta^{4}\parens{x-X\parens{\lambda}}$
over the particle's four-dimensional worldline is manifestly Lorentz
invariant: 
\begin{equation}
\int\frac{dt}{\gamma}\,\delta^{4}\parens{x-X}.\label{eq:LorentzInvWorldlineIntegral}
\end{equation}
  Evaluating this worldline integral explicitly yields a Lorentz-invariant
version of the three-dimensional delta function $\delta^{3}\parens{\vec x-\vec X\parens{\lambda}}$:
\begin{align}
 & \frac{1}{\gamma}\delta^{3}\parens{\vec x-\vec X}\nonumber \\
 & \qquad=\frac{1}{\gamma}\delta\parens{x-X}\,\delta\parens{y-Y}\,\delta\parens{z-Z}.\label{eq:LorentzInvSphSymmetricDensityDeltaFunction}
\end{align}

Because the special combination of $1/\gamma$ and $\delta^{3}\parens{\vec x-\vec X\parens{\lambda}}$
appearing in \eqref{eq:LorentzInvSphSymmetricDensityDeltaFunction}
maintains its form under Lorentz transformations, it represents the
appropriate Lorentz-invariant generalization of a pointlike volume
density. We can also understand the Lorentz invariance of \eqref{eq:LorentzInvSphSymmetricDensityDeltaFunction}
from the fact that under Lorentz transformations, $\delta^{3}\parens{\vec x-\vec X\parens{\lambda}}$
transforms like the inverse of a three-dimensional volume element
$d^{3}x$, and because $d^{3}x$ experiences Lorentz contractions
by $1/\gamma$, the three-dimensional delta function $\delta^{3}\parens{\vec x-\vec X\parens{\lambda}}$
\emph{grows} by a factor of $\gamma$, which is then compensated by
the $1/\gamma$ appearing in \eqref{eq:LorentzInvSphSymmetricDensityDeltaFunction}.

Notice that in the limiting case $\vec X\parens{\lambda}\to\vec x^{\prime}$
and $\vec v\to0$ in which the particle is at rest, we have $1/\gamma\to1$.
In this limit, \eqref{eq:LorentzInvSphSymmetricDensityDeltaFunction}
therefore reduces to the static three-dimensional delta function $\delta^{3}\parens{\vec x-\vec x^{\prime}}$
that we originally introduced in \eqref{eq:3DProdDiracDeltas}.

\subsection{Electric Monopoles}

We now have the tools necessary to model various pointlike sources
more precisely.

To start, we consider a pointlike electric monopole of charge $q$
at rest at a location $\vec x^{\prime}=\parens{x^{\prime},y^{\prime},z^{\prime}}$.
The electric monopole has charge density 
\begin{equation}
\rho_{\textrm{e}}\parens{\vec x}=q\,\delta^{3}\parens{\vec x-\vec x^{\prime}}\label{eq:ChargeDensityStaticElectricMonopole}
\end{equation}
 and vanishing current density 
\begin{equation}
\vec J_{\textrm{e}}\parens{\vec x}=0.\label{eq:CurrentDensityStaticElectricMonopole}
\end{equation}
 An elementary calculation using the Maxwell equations \eqref{eq:GaussEq}\textendash \eqref{eq:AmpereEq}
shows that the resulting electric field for all $\vec x\ne\vec x^{\prime}$
is directed outward (for $q>0$) or inward (for $q<0$) from the point
$\vec x^{\prime}$, with an inverse-square dependence on the distance
$\verts{\vec x-\vec x^{\prime}}$ from $\vec x^{\prime}$, whereas
the magnetic field vanishes: 
\begin{align}
\vec E & =\frac{1}{4\pi\epsilon_{0}}\frac{q}{\verts{\vec x-\vec x^{\prime}}^{2}}\,\vec e_{\vec x-\vec x^{\prime}},\label{eq:StaticElectricFieldFromElectricMonopole}\\
\vec B & =0.\label{eq:StaticMagneticFieldFromElectricMonopole}
\end{align}
 Here $\vec e_{\vec x-\vec x^{\prime}}$ is a three-dimensional unit
vector pointing in the direction from the source point $\vec x^{\prime}$
to the field point $\vec x$: 
\begin{equation}
\vec e_{\vec x-\vec x^{\prime}}\defeq\frac{\vec x-\vec x^{\prime}}{\verts{\vec x-\vec x^{\prime}}}.\label{eq:DefUnitVectorFromSourceToFieldPoint}
\end{equation}
We can therefore conclude that this source distribution describes
an electric monopole at rest at $\vec x^{\prime}$, as claimed.

The electric monopole has Lorentz-covariant current density 
\begin{align}
j_{\textrm{e}}^{\nu} & =\parens{\rho_{\textrm{e}}c,\vec J_{\textrm{e}}}^{\nu}\nonumber \\
 & =\parens{q\,\delta^{3}\parens{\vec x-\vec x^{\prime}}\,c,\vec 0}^{\nu}\nonumber \\
 & =q\,\parens{c,\vec 0}^{\nu}\,\delta^{3}\parens{\vec x-\vec x^{\prime}}.\label{eq:4DElectricMonopoleRestCurrentDensity}
\end{align}
 Identifying $u_{\textrm{rest}}^{\nu}=\parens{c,\vec 0}^{\nu}$ as
the electric monopole's normalized ($u_{\textrm{rest}}^{2}=-c^{2}$)
four-velocity \eqref{eq:ParticleFourVelocity} in its own rest frame,
and recalling our formula \eqref{eq:LorentzInvSphSymmetricDensityDeltaFunction}
for the Lorentz-invariant generalization of a three-dimensional delta
function, we can immediately write down the Lorentz-covariant current
density of a pointlike electric monopole of charge $q$ moving along
a trajectory $\vec X\parens t=\parens{X\parens t,Y\parens t,Z\parens t}$:
\begin{equation}
j_{\textrm{e}}^{\nu}\parens{\vec x,t}=qu^{\nu}\frac{1}{\gamma}\delta^{3}\parens{\vec x-\vec X}.\label{eq:4DElectricMonopoleCurrentDensity}
\end{equation}
 Observe that $q$ is a Lorentz scalar, $u^{\nu}$ is a Lorentz four-vector,
and the combination of $1/\gamma$ together with the three-dimensional
delta function $\delta^{3}\parens{\vec x-\vec X\parens t}$ is Lorentz
invariant, so \eqref{eq:4DElectricMonopoleCurrentDensity} is indeed
a Lorentz four-vector, as required.

Using the formula \eqref{eq:ParticleFourVelocity} for the electric
monopole's normalized four-velocity when it is in motion at a three-velocity
$\vec v=\parens{v_{x},v_{y},v_{z}}$, 
\begin{equation}
u^{\nu}=\parens{\gamma c,\gamma\vec v}^{\nu}=\gamma\frac{dX^{\nu}}{dt},\label{eq:4DElectricMonopole4VelFromCoordTimeDeriv}
\end{equation}
 where the derivative of $X^{\nu}$ is taken with respect to the coordinate
time $t$, we see that the factors of $\gamma$ in \eqref{eq:4DElectricMonopoleCurrentDensity}
cancel out, and thus our formula for the current density becomes 
\begin{align}
j_{\textrm{e}}^{\nu}\parens{\vec x,t} & =\parens{qc\,\delta^{3}\parens{\vec x-\vec X},q\vec v\,\delta^{3}\parens{\vec x-\vec X}}\nonumber \\
 & =q\frac{dX^{\nu}}{dt}\delta^{3}\parens{\vec x-\vec X},\label{eq:4DElectricMonopoleTravelingCurrentDensity}
\end{align}
 meaning that the charge density and current density are given respectively
by 
\begin{align}
\rho_{\textrm{e}}\parens{\vec x,t} & =q\,\delta^{3}\parens{\vec x-\vec X},\label{eq:ElectricMonopoleTravelingChargeDensity}\\
\vec J_{\textrm{e}}\parens{\vec x,t} & =q\vec v\,\delta^{3}\parens{\vec x-\vec X}.\label{eq:ElectricMonopoleTravelingCurrentDensity}
\end{align}
 In particular, these two functions are related according to 
\begin{equation}
\vec J_{\textrm{e}}=\rho_{\textrm{e}}\vec v.\label{eq:ElectricMonopoleRelationChargeCurrentDensityVelocity}
\end{equation}

\subsection{The Dipole-Density Tensor}

In contrast with the case of electric monopoles, we will see that
the formula \eqref{eq:ElectricMonopoleRelationChargeCurrentDensityVelocity}
does not hold for elementary dipoles and higher multipoles. This fact
that will turn out to have important implications for magnetic forces
and mechanical work.

We will be particularly interested in studying elementary dipoles.
To begin, we give names to the various components of the dipole-density
tensor $M^{\mu\nu}$ appearing in our expression \eqref{eq:4DDipoleCurrentDensity},
$j_{\textrm{d}}^{\nu}=\partial_{\mu}M^{\mu\nu}$, for the dipole-current
density. Remembering from \eqref{eq:4DDipoleTensorAntisymmetry} that
the dipole-density tensor is antisymmetric on its two indices, $M^{\mu\nu}=-M^{\nu\mu}$,
we name its components according to 
\begin{equation}
M^{\mu\nu}=\begin{pmatrix}0 & cP_{x} & cP_{y} & cP_{z}\\
-cP_{x} & 0 & -M_{z} & M_{y}\\
-cP_{y} & M_{z} & 0 & -M_{x}\\
-cP_{z} & -M_{y} & M_{x} & 0
\end{pmatrix}^{\mathclap{\mu\nu}}.\label{eq:4DipoleTensorFromPolarizationMagnetization}
\end{equation}
  Here $\vec P=\parens{M^{tx}/c,M^{ty}/c,M^{tz}/c}$ defines a three-vector
field called the polarization, which we will see describes the volume
density of electric dipoles, and $\vec M=\parens{M^{yz},M^{zx},M^{xy}}$
defines a three-vector field called the magnetization, which describes
the volume density of magnetic dipoles. The component combinations
that define $\vec P$ and $\vec M$ transform as three-vectors under
rotations, but transform as parts of the full antisymmetric tensor
$M^{\mu\nu}$ under Lorentz boosts.

Assuming that our sources include no multipole moments higher than
dipoles, the electric displacement field $\vec D$ and the auxiliary
magnetic field $\vec H$ introduced in \eqref{eq:4DAuxiliaryFaradayTensorFrom3VecFields}
are related to the electric field $\vec E$, the magnetic field $\vec B$,
the polarization $\vec P$, and the magnetization $\vec M$ through
the equations 
\begin{align}
\vec D & =\epsilon_{0}\vec E+\vec P,\label{eq:ElectricDisplacementFromElectricAndPolarization}\\
\vec H & =\frac{1}{\mu_{0}}\vec B-\vec M.\label{eq:AuxiliaryMagneticFromMagneticAndMagnetization}
\end{align}

Defining a charge density $\rho_{\textrm{d}}$ and three-vector current
density $\vec J_{\textrm{d}}=\parens{J_{\textrm{d},x},J_{\textrm{d},y},J_{\textrm{d},z}}$
from the components of the Lorentz-covariant dipole-current density
$j_{\textrm{d}}^{\nu}$ according to 
\begin{equation}
j_{\textrm{d}}^{\nu}=\parens{\rho_{\textrm{d}}c,\vec J_{\textrm{d}}}^{\mu},\label{eq:4DCurrentDensityDipoles}
\end{equation}
 it follows from a straightforward calculation starting with \eqref{eq:4DDipoleCurrentDensity},
$j_{\textrm{d}}^{\nu}=\partial_{\mu}M^{\mu\nu}$, that $\rho_{\textrm{d}}$
and $\vec J_{\textrm{d}}$ are related to the polarization $\vec P$
and magnetization $\vec M$ according to the pair of equations 
\begin{align}
\rho_{\textrm{d}} & =-\nabla\dotprod\vec P,\label{eq:DipoleChargeDensityFromPolarization}\\
\vec J_{\textrm{d}} & =\frac{\partial\vec P}{\partial t}+\nabla\crossprod\vec M.\label{eq:DipoleCurrentDensityFromMagnetization}
\end{align}
  Notice that these two formulas imply that $\rho_{\textrm{d}}$
and $\vec J_{\textrm{d}}$ automatically satisfy the continuity equation
\begin{equation}
\frac{\partial\rho_{\textrm{d}}}{\partial t}=-\nabla\dotprod\vec J_{\textrm{d}},\label{eq:3DMultipoleContinuityEq}
\end{equation}
 as was ultimately ensured by the local conservation equation \eqref{eq:4DLocalDipoleConservation}.
Observe also that $\rho_{\textrm{d}}$ and $\vec J_{\textrm{d}}$
are not related by a formula analogous to the equation \eqref{eq:ElectricMonopoleRelationChargeCurrentDensityVelocity},
$\vec J_{\textrm{e}}=\rho_{\textrm{e}}\vec v$, that held for the
case of electric monopoles.

\subsection{Composite Dipoles}

We can provide an intuitive explanation for why the formulas \eqref{eq:DipoleChargeDensityFromPolarization}
for $\rho_{\textrm{d}}$ and \eqref{eq:DipoleCurrentDensityFromMagnetization}
for $\vec J_{\textrm{d}}$ indeed describe dipoles, as claimed. For
this purpose, we momentarily put aside the case of \emph{elementary}
dipoles and consider instead a \emph{composite} electric dipole consisting
more fundamentally of a pair of electric monopoles with respective
charges $q>0$ and $-q<0$, located respectively at positions $x=d>0$
and $x=0$ on the $x$ axis. The charge density is then 
\[
\rho\parens{x,y,z}=\parens{+q}\,\delta\parens{x-d}\,\delta\parens y\,\delta\parens z+\parens{-q}\,\delta\parens x\,\delta\parens y\,\delta\parens z.
\]
 Letting $\vec d=\parens{d,0,0}$ denote the spatial displacement
vector extending from the negative electric monopole to the positive
electric monopole, we define the system's electric dipole moment by
\begin{equation}
\vecgreek{\pi}\defeq q\vec d.\label{eq:DefCompositeElectricDipoleMoment}
\end{equation}
 Taking the limit in which $d\to0$ and $q\to\infty$ with $\vecgreek{\pi}\defeq q\vec d$
held fixed at finite magnitude and direction, we can write our expression
for the charge density as 
\begin{align}
\rho\parens{x,y,z} & =qd\,\parens{-\delta^{\prime}\parens x}\,\delta\parens y\,\delta\parens z\nonumber \\
 & =-q\vec d\dotprod\nabla\parens{\delta\parens x\,\delta\parens y\,\delta\parens z}\vphantom{\frac{1}{1}}\nonumber \\
 & =-\nabla\dotprod\parens{\vecgreek{\pi}\,\delta^{3}\parens{\vec x}},\label{eq:CompositeElectricDipoleChargeDensity}
\end{align}
 which replicates \eqref{eq:DipoleChargeDensityFromPolarization},
$\rho_{\textrm{d}}=-\nabla\dotprod\vec P$, for a polarization $\vec P$
defined as the pointlike density $\vecgreek{\pi}\,\delta^{3}\parens{\vec x}$
that corresponds to the dipole moment $\vecgreek{\pi}=q\vec d$ of
the pair of electric point charges.

Under Lorentz boosts, the polarization transforms as part of the antisymmetric
Lorentz tensor $M^{\mu\nu}$ in \eqref{eq:4DipoleTensorFromPolarizationMagnetization},
whose form then dictates the formula \eqref{eq:DipoleCurrentDensityFromMagnetization}
for the current density $\vec J_{\textrm{d}}$. We can alternatively
understand the form of $\vec J_{\textrm{d}}$ by analogy with \emph{composite}
electric dipoles consisting of time-dependent pairs of electric monopoles
and \emph{composite} magnetic dipoles consisting of circulating loops
of electric current.

\subsection{Elementary Dipoles}

We can also study the case of a pointlike \emph{elementary} dipole
at rest at a position $\vec x^{\prime}=\parens{x^{\prime},y^{\prime},z^{\prime}}$.
We define the particle's elementary electric dipole moment $\vecgreek{\pi}$
and elementary magnetic dipole moment $\vecgreek{\mu}$ in terms of
the polarization $\vec P$ and magnetization $\vec M$ in the delta-function
limit as 
\begin{align}
\vec P\parens{\vec x} & =\vecgreek{\pi}\,\delta^{3}\parens{\vec x-\vec x^{\prime}},\label{eq:PolarizationFromPointlikeElectricDipole}\\
\vec M\parens{\vec x} & =\vecgreek{\mu}\,\delta^{3}\parens{\vec x-\vec x^{\prime}}.\label{eq:MagnetizationFromPointlikeMagneticDipole}
\end{align}
 From \eqref{eq:DipoleChargeDensityFromPolarization}, $\rho_{\textrm{d}}=-\nabla\dotprod\vec P$,
the corresponding charge density is precisely as in \eqref{eq:CompositeElectricDipoleChargeDensity}
from the composite case, 
\begin{align}
\rho_{\textrm{d}}\parens{\vec x} & =-\nabla\dotprod\parens{\vecgreek{\pi}\,\delta^{3}\parens{\vec x-\vec x^{\prime}}}\nonumber \\
 & =-\vecgreek{\pi}\dotprod\nabla\delta^{3}\parens{\vec x-\vec x^{\prime}}.\label{eq:ChargeDensityStaticElectricDipole}
\end{align}
 From \eqref{eq:DipoleCurrentDensityFromMagnetization}, the current
density $\vec J_{\textrm{d}}$ is 
\begin{align}
\vec J_{\textrm{d}}\parens{\vec x} & =\nabla\crossprod\parens{\vecgreek{\mu}\,\delta^{3}\parens{\vec x-\vec x^{\prime}}}\nonumber \\
 & =-\vecgreek{\mu}\crossprod\nabla\delta^{3}\parens{\vec x-\vec x^{\prime}}.\label{eq:CurrentDensityStaticMagneticDipole}
\end{align}
 Another elementary calculation using the Maxwell equations \eqref{eq:GaussEq}\textendash \eqref{eq:AmpereEq}
shows that the resulting electric field and magnetic field for all
$\vec x\ne\vec x^{\prime}$ have the standard inverse-cube dependence
that is characteristic of dipoles: 
\begin{align}
\vec E\parens{\vec x} & =\frac{1}{4\pi\epsilon_{0}}\frac{3\parens{\vecgreek{\pi}\dotprod\vec e_{\vec x-\vec x^{\prime}}}\vec e_{\vec x-\vec x^{\prime}}-\vecgreek{\pi}}{\verts{\vec x-\vec x^{\prime}}^{3}}-\frac{\vecgreek{\pi}}{3\epsilon_{0}}\delta^{3}\parens{\vec x-\vec x^{\prime}},\label{eq:StaticElectricFieldFromDipole}\\
\vec B\parens{\vec x} & =\,\,\frac{\mu_{0}}{4\pi}\,\,\,\frac{3\parens{\vecgreek{\mu}\dotprod\vec e_{\vec x-\vec x^{\prime}}}\vec e_{\vec x-\vec x^{\prime}}-\vecgreek{\mu}}{\verts{\vec x-\vec x^{\prime}}^{3}}+\frac{2\mu_{0}\vecgreek{\mu}}{3}\delta^{3}\parens{\vec x-\vec x^{\prime}}.\label{eq:StaticMagneticFieldFromDipole}
\end{align}
 Here the unit vector $\vec e_{\vec x-\vec x^{\prime}}$, originally
defined in \eqref{eq:DefUnitVectorFromSourceToFieldPoint}, is directed
from the source point $\vec x^{\prime}$ toward the field point $\vec x$,
and the delta-function contact terms ensure agreement with the homogeneous
Maxwell equations \eqref{eq:NoNameEq} and \eqref{eq:FaradayEq}.
We conclude that this source distribution indeed describes an elementary
dipole at rest at $\vec x^{\prime}$.

\section{Classical Electromagnetism with Elementary Dipoles}

Now that we have introduced sources into classical electromagnetism\textemdash namely,
electric monopoles, elementary dipoles, and higher multipoles\textemdash we
will need to determine the resulting dynamics. We will start by characterizing
the electromagnetic properties of elementary dipoles. We will later
move on to the Lagrangian formulation of the theory.

\subsection{Dynamical Elementary Dipoles}

Recall from its definition \eqref{eq:4DAbbrevMultipoleTensor} that
the multipole-density tensor $Q^{\mu\nu}=-Q^{\nu\mu}$ is given in
terms of the tensors $M^{\mu\nu},N^{\mu\rho\nu},\dotsc$ respectively
describing the densities of dipoles, quadrupoles, and higher multipoles
by 
\[
Q^{\mu\nu}\defeq M^{\mu\nu}+\partial_{\rho}N^{\mu\rho\nu}+\cdots.
\]
 For a pointlike charged particle with position $\vec X$, recall
that the electric-monopole current density $j_{\textrm{e}}^{\nu}$
is given in terms of the Lorentz four-vector $qu^{\nu}$ and the Lorentz-invariant,
three-dimensional delta function \eqref{eq:LorentzInvSphSymmetricDensityDeltaFunction}
according to \eqref{eq:4DElectricMonopoleCurrentDensity}, 
\[
j_{\textrm{e}}^{\nu}\parens{\vec x,t}=qu^{\nu}\frac{1}{\gamma}\delta^{3}\parens{\vec x-\vec X}.
\]
 Similarly, the density tensors $M^{\mu\nu},N^{\mu\rho\nu},\dotsc$
for such a particle are given in terms of Lorentz tensors $m^{\mu\nu},n^{\mu\rho\nu},\dotsc$
and the Lorentz-invariant, three-dimensional delta function according
to 
\begin{align}
M^{\mu\nu} & =m^{\mu\nu}\frac{1}{\gamma}\delta^{3}\parens{\vec x-\vec X},\label{eq:ElemDipoleTensorFromDensityDeltaFunc}\\
N^{\mu\rho\nu} & =n^{\mu\rho\nu}\frac{1}{\gamma}\delta^{3}\parens{\vec x-\vec X},\label{eq:ElemQuadrupoleTensorFromDensityDeltaFunc}
\end{align}
 and so forth. It follows that the particle's total multipole-density
tensor would be 
\begin{equation}
Q^{\mu\nu}=\frac{1}{\gamma}\parens{m^{\mu\nu}+n^{\mu\rho\nu}\partial_{\rho}+\cdots}\delta^{3}\parens{\vec x-\vec X}.\label{eq:ElemMultipoleTensorFromDensityDeltaFunc}
\end{equation}

To simplify our work ahead, we will assume that the particle does
not have elementary quadrupole or higher multipole moments.\footnote{See also \citep{GerochWeatherall:2018msbst} for supporting physical
arguments for this assumption.} In that case, $Q^{\mu\nu}$ reduces to the dipole-density tensor
\eqref{eq:ElemDipoleTensorFromDensityDeltaFunc} alone, 
\begin{equation}
Q^{\mu\nu}=M^{\mu\nu}=m^{\mu\nu}\frac{1}{\gamma}\delta^{3}\parens{\vec x-\vec X},\label{eq:ElemMultipoleTensorSimplifyToDipole}
\end{equation}
 where we will call the antisymmetric tensor $m^{\mu\nu}=-m^{\nu\mu}$
the particle's elementary dipole tensor.

Mimicking our formula \eqref{eq:4DipoleTensorFromPolarizationMagnetization}
relating the dipole-density tensor $M^{\mu\nu}$ to the polarization
$\vec P$ and magnetization $\vec M$, we define the particle's elementary electric-dipole moment
as $\vecgreek{\pi}\defeq\parens{m^{tx}/c,m^{ty}/c,m^{tz}/c}$ and
its elementary magnetic-dipole moment as $\vecgreek{\mu}\defeq\parens{m^{yz},m^{zx},m^{xy}}$,
so that these three-vectors are related to the particle's elementary
dipole tensor $m^{\mu\nu}$ according to 
\begin{equation}
m^{\mu\nu}\defeq\begin{pmatrix}0 & c\pi_{x} & c\pi_{y} & c\pi_{z}\\
-c\pi_{x} & 0 & -\mu_{z} & \mu_{y}\\
-c\pi_{y} & \mu_{z} & 0 & -\mu_{x}\\
-c\pi_{z} & -\mu_{y} & \mu_{x} & 0
\end{pmatrix}^{\mathclap{\mu\nu}}.\label{eq:4DElementaryDipoleTensorFromMoments}
\end{equation}
 In the particle's reference state, for which its four-momentum $p_{\refvalue}^{\mu}$
is \eqref{eq:4MomRef} and its spin tensor $S_{\refvalue}^{\mu\nu}$
is \eqref{eq:MassivePosEnergyRefSpinTensor}, we can introduce a pair
of purely spacelike four-vectors defined by 
\begin{align}
\pi_{\refvalue}^{\mu} & \defeq\parens{0,\vecgreek{\pi}_{\refvalue}}^{\mu},\label{eq:DefRefElectricDipole4VecFrom3Vec}\\
\mu_{\refvalue}^{\mu} & \defeq\parens{0,\vecgreek{\mu}_{\refvalue}}^{\mu}.\label{eq:DefRefMagneticDipole4VecFrom3Vec}
\end{align}
 As in \citep{GrallaHarteWald:2009rdesf}, we can then write the particle's
elementary dipole tensor in general as 
\begin{equation}
m^{\mu\nu}=\pi^{\mu\nu}+\mu^{\mu\nu},\label{eq:4DElementaryDipoleTensorFromElMagTensors}
\end{equation}
 with 
\begin{align}
\pi^{\mu\nu} & \defeq\frac{1}{mc}\parens{p^{\mu}\pi^{\nu}-p^{\nu}\pi^{\mu}},\label{eq:4DElementaryElDipoleTensorFrom4Vecs}\\
\mu^{\mu\nu} & \defeq\frac{1}{mc}\epsilon^{\mu\nu\rho\sigma}p_{\rho}\mu_{\sigma},\label{eq:4DElementaryMagDipoleTensorFrom4Vecs}
\end{align}
 where $\pi^{\nu}\parens{\lambda}$ and $\mu^{\mu}\parens{\lambda}$
are related to their reference values $\pi_{\refvalue}^{\nu}$ and
$\mu_{\refvalue}^{\nu}$ and to the particle's variable Lorentz-transformation
matrix $\tud{\Lambda}{\mu}{\nu}\parens{\lambda}$ according to 
\begin{align}
\pi^{\mu}\parens{\lambda} & \defeq\tud{\Lambda}{\mu}{\nu}\parens{\lambda}\pi_{\refvalue}^{\nu},\label{eq:ElectricDipole4VecFromRef}\\
\mu^{\mu}\parens{\lambda} & \defeq\tud{\Lambda}{\mu}{\nu}\parens{\lambda}\mu_{\refvalue}^{\nu}.\label{eq:MagneticDipole4VecFromRef}
\end{align}

By construction, the particle's phase space includes a single state
of the form \eqref{eq:ReferenceState}, $\parens{X_{0},p_{0},S_{\refvalue}}\defeq\parens{0,\parens{mc,\vec 0},S_{\refvalue}}$,
which defines the particle's reference state. It follows that the
particle cannot have any physical attributes that violate the uniqueness
of this reference state, such as vector properties that transform
nontrivially under rotations that otherwise preserve the particle's
reference state.

As a direct consequence of this group-theoretic analysis, the reference
value $\vecgreek{\pi}_{\refvalue}$ of the particle's elementary electric-dipole
moment appearing in \eqref{eq:DefRefElectricDipole4VecFrom3Vec} and
the reference value $\vecgreek{\mu}_{\refvalue}$ of the particle's
elementary magnetic-dipole moment appearing in \eqref{eq:DefRefMagneticDipole4VecFrom3Vec}
must both be collinear with the reference value $\vec S_{0}$ of the
particle's spin three-vector. In particular, $\vecgreek{\pi}_{\refvalue}$
and $\vecgreek{\mu}_{\refvalue}$ must be collinear with each other.

Thus, there must exist two constants of proportionality, $\Xi$ and
$\Gamma$, such that\footnote{The quantum-mechanical analogues of these classical collinearity
conditions follow from the Wigner-Eckart theorem.} 
\begin{align}
\vecgreek{\pi}_{\refvalue} & =\frac{1}{c}\,\Xi\,\vec S_{\refvalue},\label{eq:RefElectricDipole3VecParallelSpin3VecCondition}\\
\vecgreek{\mu}_{\refvalue} & =\Gamma\,\vec S_{\refvalue},\label{eq:RefMagneticDipole3VecParallelSpin3VecCondition}
\end{align}
 where the factor of $1/c$ in the first of these two equations compensates
for the factors of $c$ in \eqref{eq:4DElementaryDipoleTensorFromMoments}.
The particle's elementary electric-dipole moment $\vecgreek{\pi}_{\refvalue}$
is a proper vector, whereas its elementary magnetic-dipole moment
$\vecgreek{\mu}_{0}$ and its spin three-vector $\vec S_{0}$ are
both pseudovectors. Hence, $\Xi$ must be a pseudoscalar, whereas
$\Gamma$ must be a proper scalar.

Because $\Gamma$ is the proportionality constant relating $\vecgreek{\mu}_{0}$
to $\vec S_{0}$, we can physically interpret it as the particle's
gyromagnetic ratio. Note, however, that the specific value of $\Gamma$
is not fixed by the group-theoretic considerations here.

\subsection{The Maxwell Action Functional with Sources}

If our particle carries an electric-monopole charge $q$ in addition
to its elementary dipole tensor $m^{\mu\nu}$, then coupling the particle
to the electromagnetic field leads immediately to the following generalization
of the particle's action functional \eqref{eq:FreeParticleActionFunctionalWithSpin}
and the Maxwell action functional \eqref{eq:MaxwellActionInVacuum},
and thereby provides a classical extension of Maxwell's original theory
of electromagnetism: 
\begin{align}
 & S\bracks{X,\Lambda,A}\defeq S_{\textrm{particle}}\bracks{X,\Lambda}+S_{\textrm{field}}\bracks A+S_{\textrm{int}}\bracks{X,\Lambda,A}\nonumber \\
\nonumber \\
 & \begin{aligned} & =\int d\lambda\,\biggparens{p_{\mu}\dot{X}^{\mu}+\frac{1}{2}\Trace\bracks{S\dot{\Lambda}\Lambda^{-1}}} & \quad & \parens{S_{\textrm{particle}}}\\
 & \qquad+\int dt\int d^{3}x\,\biggparens{-\frac{1}{4\mu_{0}}F^{\mu\nu}F_{\mu\nu}} &  & \parens{S_{\textrm{field}}}\\
 & \qquad+\int dt\int d^{3}x\,j^{\nu}A_{\nu} &  & \parens{S_{\textrm{int}}}.
\end{aligned}
\label{eq:MaxwellActionWithParticle}
\end{align}
 Here we have included an important new contribution $S_{\textrm{int}}\bracks{X,\Lambda,A}$
that describes interactions between the particle and the electromagnetic
field: 
\begin{equation}
S_{\textrm{int}}\bracks{X,\Lambda,A}\defeq\int dt\int d^{3}x\,j^{\nu}A_{\nu}.\label{eq:EMParticleInteractionActionFunctional}
\end{equation}

The terms in the action functional \eqref{eq:MaxwellActionWithParticle}
that contain a dependence on the field degrees of freedom $A_{\mu}$
have the standard form \eqref{eq:ClassicalFieldActionFromLagrangianDensity},
$S\defeq\int dt\int d^{3}x\,\mathcal{L}$, for a Lagrangian density
$\mathcal{L}$ given by 
\begin{align}
 & \mathcal{L}=\mathcal{L}_{\textrm{field}}+\mathcal{L}_{\textrm{int}}\nonumber \\
 & \begin{aligned}\qquad & =-\frac{1}{4\mu_{0}}F^{\mu\nu}F_{\mu\nu} & \quad & \parens{\mathcal{L}_{\textrm{field}}}\\
 & \qquad+j^{\nu}A_{\nu} &  & \parens{\mathcal{L}_{\textrm{int}}}.
\end{aligned}
\label{eq:MaxwellLagrangianWithParticle}
\end{align}
 Using this Lagrangian density, the field-theoretic Euler-Lagrange
equations \eqref{eq:EulerLagrangeEquationsField} yield 
\begin{align*}
 & \frac{\partial\mathcal{L}}{\partial A_{\nu}}-\partial_{\mu}\biggparens{\frac{\partial\mathcal{L}}{\partial\parens{\partial_{\mu}A_{\nu}}}}\\
 & \qquad=j^{\nu}-\partial_{\mu}\biggparens{-\frac{1}{\mu_{0}}F^{\mu\nu}}=0,
\end{align*}
 thereby giving us the inhomogeneous Maxwell equation \eqref{eq:4DInhomogMaxwellEq},
\[
\partial_{\mu}F^{\mu\nu}=-\mu_{0}j^{\nu}.
\]
 As was true for the free electromagnetic field, the homogeneous Maxwell
equation \eqref{eq:4DHomogMaxwellEq} is already encoded into the
formula $F_{\mu\nu}=\partial_{\mu}A_{\nu}-\partial_{\nu}A_{\mu}$
for the Faraday tensor from \eqref{eq:FaradayTensorFromDerivsGaugePot},
together with the definition \eqref{eq:DualFaradayTensor} of the
dual Faraday tensor $\tilde{F}_{\mu\nu}\defeq\parens{1/2}\epsilon_{\mu\nu\rho\sigma}F^{\rho\sigma}$:
\[
\partial_{\mu}\tilde{F}^{\mu\nu}=0.
\]

The interaction term $j^{\nu}A_{\nu}$ appearing in the action functional
\eqref{eq:MaxwellActionWithParticle} may not look gauge invariant,
but under a gauge transformation \eqref{eq:4DGaugeTransformation},
\[
A_{\nu}\mapsto A_{\nu}+\partial_{\nu}f,
\]
 the interaction term changes according to 
\begin{equation}
j^{\nu}A_{\nu}\mapsto j^{\nu}A_{\nu}+j^{\nu}\parens{\partial_{\nu}f}.\label{eq:EMParticleInteractionTermGaugeTransf}
\end{equation}
 Using the product rule in reverse (again, ``integration by parts''
without an actual integration) to move the spacetime derivative from
$f$ to $j^{\mu}$ at the cost of a minus sign, we end up with 
\begin{equation}
j^{\nu}A_{\nu}-\parens{\partial_{\nu}j^{\nu}}f+\left(\substack{{\displaystyle \textrm{total spacetime}}\\
{\displaystyle \textrm{divergence}}
}
\right).\label{eq:EMParticleInteractionTermGaugeTransfAfterIntByParts}
\end{equation}
 The second term vanishes by local current conservation \eqref{eq:4DLocalCurrentConservationEq},
$\partial_{\nu}j^{\nu}=0$, when the system's equations of motion
are imposed, and the total spacetime divergence disappears from the
action functional by the four-dimensional divergence theorem, under
the assumption that our fields go to zero sufficiently rapidly at
infinity. The action functional \eqref{eq:MaxwellActionWithParticle}
is therefore effectively unchanged by gauge transformations, as required.

Before we can discuss the equations of motion for the particle, or
the total energy and momentum of the overall system consisting of
the particle together with the electromagnetic field, we will need
to begin by recalling the multipole expansion \eqref{eq:4DCurrentDensityMultipoleExpansion}:
\begin{align*}
j^{\nu} & =j_{\textrm{e}}^{\nu}+j_{\textrm{d}}^{\nu}+j_{\textrm{q}}^{\nu}+\dotsb\\
 & =j_{\textrm{e}}^{\nu}+\partial_{\mu}M^{\mu\nu}+\partial_{\mu}\partial_{\rho}N^{\mu\rho\nu}+\dotsb.
\end{align*}
 Dropping quadrupole and higher multipole moments, in line with our
assumptions about the particle, this expansion truncates to just its
electric-monopole and elementary-dipole terms: 
\begin{equation}
j^{\nu}=j_{\textrm{e}}^{\nu}+\partial_{\mu}M^{\mu\nu}.\label{eq:4DCurrentDensityDipoleTruncate}
\end{equation}
 Substituting this expression into the interaction term $j^{\nu}A_{\nu}$
yields 
\begin{equation}
j^{\nu}A_{\nu}=j_{\textrm{e}}^{\nu}A_{\nu}+\parens{\partial_{\mu}M^{\mu\nu}}A_{\nu},\label{eq:EMParticleInteractionTermFromElMonopoleAndDipoleTensors}
\end{equation}
 so the overall system's action functional \eqref{eq:MaxwellActionWithParticle}
becomes 
\begin{align}
 & S\bracks{X,\Lambda,A}\defeq S_{\textrm{particle}}\bracks{X,\Lambda}+S_{\textrm{field}}\bracks A+S_{\textrm{int}}\bracks{X,\Lambda,A}\nonumber \\
\nonumber \\
 & \begin{aligned} & =\int d\lambda\,\biggparens{p_{\mu}\dot{X}^{\mu}+\frac{1}{2}\Trace\bracks{S\dot{\Lambda}\Lambda^{-1}}} &  & \parens{S_{\textrm{particle}}}\\
 & \qquad+\int dt\int d^{3}x\,\biggparens{-\frac{1}{4\mu_{0}}F^{\mu\nu}F_{\mu\nu}} &  & \parens{S_{\textrm{field}}}\\
 & \qquad+\int dt\int d^{3}x\,\parens{j_{\textrm{e}}^{\nu}A_{\nu}+\parens{\partial_{\mu}M^{\mu\nu}}A_{\nu}} &  & \parens{S_{\textrm{int}}}.
\end{aligned}
\label{eq:MaxwellActionWithParticleDipoleIntsBeforeIBP}
\end{align}

Recall the Lagrangian \eqref{eq:2DOFDerivExampleLagrangian} for our
$xy$ system consisting of a pair of subsystems with individual degrees
of freedom $x$ and $y$: 
\[
L\defeq\frac{1}{2}m\dot{x}^{2}+\frac{1}{2}M\dot{y}^{2}-ay^{2}-bxy+c\dot{x}y.
\]
 We then have an analogy in which the $x$ subsystem plays the role
of our relativistic particle and the $y$ subsystem plays the role
of the electromagnetic field, with the following detailed correspondences:
\begin{equation}
\left.\begin{aligned}\frac{1}{2}m\dot{x}^{2} & \quad\iff\quad p_{\mu}\dot{X}^{\mu}+\frac{1}{2}\Trace\bracks{S\dot{\Lambda}\Lambda^{-1}},\\
\frac{1}{2}M\dot{y}^{2}-ay^{2} & \quad\iff\quad\int d^{3}x\,\biggparens{-\frac{1}{4\mu_{0}}F^{\mu\nu}F_{\mu\nu}},\\
-bxy & \quad\iff\quad\int d^{3}x\,\parens{j_{\textrm{e}}^{\nu}A_{\nu}},\\
c\dot{x}y & \quad\iff\quad\int d^{3}x\,\parens{\partial_{\mu}M^{\mu\nu}}A_{\nu}.
\end{aligned}
\quad\right\} \label{eq:2DOFDerivExampleMaxwellActionAnalogy}
\end{equation}
 We will find it useful to refer back to this analogy on several more
occasions in our work ahead.

At the cost of a minus sign and an irrelevant additive total spacetime
divergence, we are free to use the product rule in reverse to rewrite
the final interaction term $\parens{\partial_{\mu}M^{\mu\nu}}A_{\nu}$
in the integrand of the action functional \eqref{eq:MaxwellActionWithParticleDipoleIntsBeforeIBP}
as 
\begin{equation}
\parens{\partial_{\mu}M^{\mu\nu}}A_{\nu}=-M^{\mu\nu}\parens{\partial_{\mu}A_{\nu}}+\left(\substack{{\displaystyle \textrm{total spacetime}}\\
{\displaystyle \textrm{divergence}}
}
\right).\label{eq:EMDipoleInteractionTermAfterIntByParts}
\end{equation}
 Taking advantage of the antisymmetry of the dipole-density tensor
$M^{\mu\nu}$, we can write the first term on the right-hand side
of this equation as 
\begin{equation}
-M^{\mu\nu}\parens{\partial_{\mu}A_{\nu}}=-\frac{1}{2}M^{\mu\nu}\parens{\partial_{\mu}A_{\nu}-\partial_{\nu}A_{\mu}}.\label{eq:EMDipoleInteractionTermAntisymmetryTrick}
\end{equation}
 Remembering again the formula \eqref{eq:FaradayTensorFromDerivsGaugePot}
relating the Faraday tensor $F_{\mu\nu}$ to the gauge potential $A_{\mu}$,
we have 
\begin{equation}
-\frac{1}{2}M^{\mu\nu}\parens{\partial_{\mu}A_{\nu}-\partial_{\nu}A_{\mu}}=-\frac{1}{2}M^{\mu\nu}F_{\mu\nu}.\label{eq:EMDipoleInteractionTermFromFaradayTensor}
\end{equation}
 We can therefore write the overall system's action functional \eqref{eq:MaxwellActionWithParticleDipoleIntsBeforeIBP}
in the alternative but physically equivalent form 
\begin{align}
 & S\bracks{X,\Lambda,A}\defeq S_{\textrm{particle}}\bracks{X,\Lambda}+S_{\textrm{field}}\bracks A+S_{\textrm{int}}\bracks{X,\Lambda,A}\nonumber \\
\nonumber \\
 & \begin{aligned} & =\int d\lambda\,\biggparens{p_{\mu}\dot{X}^{\mu}+\frac{1}{2}\Trace\bracks{S\dot{\Lambda}\Lambda^{-1}}} & \quad & \parens{S_{\textrm{particle}}}\\
 & \qquad+\int dt\int d^{3}x\,\biggparens{-\frac{1}{4\mu_{0}}F^{\mu\nu}F_{\mu\nu}} &  & \parens{S_{\textrm{field}}}\\
 & \qquad+\int dt\int d^{3}x\,\biggparens{j_{\textrm{e}}^{\nu}A_{\nu}-\frac{1}{2}M^{\mu\nu}F_{\mu\nu}} &  & \parens{S_{\textrm{int}}}.
\end{aligned}
\label{eq:MaxwellActionWithParticleDipoleIntsAfterIBP}
\end{align}

This last step of using the product rule in reverse to replace $\parens{\partial_{\mu}M^{\mu\nu}}A_{\nu}$
with $-\parens{1/2}M^{\mu\nu}F_{\mu\nu}$ is analogous to our use
of the product rule in reverse in \eqref{eq:2DOFDerivExampleVelDepIntIBP}
to replace $c\dot{x}y$ with $-cx\dot{y}$ for the $xy$ system. As
was true in that example, this manipulation has no physical consequences,
but we will find that our calculations ahead will be easier if we
use \eqref{eq:MaxwellActionWithParticleDipoleIntsAfterIBP} rather
than \eqref{eq:MaxwellActionWithParticleDipoleIntsBeforeIBP} as our
system's action functional, as the former version ends up requiring
fewer computations that explicitly involve delta functions than the
latter version.

\subsection{The Action Functional for a Charged Particle with an Elementary Dipole
Moment}

Gathering together all the terms in the action functional \eqref{eq:MaxwellActionWithParticleDipoleIntsAfterIBP}
that involve the particle's degrees of freedom $X^{\mu}\parens{\lambda}=\parens{cT\parens{\lambda},\vec X\parens{\lambda}}^{\mu}$
and $\tud{\Lambda}{\mu}{\nu}\parens{\lambda}$, we obtain 
\begin{align}
 & S_{\textrm{particle+int}}\bracks{X,\Lambda,A}=\int d\lambda\,\biggparens{p_{\mu}\dot{X}^{\mu}+\frac{1}{2}\Trace\bracks{S\dot{\Lambda}\Lambda^{-1}}}\nonumber \\
 & \qquad+\int dt\int d^{3}x\,j_{\textrm{e}}^{\nu}A_{\nu}+\int dt\int d^{3}x\,\biggparens{-\frac{1}{2}}M^{\mu\nu}F_{\mu\nu}.\label{eq:ParticleActionWithDipoleMomentsInitial}
\end{align}
 Before we can compute the system's Euler-Lagrangian equations, we
will need to replace the integrals $\int dt\int d^{3}x\,\parens{\cdots}$
over time and space with appropriate integrals $\int d\lambda\,\parens{\cdots}$
over the particle's worldline parameter $\lambda$, and we will need
to make the particle's worldline degrees of freedom $X^{\mu}\parens{\lambda}$
and $\tud{\Lambda}{\mu}{\nu}\parens{\lambda}$ more manifest.

Under the assumption that our particle has charge $q$, the electric-monopole
current density is given by \eqref{eq:4DElectricMonopoleTravelingCurrentDensity}:
\begin{align*}
 & j_{\textrm{e}}^{\nu}=q\frac{dX^{\nu}\parens t}{dt}\delta^{3}\parens{\vec x-\vec X\parens t}\\
 & =q\int dT\,\frac{dX^{\nu}\parens T}{dT}\delta\parens{t-T}\,\delta^{3}\parens{\vec x-\vec X\parens T}\\
 & =q\int d\lambda\,\frac{dT\parens{\lambda}}{d\lambda}\frac{dX^{\nu}\parens{T\parens{\lambda}}}{dT\parens{\lambda}}\delta\parens{t-T\parens{\lambda}}\,\delta^{3}\parens{\vec x-\vec X\parens{T\parens{\lambda}}}\\
 & =\int d\lambda\,q\frac{dX^{\nu}\parens{T\parens{\lambda}}}{d\lambda}\delta\parens{t-T\parens{\lambda}}\,\delta^{3}\parens{\vec x-\vec X\parens{T\parens{\lambda}}}.
\end{align*}
 We can write this formula more succinctly as 
\[
j_{\textrm{e}}^{\nu}=\int d\lambda\,q\dot{X}^{\nu}\delta\parens{t-T}\delta^{3}\parens{\vec x-\vec X},
\]
 where, as usual, dots denote derivatives with respect to the particle's
worldline parameter $\lambda$. We can therefore express the first
interaction term in the particle's action functional \eqref{eq:ParticleActionWithDipoleMomentsInitial}
as 
\begin{align}
 & \int dt\int d^{3}x\,j_{\textrm{e}}^{\nu}A_{\nu}\nonumber \\
 & \qquad=\int d\lambda\,q\dot{X}^{\nu}A_{\nu}.\label{eq:IntegratedElectricMonopoleTerm}
\end{align}

Similarly, we can write the dipole-density tensor $M^{\mu\nu}$ in
terms of the particle's elementary dipole tensor $m^{\mu\nu}$ and
the Lorentz-invariant three-dimensional delta function \eqref{eq:LorentzInvSphSymmetricDensityDeltaFunction}
as in \eqref{eq:ElemDipoleTensorFromDensityDeltaFunc}: 
\begin{align}
M^{\mu\nu} & =m^{\mu\nu}\frac{1}{\gamma}\delta^{3}\parens{\vec x-\vec X}\nonumber \\
 & =\int d\lambda\,\frac{dT}{d\lambda}m^{\mu\nu}\frac{1}{\gamma}\delta\parens{t-T}\delta^{3}\parens{\vec x-\vec X}.\label{eq:MultipoleDensityFromElemMultipoleTensor}
\end{align}
 Combining the factor of $dT/d\lambda$ with the reciprocal Lorentz
factor $1/\gamma$ to obtain 
\begin{align*}
\frac{dT}{d\lambda}\frac{1}{\gamma} & =\frac{dT}{d\lambda}\sqrt{1-\biggparens{\frac{d\vec X}{dT}}^{2}/c^{2}}\\
 & \sqrt{\biggparens{\frac{dT}{d\lambda}}^{2}-\biggparens{\frac{d\vec X}{d\lambda}}^{2}/c^{2}}\\
 & =\frac{1}{c}\sqrt{-\dot{X}^{2}},
\end{align*}
 the second interaction term becomes 
\begin{align}
 & \int dt\int d^{3}x\,\biggparens{-\frac{1}{2}}M^{\mu\nu}F_{\mu\nu}\nonumber \\
 & \qquad=\int d\lambda\,\biggparens{-\frac{1}{2c}}\sqrt{-\dot{X}^{2}}\,m^{\mu\nu}F_{\mu\nu}.\label{eq:IntegratedDipoleInteractionTerm}
\end{align}
 Putting everything together, we see that the particle's action functional
is of the manifestly covariant form described in \citep{Barandes:2019mcl},
\begin{equation}
S_{\textrm{particle+int}}\bracks{X,\Lambda,A}=\int d\lambda\,\mathscr{L}_{\textrm{particle+int}},\label{eq:ParticleActionWithDipoleMomentsFinal}
\end{equation}
 for a manifestly covariant Lagrangian defined by 
\begin{align}
\mathscr{L}_{\textrm{particle+int}} & \defeq p_{\mu}\dot{X}^{\mu}+\frac{1}{2}\Trace\bracks{S\dot{\Lambda}\Lambda^{-1}}\nonumber \\
 & \qquad+q\dot{X}^{\nu}A_{\nu}-\frac{1}{2c}\sqrt{-\dot{X}^{2}}m^{\mu\nu}F_{\mu\nu}.\label{eq:ParticleLagrangianWithDipoleMoments}
\end{align}

\subsection{The Dynamics of the Canonical Momentum of an Elementary Dipole}

Now we are ready to calculate the particle's canonical momenta and
its equations of motion. As we proceed, we will need to keep in mind
that the gauge potential $A_{\nu}=A_{\nu}\parens{X\parens{\lambda}}$
and the Faraday tensor $F_{\mu\nu}=F_{\mu\nu}\parens{X\parens{\lambda}}$
depend on the particle's spacetime degrees of freedom $X^{\mu}\parens{\lambda}$,
as well as remember from \eqref{eq:FourMomentumFromRest} that the
particle's four-momentum $p^{\mu}\parens{\lambda}=mc\,\tud{\Lambda}{\mu}t\parens{\lambda}$
does not depend on $X^{\mu}\parens{\lambda}$ before the equations
of motion have been imposed.

Following the manifestly covariant formalism presented in \citep{Barandes:2019mcl},
the covariant canonical four-momentum conjugate to $X^{\mu}\parens{\lambda}$
is given by 
\begin{align}
p_{\textrm{can},\mu} & \defeq\frac{\partial\mathscr{L}_{\textrm{particle+int}}}{\partial\dot{X}^{\mu}}\nonumber \\
 & =p_{\mu}+qA_{\mu}+\frac{1}{2c}\frac{\dot{X}_{\mu}}{\sqrt{-\dot{X}^{2}}}m^{\rho\sigma}F_{\rho\sigma}.\label{eq:ParticleWithDipoleMomentsCanonical4Momentum}
\end{align}
 Using the chain rule to write $d/d\lambda=\dot{X}^{\nu}\partial_{\nu}$
as needed, the covariant Euler-Lagrange equation for $X^{\mu}\parens{\lambda}$,
\begin{equation}
\frac{\partial\mathscr{L}_{\textrm{particle+int}}}{\partial X^{\mu}}-\frac{d}{d\lambda}\biggparens{\frac{\partial\mathscr{L}_{\textrm{particle+int}}}{\partial\dot{X}^{\mu}}}=0,\label{eq:ParticleEulerLagrangeCoords}
\end{equation}
 yields the following equation of motion for the particle's four-momentum
$p^{\mu}$: 
\begin{align}
\dot{p}^{\mu} & =-q\dot{X}_{\nu}F^{\nu\mu}-\frac{1}{2}\sqrt{-\dot{X}^{2}}m^{\rho\sigma}\partial^{\mu}F_{\rho\sigma}\nonumber \\
 & \qquad\qquad-\frac{1}{2c}\frac{d}{d\lambda}\biggparens{\frac{\dot{X}^{\mu}}{\sqrt{-\dot{X}^{2}}}m^{\rho\sigma}F_{\rho\sigma}}.\label{eq:ParticleWithDipoleMomentsCoordEOM}
\end{align}
 This equation simplifies if we choose our worldline parameter $\lambda$
to be the particle's proper time $\tau$, in which case 
\begin{equation}
\sqrt{-\dot{X}^{2}}\mapsto\sqrt{-\parens{dX/d\tau}^{2}}=c.\label{eq:SqrtRootXDotSquaredForProperTime}
\end{equation}
 The particle's normalized four-velocity \eqref{eq:ParticleFourVelocity}
then takes the form \eqref{eq:ParticleFourVelocityFromProperTime},
\[
u^{\mu}=\frac{dX^{\mu}}{d\tau},
\]
 and so the equation of motion \eqref{eq:ParticleWithDipoleMomentsCoordEOM}
becomes 
\begin{align}
\frac{dp^{\mu}}{d\tau} & =-qu_{\nu}F^{\nu\mu}-\frac{1}{2}m^{\rho\sigma}\partial^{\mu}F_{\rho\sigma}-\frac{1}{2c^{2}}\frac{d}{d\tau}\parens{u^{\mu}m^{\rho\sigma}F_{\rho\sigma}}\nonumber \\
 & =-qu_{\nu}F^{\nu\mu}-\frac{1}{2}m^{\rho\sigma}\parens{\eta^{\mu\nu}+\frac{1}{c^{2}}u^{\mu}u^{\nu}}\partial_{\nu}F_{\rho\sigma}\nonumber \\
 & \qquad\qquad-\frac{1}{2c^{2}}\frac{d}{d\tau}\parens{u^{\mu}m^{\rho\sigma}}F_{\rho\sigma},\label{eq:ParticleWithDipoleMomentsCoordEOMProperTime}
\end{align}
 as obtained in \citep{VanDamRuijgrok:1980crepsmef,SkagerstamStern:1981ldccps,GerochWeatherall:2018msbst}. 

\subsection{The Non-Relativistic Limit with Time-Independent External Fields}

We will now examine the equation of motion \eqref{eq:ParticleWithDipoleMomentsCoordEOMProperTime}
in the non-relativistic limit. In that limit, the particle's proper
time $\tau$ reduces to the coordinate time $t$, and the particle's
four-velocity $u^{\nu}$ reduces to a four-vector consisting of the
speed of light $c$ and the particle's three-dimensional velocity
$\vec v$: 
\begin{align}
\tau & \approx t,\nonumber \\
u^{\nu} & \approx\parens{c,\vec v}^{\nu}.\label{eq:NonrelativisticLimitProperTime4Velocity}
\end{align}

In the calculations ahead, we will find it useful to make use of the
tensor-contraction identity 
\begin{equation}
m^{\rho\sigma}\parens{\cdots}F_{\rho\sigma}=-2\parens{\parens{\cdots}\vec E}\dotprod\vecgreek{\pi}-2\parens{\parens{\cdots}\vec B}\dotprod\vecgreek{\mu}.\label{eq:ElementaryDipoleTensorFaradayIdentity}
\end{equation}
 Here $\parens{\cdots}$ represents numerical quantities or derivative
operators, and the particle's elementary dipole moments $\vecgreek{\pi}$
and $\vecgreek{\mu}$ are defined in terms of $m^{\rho\sigma}$ according
to \eqref{eq:4DElementaryDipoleTensorFromMoments}.

In keeping with the typical background assumptions underlying the
Lorentz force law, we will assume that the particle's three-velocity
$\vec v$ changes slowly enough that we can neglect radiative effects.
We will accordingly drop contributions to the electromagnetic fields
from the particle itself. Then $\vec E\mapsto\vec E_{\textrm{ext}}$
and $\vec B\mapsto\vec B_{\textrm{ext}}$ will both refer to external
fields, where we will also assume that these external fields are time-independent
(but not necessarily uniform in space) in the given inertial reference
frame.\footnote{For a rigorous treatment of self-forces and self-energies, see \citep{GrallaHarteWald:2009rdesf}.} 

We start with the $\mu=t$ case of the particle's four-dimensional
equation of motion \eqref{eq:ParticleWithDipoleMomentsCoordEOMProperTime},
introducing an overall factor of $c$ for later convenience: 
\begin{align*}
\frac{dE}{dt} & \approx-qu_{\nu}cF^{\nu t}-\frac{c}{2}m^{\rho\sigma}\partial^{t}F_{\rho\sigma}-\frac{1}{2c}\frac{d}{dt}\parens{u^{t}m^{\rho\sigma}F_{\rho\sigma}}\\
 & \approx q\vec v\dotprod\vec E_{\textrm{ext}}+\frac{d}{dt}\parens{\vecgreek{\pi}\dotprod\vec E_{\textrm{ext}}+\vecgreek{\mu}\dotprod\vec B_{\textrm{ext}}}\\
 & =\frac{d}{dt}\parens{-q\Phi_{\textrm{ext}}+\vecgreek{\pi}\dotprod\vec E_{\textrm{ext}}+\vecgreek{\mu}\dotprod\vec B_{\textrm{ext}}}.
\end{align*}
 In going from the first line to the second line here, we used our
assumption of time-independent external fields to drop the second
term, and we used the tensor-contraction identity \eqref{eq:ElementaryDipoleTensorFaradayIdentity}
to simplify the third term. To obtain the third equality, we used
the usual formula $\vec E_{\textrm{ext}}=-\nabla\Phi_{\textrm{ext}}$,
appropriate to the static-field case, where $\Phi_{\textrm{ext}}$
is the electrostatic scalar potential, and then we used the chain
rule to replace $\vec v\dotprod\nabla\Phi_{\textrm{ext}}$ with $d\Phi_{\textrm{ext}}/dt$.

Looking back at the four-dimensional equation of motion \eqref{eq:ParticleWithDipoleMomentsCoordEOMProperTime}
for $dp^{\mu}/d\tau$, notice that although its final term includes
an explicit factor of $1/c^{2}$, the calculation of $dE/dt$ above
shows that it is not a purely relativistic correction. That final
term was necessary for getting the correct non-relativistic formula
for $dE/dt$.

Turning next to the $\mu=i$ case of the particle's four-dimensional
equation of motion \eqref{eq:ParticleWithDipoleMomentsCoordEOMProperTime},
where $i$ refers to any one of the spatial coordinates $x$, $y$,
or $z$, we find 

\begin{align*}
\frac{dp^{i}}{dt} & \approx-qu_{\nu}F^{\nu i}-\frac{1}{2}m^{\rho\sigma}\partial^{i}F_{\rho\sigma}-\frac{1}{2c^{2}}\frac{d}{dt}\parens{u^{i}m^{\rho\sigma}F_{\rho\sigma}}\\
 & \approx q\parens{\vec E_{\textrm{ext}}+\vec v\crossprod\vec B_{\textrm{ext}}}_{i}+\partial_{i}\parens{\vecgreek{\pi}\dotprod\vec E_{\textrm{ext}}+\vecgreek{\mu}\dotprod\vec B_{\textrm{ext}}}.
\end{align*}
 In going from the first line to the second line here, we used the
tensor-contraction identity \eqref{eq:ElementaryDipoleTensorFaradayIdentity}
to simplify the second term, and the third term dropped out at zeroth
order in $c$.

Putting everything together, we see that in the non-relativistic limit,
with time-independent external fields, the particle's four-dimensional
equation of motion \eqref{eq:ParticleWithDipoleMomentsCoordEOMProperTime}
reduces to the following pair of three-dimensional equations: 
\begin{align}
\frac{dE}{dt} & \approx\frac{d}{dt}\parens{-q\Phi_{\textrm{ext}}+\vecgreek{\pi}\dotprod\vec E_{\textrm{ext}}+\vecgreek{\mu}\dotprod\vec B_{\textrm{ext}}},\label{eq:NonrelWorkLawOnDipole}\\
\frac{d\vec p}{dt} & \approx q\parens{\vec E_{\textrm{ext}}+\vec v\crossprod\vec B_{\textrm{ext}}}+\nabla\parens{\vecgreek{\pi}\dotprod\vec E_{\textrm{ext}}+\vecgreek{\mu}\dotprod\vec B_{\textrm{ext}}}.\label{eq:NonrelForceLawOnDipole}
\end{align}
 Here $E=\sqrt{\vec p^{2}c^{2}+m^{2}c^{4}}$ is the particle's relativistic-kinetic
energy, as in \eqref{eq:ReparamExampleHamiltonian}, and $\vec p$
is the particle's three-dimensional momentum, where the non-relativistic
limit gives the approximation 
\begin{equation}
p^{\mu}=\parens{E/c,\vec p}^{\mu}\approx\parens{mc+\parens{1/2}m\vec v^{2}/c,\vec p}^{\mu}.\label{eq:NonrelParticleFourMom}
\end{equation}

\subsection{The Generalized Lorentz Force Law for Elementary Dipoles}

Comparing the second of the particle's two non-relativistic equations,
\eqref{eq:NonrelForceLawOnDipole}, with the general force-momentum
relationship $\vec F=d\vec p/dt$ from \eqref{eq:DefForceFromMomentum},
we can identify the electromagnetic force on the particle as 
\begin{equation}
\vec F=q\vec E_{\textrm{ext}}+q\vec v\crossprod\vec B_{\textrm{ext}}+\nabla\parens{\vecgreek{\pi}\dotprod\vec E_{\textrm{ext}}}+\nabla\parens{\vecgreek{\mu}\dotprod\vec B_{\textrm{ext}}}.\label{eq:LorentzForceLawWithDipoles}
\end{equation}
 This formula agrees with our claimed generalization \eqref{eq:LorentzForceLawNonRelApproxIntro}
of the Lorentz force law.

Notice that the magnetic field participates in the dipole terms $\nabla\parens{\vecgreek{\pi}\dotprod\vec E_{\textrm{ext}}}+\nabla\parens{\vecgreek{\mu}\dotprod\vec B_{\textrm{ext}}}$
of this force law on an equal footing with the electric field.  Furthermore,
if the particle moves at a constant velocity $\vec v$ through incremental
spatial displacements $d\vec X=\vec vdt$ over infinitesimal time
intervals $dt$, then the work \eqref{eq:DefWork} done by the electromagnetic
field on the particle as it travels from an initial location $A$
to a final location $B$ is 
\begin{align}
W & =\int_{A}^{B}d\vec X\dotprod\vec F=\int_{A}^{B}dt\,\vec v\dotprod\vec F\nonumber \\
 & =\int_{A}^{B}dt\,\vec v\dotprod\parens{q\vec E_{\textrm{ext}}+q\vec v\crossprod\vec B_{\textrm{ext}}}\nonumber \\
 & \qquad+\int_{A}^{B}dt\,\vec v\dotprod\parens{\nabla\parens{\vecgreek{\pi}\dotprod\vec E_{\textrm{ext}}}+\nabla\parens{\vecgreek{\mu}\dotprod\vec B_{\textrm{ext}}}}\nonumber \\
 & =\int_{A}^{B}dt\,\parens{q\vec v\dotprod\vec E_{\textrm{ext}}}+\Delta\parens{\vecgreek{\pi}\dotprod\vec E_{\textrm{ext}}}+\Delta\parens{\vecgreek{\mu}\dotprod\vec B_{\textrm{ext}}}.\label{eq:EMWorkDoneOnParticle}
\end{align}
 Here we have dropped the $q\vec v\crossprod\vec B_{\textrm{ext}}$
term because its dot product with $\vec v$ vanishes, and $\Delta$
denotes a total change over the particle's full displacement from
$A$ to $B$.

We have therefore arrived at the key conclusion of this paper\textemdash namely,
that although magnetic forces do not do work on \emph{electric monopoles},
they are entirely capable of doing work on \emph{elementary magnetic
dipoles}.

Turning now to the first of the particle's two non-relativistic equations,
\eqref{eq:NonrelWorkLawOnDipole}, we observe that the following quantity
is conserved: 
\begin{align}
 & E+q\Phi_{\textrm{ext}}-\vecgreek{\pi}\dotprod\vec E_{\textrm{ext}}-\vecgreek{\mu}\dotprod\vec B_{\textrm{ext}}\nonumber \\
 & \qquad\approx mc^{2}+\frac{1}{2}m\vec v^{2}+q\Phi_{\textrm{ext}}-\vecgreek{\pi}\dotprod\vec E_{\textrm{ext}}-\vecgreek{\mu}\dotprod\vec B_{\textrm{ext}}.\label{eq:NonrelParticleConservedEnergy}
\end{align}
 We naturally identify the combination 
\begin{equation}
V=q\Phi_{\textrm{ext}}-\vecgreek{\pi}\dotprod\vec E_{\textrm{ext}}-\vecgreek{\mu}\dotprod\vec B_{\textrm{ext}}\label{eq:NonrelPotentialEnergy}
\end{equation}
 as the particle's potential energy in the time-independent external
fields. Indeed, the components of the electromagnetic force \eqref{eq:LorentzForceLawWithDipoles}
on the particle that capable of doing work are conservative in the
sense of \eqref{eq:DefConservativeForceFromPotentialEnergy}, 
\begin{align*}
\vec F\ \left[\textrm{work-capable}\right] & =q\vec E_{\textrm{ext}}+\nabla\parens{\vecgreek{\pi}\dotprod\vec E_{\textrm{ext}}}+\nabla\parens{\vecgreek{\mu}\dotprod\vec B_{\textrm{ext}}}\\
 & =-\nabla\parens{q\Phi_{\textrm{ext}}-\vecgreek{\pi}\dotprod\vec E_{\textrm{ext}}-\vecgreek{\mu}\dotprod\vec B_{\textrm{ext}}}\\
 & =-\nabla V,
\end{align*}
 so the work \eqref{eq:EMWorkDoneOnParticle} done by the electromagnetic
field on the particle simplifies to 
\[
W=-\Delta V,
\]
 in accordance with the general relationship \eqref{eq:WorkEqualsMinusChangePotential}
between the work $W$ done on a mechanical object and the object's
corresponding potential energy $V$.

Finally, the \emph{rate} at which electromagnetic forces do work on
the particle is 
\begin{align}
\frac{dW}{dt} & =\frac{d}{dt}\int^{t}d\vec X\dotprod\vec F=-\frac{d}{dt}\int^{t}d\vec X\dotprod\nabla V=-\frac{dV}{dt}\nonumber \\
 & =\frac{d}{dt}\parens{-q\Phi_{\textrm{ext}}+\vecgreek{\pi}\dotprod\vec E_{\textrm{ext}}+\vecgreek{\mu}\dotprod\vec B_{\textrm{ext}}}.\label{eq:LorentzPowerLawWithDipoles}
\end{align}
 This formula precisely agrees with our non-relativistic dynamical
equation \eqref{eq:NonrelWorkLawOnDipole} for the rate $dE/dt$ at
which the particle's kinetic energy is changing, so the work being
done on the particle by the electromagnetic field is translating directly
into the particle's kinetic energy.

\subsection{The Dynamics of the Intrinsic Spin}

Next, we will use the particle's action functional \eqref{eq:ParticleActionWithDipoleMomentsFinal}
to calculate the equation of motion for the particle's spin tensor
$S^{\mu\nu}\parens{\lambda}$. Varying the action functional with
respect to the variable Lorentz-transformation matrix $\tud{\Lambda}{\mu}{\nu}\parens{\lambda}$,
we obtain 
\begin{align}
 & \delta S_{\textrm{particle+int}}=\int d\lambda\,\bigg(\delta p^{\mu}\dot{X}_{\mu}+\frac{1}{2}\Trace\bracks{\delta\parens{S\dot{\Lambda}\Lambda^{-1}}}\nonumber \\
 & \quad\qquad\qquad\qquad\qquad\qquad-\frac{1}{2c}\sqrt{-\dot{X}^{2}}\delta m^{\mu\nu}F_{\mu\nu}\bigg).\label{eq:VariationActionLorentzDOF}
\end{align}
 As derived in \citep{Barandes:2019mcl}, the first two terms yield
\begin{align*}
\delta p^{\mu}\dot{X}_{\mu} & =\frac{1}{2}\parens{-\dot{X}_{\rho}p_{\sigma}+\dot{X}_{\sigma}p_{\rho}}\delta\theta^{\rho\sigma},\\
\frac{1}{2}\Trace\bracks{\delta\parens{S\dot{\Lambda}\Lambda^{-1}}} & =\frac{1}{2}S_{\rho\sigma}\frac{d}{d\lambda}\delta\theta^{\rho\sigma},
\end{align*}
 where $\delta\theta^{\rho\sigma}$ is an array of small boost and
rotation parameters corresponding to the infinitesimal variation in
$\tud{\Lambda}{\mu}{\nu}\parens{\lambda}$. Meanwhile, using the commutation
relations \eqref{eq:LorentzGeneratorsCommutator} satisfied by the
Lorentz generators, together with\footnote{Keep in mind the suppressed indices in the first three lines of this
calculation.} 
\begin{align}
 & \delta m^{\mu\nu}=-\frac{i}{2}\delta\parens{\Trace\bracks{\Lambda m_{\refvalue}\Lambda^{-1}\sigma^{\mu\nu}}})\nonumber \\
 & =-\frac{i}{2}\parens{\Trace\bracks{m_{\refvalue}\Lambda^{-1}\sigma^{\mu\nu}\delta\Lambda+m_{\refvalue}\parens{\delta\Lambda^{-1}}\sigma^{\mu\nu}\Lambda}})\nonumber \\
 & =-\frac{1}{4}\Trace\bracks{m\parens{\sigma^{\mu\nu}\sigma^{\rho\sigma}-\sigma^{\rho\sigma}\sigma^{\mu\nu}}}\delta\theta_{\rho\sigma}\nonumber \\
 & =\frac{1}{2}\parens{-m^{\nu\rho}\eta^{\mu\sigma}-m^{\mu\sigma}\eta^{\nu\rho}+m^{\nu\sigma}\eta^{\mu\rho}+m^{\mu\rho}\eta^{\nu\sigma}}\delta\theta_{\rho\sigma},\label{eq:VariationMultipoleTensorLorentzDOF}
\end{align}
 the third term in the varied action functional \eqref{eq:VariationActionLorentzDOF}
gives 
\begin{align}
 & -\frac{1}{2c}\sqrt{-\dot{X}^{2}}\delta m^{\mu\nu}F_{\mu\nu}\nonumber \\
 & \qquad=-\frac{1}{2c}\sqrt{-\dot{X}^{2}}\parens{m^{\rho\mu}\tud F{\sigma}{\mu}-m^{\sigma\mu}\tud F{\rho}{\mu}}\delta\theta_{\rho\sigma}.\label{eq:VariationSpinInteractionTermLorentzDOF}
\end{align}

Combining all these results and setting the overall variation \eqref{eq:VariationActionLorentzDOF}
in the particle's action functional to zero, in accordance with the
usual extremization condition \eqref{eq:VariationActionFunctional},
we obtain 
\begin{align*}
 & \delta S_{\textrm{particle+int}}\\
 & =\int d\lambda\,\frac{1}{2}\bigg(-\parens{\dot{X}^{\rho}p^{\sigma}-\dot{X}^{\sigma}p^{\rho}}-\dot{S}^{\rho\sigma}\\
 & \qquad\qquad\quad-\frac{1}{c}\sqrt{-\dot{X}^{2}}\parens{m^{\rho\mu}\tud F{\sigma}{\mu}-m^{\sigma\mu}\tud F{\rho}{\mu}}\bigg)\delta\theta_{\rho\sigma}=0,
\end{align*}
 where we have dropped a total derivative $d\parens{S^{\rho\sigma}\delta\theta_{\rho\sigma}}/d\lambda$
in writing down the middle term. We therefore find the following equation
of motion for the particle's spin tensor $S^{\mu\nu}$: 
\begin{align}
\dot{S}^{\mu\nu} & =-\parens{\dot{X}^{\mu}p^{\nu}-\dot{X}^{\nu}p^{\mu}}\nonumber \\
 & \qquad-\frac{1}{c}\sqrt{-\dot{X}^{2}}\parens{m^{\mu\rho}\tud F{\nu}{\rho}-m^{\nu\rho}\tud F{\mu}{\rho}}.\label{eq:ParticleSpinTensorEOM}
\end{align}
 Once again, we simplify this equation by choosing the worldline parameter
$\lambda$ to be the particle's proper time, so that from \eqref{eq:SqrtRootXDotSquaredForProperTime},
we have 
\[
\sqrt{-\dot{X}^{2}}\mapsto c,
\]
 and from \eqref{eq:ParticleFourVelocityFromProperTime}, we have
\[
\dot{X}^{\mu}\mapsto u^{\mu}.
\]
 The equation of motion for $S^{\mu\nu}$ then becomes 
\begin{equation}
\frac{dS^{\mu\nu}}{d\tau}=-\parens{u^{\mu}p^{\nu}-u^{\nu}p^{\mu}}-\parens{m^{\mu\rho}\tud F{\nu}{\rho}-m^{\nu\rho}\tud F{\mu}{\rho}},\label{eq:ParticleSpinTensorEOMProperTime}
\end{equation}
 which generalizes the results of \citep{BargmannMichelTelegdi:1959pppmhef,VanDamRuijgrok:1980crepsmef,SkagerstamStern:1981ldccps}.

The particle's orbital angular-momentum tensor is defined as in \eqref{eq:FreeParticleOrbAngMomTensor}
by 
\[
L^{\mu\nu}\defeq X^{\mu}p^{\nu}-X^{\nu}p^{\mu},
\]
 with $\vec L\defeq\parens{L^{yz},L^{zx},L^{xy}}=\vec X\crossprod\vec p$
the particle's orbital angular-momentum pseudovector. Using 
\begin{equation}
\frac{dL^{\mu\nu}}{d\tau}=u^{\mu}p^{\nu}-u^{\nu}p^{\mu}+X^{\mu}\frac{dp^{\nu}}{d\tau}-X^{\nu}\frac{dp^{\mu}}{d\tau},\label{eq:DerivParticleOrbAngMomTensor}
\end{equation}
 we have in the non-relativistic limit that 
\begin{equation}
\frac{d\vec L}{dt}\approx\vec v\crossprod\vec p+\vec X\crossprod\frac{d\vec p}{dt}.\label{eq:NonrelTimeDerivParticleOrbAngMomTensor}
\end{equation}
It follows from a straightforward calculation that if we ignore self-field
effects, then the non-relativistic limit of the spin tensor's equation
of motion \eqref{eq:ParticleSpinTensorEOMProperTime} is 
\begin{equation}
\frac{d\vec S}{dt}\approx-\vec v\crossprod\vec p+\vec{\vecgreek{\pi}}\crossprod\vec E_{\textrm{ext}}+\vec{\vecgreek{\mu}}\crossprod\vec B_{\textrm{ext}}.\label{eq:NonrelTimeDerivParticleSpinTensor}
\end{equation}
 Combining these last two equations, we obtain the non-relativistic
equation 
\begin{equation}
\frac{d\vec J}{dt}\approx\vec X\crossprod\frac{d\vec p}{dt}+\vec{\vecgreek{\pi}}\crossprod\vec E_{\textrm{ext}}+\vec{\vecgreek{\mu}}\crossprod\vec B_{\textrm{ext}},\label{eq:NonrelTorqueEqOnDipole}
\end{equation}
 where $\vec J\defeq\vec L+\vec S$ is the particle's total angular
momentum. This last equation describes a net torque on the particle
given by the sum of orbital and dipole contributions.

\subsection{Self-Consistency Conditions}

Now that we have obtained the particle's equations of motion, we will
need to ensure that they are compatible with the fundamental structure
of the particle's phase space\textemdash specifically, that they are
consistent with the constancy of the invariant quantities $m^{2}$,
$w^{2}$, $s^{2}$, and $\tilde{s}^{2}$ defined \eqref{eq:Def4DMassSquaredAsInvariant}\textendash \eqref{eq:Def4DDualSpinSquaredAsInvariant},
as well as with the condition $p_{\mu}S^{\mu\nu}=0$ from \eqref{eq:FourMomSpinTensorZeroPhysicalCondition}.

We will start by examining the condition $p_{\mu}S^{\mu\nu}=0$. Taking
its derivative with respect to the proper time $\tau$, we find 
\[
\frac{dp_{\mu}}{d\tau}S^{\mu\nu}+p_{\mu}\frac{dS^{\mu\nu}}{d\tau}=0,
\]
 which yields an equation of the form 
\begin{equation}
p^{\mu}=m_{\textrm{eff}}u^{\mu}+b^{\mu}.\label{eq:FourMomFrom4VelPlusDiscrep}
\end{equation}
 Here the coefficient function $m_{\textrm{eff}}\parens{\lambda}$
is defined by 
\begin{equation}
m_{\textrm{eff}}\defeq-\frac{m^{2}c^{2}}{p\dotprod u}.\label{eq:DefEffectiveMass}
\end{equation}
 Roughly speaking, $m_{\textrm{eff}}\parens{\lambda}$ plays the functional
role of an ``effective'' inertial mass, in the sense of how it appears
in the relationship \eqref{eq:FourMomFrom4VelPlusDiscrep} between
the particle's four-momentum $p^{\mu}$ and its four-velocity $u^{\mu}$.
However, it is important to keep in mind that the particle still adheres
to the invariant condition \eqref{eq:Def4DMassSquaredAsInvariant},
$p^{2}=-m^{2}c^{2}$, for the same fixed value of $m$ as before.

The four-vector $b^{\mu}\parens{\lambda}$, which represents the discrepancy
between $p^{\mu}\parens{\lambda}$ and $m_{\textrm{eff}}\parens{\lambda}u^{\mu}\parens{\lambda}$,
is defined by 
\begin{equation}
b^{\mu}\defeq\frac{1}{p\dotprod u}\biggparens{\frac{dp_{\nu}}{d\tau}S^{\nu\mu}-p_{\nu}\parens{m^{\nu\rho}\tud F{\mu}{\rho}-m^{\mu\rho}\tud F{\nu}{\rho}}}.\label{eq:Def4VecDiscrep}
\end{equation}
 Following \citep{SkagerstamStern:1981ldccps}, we regard \eqref{eq:FourMomFrom4VelPlusDiscrep}
as an \emph{implicit formula} for determining the behavior of the
particle's four-velocity $u^{\mu}$ as a function of the proper time
$\tau$ along the particle's worldline. 

Combining the condition $p_{\mu}S^{\mu\nu}=0$ with the definition
\eqref{eq:Def4VecDiscrep} of $b^{\mu}$, we see that $b^{\mu}$ has
vanishing Lorentz dot product with the particle's four-momentum $p^{\mu}$:
\begin{equation}
b\dotprod p=0.\label{eq:4VecDiscrepDotProdFourMomZero}
\end{equation}
 Contracting both sides of \eqref{eq:FourMomFrom4VelPlusDiscrep}
with $p_{\mu}$ then yields \eqref{eq:Def4DMassSquaredAsInvariant},
$p^{2}=-m^{2}c^{2}$, thereby ensuring that $p^{2}$ is constant,
as required: 
\begin{equation}
\frac{d}{dt}\parens{p^{2}}=0.\label{eq:MassSquaredConst}
\end{equation}

If the electromagnetic field is zero, $F_{\mu\nu}=0$, then it follows
from a straightforward calculation that $b^{\mu}=0$ and $m_{\textrm{eff}}=m$,
so the particle's four-momentum $p^{\mu}$ is parallel to its four-velocity,
with $m$ playing the role of the proportionality constant: 
\begin{equation}
p^{\mu}=mu^{\mu}\quad\parens{F_{\mu\nu}=0}.\label{eq:FourMomFrom4VelZeroField}
\end{equation}
  By contrast, for nonzero electromagnetic field, $F_{\mu\nu}\ne0$,
the terms in the definition \eqref{eq:Def4VecDiscrep} of $b^{\mu}$
go like $1/c^{2}$, so the discrepancy four-vector $b^{\mu}$ is a
relativistic correction. It follows that $m_{\textrm{eff}}-m$ is
likewise a relativistic correction of order $1/c^{2}$, so 
\begin{equation}
p^{\mu}=mu^{\mu}+\parens{\textrm{terms of order \ensuremath{1/c^{2}}}}.\label{eq:FourMomEqFourVelUpToRelCorrections}
\end{equation}

One key implication of these results is that when discussing work
done by electromagnetic forces on the particle in the non-relativistic
limit, as in \eqref{eq:NonrelWorkLawOnDipole} and \eqref{eq:NonrelForceLawOnDipole},
there is no ambiguity over whether we should identify $E=p^{t}c$
or $u^{t}mc^{2}$ as the particle's ``true'' relativistic kinetic
energy. Indeed, in the non-relativistic limit, they agree: 
\begin{equation}
E=p^{t}c\approx u^{t}mc^{2}\approx mc^{2}+\frac{1}{2}m\vec v^{2}.\label{eq:NonRelInternalEnergyAsNewtonianKinetic}
\end{equation}

Next, we study the invariant spin-squared scalar $s^{2}$ defined
in \eqref{eq:Def4DSpinSquaredAsInvariant}. Invoking the spin tensor's
equation of motion \eqref{eq:ParticleSpinTensorEOMProperTime} together
with the condition \eqref{eq:FourMomSpinTensorZeroPhysicalCondition},
$p_{\mu}S^{\mu\nu}=0$, we have 
\begin{align}
\frac{d}{d\tau}\parens{s^{2}} & =\frac{d}{d\tau}\biggparens{\frac{1}{2}S_{\mu\nu}S^{\mu\nu}}=2S_{\mu\nu}\frac{dS^{\mu\nu}}{d\tau}\nonumber \\
 & =\parens{\tud S{\rho}{\mu}m^{\mu\sigma}-\tud S{\sigma}{\mu}m^{\mu\rho}}F_{\rho\sigma}.\label{eq:SpinTensorSquareConst}
\end{align}
 The scalar quantity $s^{2}$ is therefore constant along the particle's
worldline for generic states of the electromagnetic field only if
the quantity in parentheses above vanishes, meaning that 
\begin{equation}
\tud S{\rho}{\mu}m^{\mu\sigma}=\tud S{\sigma}{\mu}m^{\mu\rho}.\label{eq:SpinTensorDipoleTensorCondition}
\end{equation}

The equality \eqref{eq:SpinTensorDipoleTensorCondition} implies that
in the particle's reference state, the reference values \eqref{eq:DefRefElectricDipole4VecFrom3Vec}
and \eqref{eq:DefRefMagneticDipole4VecFrom3Vec} of the particle's
three-dimensional elementary electric-dipole moment $\vecgreek{\pi}_{\refvalue}$
and magnetic-dipole moment $\vecgreek{\mu}_{\refvalue}$ must both
have vanishing cross products with the reference value $\vec S_{\refvalue}$
of the particle's spin three-vector: 
\begin{equation}
\left.\begin{aligned}\vecgreek{\pi}_{\refvalue}\crossprod\vec S_{\refvalue} & =0,\\
\vecgreek{\mu}_{\refvalue}\crossprod\vec S_{\refvalue} & =0.
\end{aligned}
\quad\right\} \label{eq:RefDipole3VecsCrossProdSpin3VecZeroCondition}
\end{equation}
  These conditions are consistent with our requirements \eqref{eq:RefElectricDipole3VecParallelSpin3VecCondition}
and \eqref{eq:RefMagneticDipole3VecParallelSpin3VecCondition} that
the particle's elementary dipole moments must be collinear with its
spin. The conditions \eqref{eq:RefDipole3VecsCrossProdSpin3VecZeroCondition}
also make physical sense, because if the particle had elementary dipole
moments that were not collinear with the particle's spin axis, then
electromagnetic torques acting on the particle's elementary dipole
moments would be capable of ``speeding up'' or ``slowing down''
the particle's total spin, thereby contravening the invariance of
$s^{2}$.

Finally, one can readily show that 
\begin{align}
w^{2} & =m^{2}c^{2}s^{2},\label{eq:PauliLubanskiSquareFromSpinSquare}\\
\tilde{s}^{2} & =0.\label{eq:DualSpinSquaredZero}
\end{align}
 Hence, $w^{2}$ and $\tilde{s}^{2}$ are likewise constant, as required:
\begin{align}
\frac{d}{d\tau}\parens{w^{2}} & =0,\label{eq:PauliLubanskiSquareConst}\\
\frac{d}{d\tau}\parens{\tilde{s}^{2}} & =0.\label{eq:DualSpinSquaredConst}
\end{align}

\section{Conservation Laws and their Implications}

To provide a crucial set of consistency checks on our results so far,
we now proceed to replicate them from the perspective of local conservation
laws. We will begin by discussing Noether's theorem, which we will
use to construct tensors that encode conserved notions of energy,
momentum, and angular momentum. After calculating these tensors for
the electromagnetic field coupled to a relativistic charged particle
with elementary electric and magnetic dipole moments, we will show
explicitly that the exchange of relevant conserved quantities precisely
accounts for the generalized Lorentz force law and the work done by
the field on the particle.

\subsection{Conservation Laws and Noether's Theorem}

In its various versions, Noether's theorem establishes a correspondence
between the symmetries of a physical system's dynamics and the quantities
that are conserved when the system evolves according to its equations
of motion. We will present and prove one version of the theorem whose
details will end up being particularly relevant to our elementary-dipole
model.

To begin, we consider a continuous symmetry of our system's dynamics,
meaning a transformation $q_{\alpha}\mapsto q_{\alpha}^{\prime}$
of the system's degrees of freedom that can be performed by an arbitrarily
small amount and that leaves the system's Euler-Lagrange equations
\eqref{eq:EulerLagrangeEquations} unchanged. More precisely, a continuous
symmetry has the following ingredients.
\begin{itemize}
\item The transformation rule can be expressed in infinitesimal form as
\begin{align}
q_{\alpha} & \mapsto q_{\alpha}^{\prime}=q_{\alpha}+\delta_{\epsilon}q_{\alpha},\nonumber \\
 & \qquad\qquad\delta_{\epsilon}q_{\alpha}=\sum_{b}g_{q_{\alpha},b}\epsilon_{b},\label{eq:InfinitesimalSymmetryTransf}
\end{align}
 where the coefficients $g_{q_{\alpha},b}$ depend on the degrees
of freedom and characterize the precise form of the transformation,
and where the parameters $\epsilon_{b}$ are constants that are assumed
to be small but are otherwise arbitrary.
\item The system's Lagrangian $L$ does not depend explicitly on the parameters
$\epsilon_{b}$, 
\begin{equation}
\frac{\partial L}{\partial\epsilon_{b}}=0,\label{eq:InfinitesimalSymmetryPartialLagrangianParamZero}
\end{equation}
 meaning that any possible dependence of $L$ on the parameters $\epsilon_{b}$
arises solely through the degrees of freedom $q_{\alpha}$.
\item The Lagrangian is invariant under the given transformation rule, up
to a possible total time derivative: 
\begin{align}
L & \mapsto L+\delta_{\epsilon}L,\nonumber \\
 & \qquad\delta_{\epsilon}L=\frac{d}{dt}\biggparens{\sum_{b}f_{b}\epsilon_{b}}=\sum_{b}\frac{df_{b}}{dt}\epsilon_{b}.\label{eq:InfinitesimalSymmetryChangeLagrangian}
\end{align}
 The functions $f_{b}$ here are zero in the simplest cases.
\end{itemize}
The condition \eqref{eq:InfinitesimalSymmetryChangeLagrangian} ensures
that the system's action functional $S\defeq\int dt\,L$ changes by
at most boundary terms that give no contribution when we apply the
extremization condition \eqref{eq:VariationActionFunctional} to obtain
the system's Euler-Lagrange equations.

It is important to keep in mind that in order for the transformation
\eqref{eq:InfinitesimalSymmetryTransf} to qualify as a symmetry of
the dynamics, the condition \eqref{eq:InfinitesimalSymmetryChangeLagrangian}
on the Lagrangian must hold \emph{before} applying the system's equations
of motion. Note also that identifying the correct functions $f_{b}$
is a crucial step, as we will see when we use Noether's theorem to
derive both the conserved energy-momentum and the conserved angular
momentum for the electromagnetic field coupled to an elementary dipole.

To prove the theorem and derive an explicit formula for the associated
conserved quantities, we begin by applying the chain rule to the variation
$\delta_{\epsilon}L$ of the Lagrangian appearing on the left-hand
side of \eqref{eq:InfinitesimalSymmetryChangeLagrangian}: 
\begin{align*}
 & \delta_{\epsilon}L-\sum_{b}\frac{df_{b}}{dt}\epsilon_{b}\\
 & \qquad=\sum_{\alpha}\frac{\partial L}{\partial q_{\alpha}}\delta_{\epsilon}q_{\alpha}+\sum_{\alpha}\frac{\partial L}{\partial\dot{q}_{\alpha}}\delta_{\epsilon}\dot{q}_{\alpha}\\
 & \qquad\qquad+\sum_{b}\frac{\partial L}{\partial\epsilon_{b}}\epsilon_{b}-\sum_{b}\frac{df_{b}}{dt}\epsilon_{b}=0.
\end{align*}
 Invoking the assumed transformation formula \eqref{eq:InfinitesimalSymmetryTransf}
together with the requirement \eqref{eq:InfinitesimalSymmetryPartialLagrangianParamZero}
that the Lagrangian has no explicit dependence on the transformation
parameters $\epsilon_{b}$, we have 
\[
\sum_{b}\biggparens{\sum_{\alpha}\frac{\partial L}{\partial q_{\alpha}}g_{q_{\alpha},b}\epsilon_{b}+\sum_{\alpha}\frac{\partial L}{\partial\dot{q}_{\alpha}}\dot{g}_{q_{\alpha},b}\epsilon_{b}-\frac{df_{b}}{dt}\epsilon_{b}}=0.
\]
 Using the product rule in reverse on the second term, we obtain 
\begin{align*}
 & \sum_{b}\sum_{\alpha}\biggparens{\frac{\partial L}{\partial q_{\alpha}}-\frac{d}{dt}\frac{\partial L}{\partial\dot{q}_{\alpha}}}g_{q_{\alpha},b}\epsilon_{b}\\
 & \qquad+\sum_{b}\frac{d}{dt}\biggparens{\sum_{\alpha}\frac{\partial L}{\partial\dot{q}_{\alpha}}g_{q_{\alpha},b}-f_{b}}\epsilon_{b}=0.
\end{align*}
 If we now consider a trajectory $q_{\alpha}\parens t$ that satisfies
the system's Euler-Lagrangian equations \eqref{eq:EulerLagrangeEquations},
then the first term above vanishes and we are left with 
\[
\sum_{b}\frac{d}{dt}\biggparens{\frac{\partial L}{\partial\dot{q}_{\alpha}}g_{q_{\alpha},b}-f_{b}}\epsilon_{b}=0.
\]
 This equation must hold for arbitrary values of the parameters $\epsilon_{b}$,
so we conclude that the quantity $Q_{b}$ defined as the terms in
parentheses for each value of $b$ is individually conserved.

We have therefore proved Noether's theorem, and obtained an explicit
formula for the conserved quantities $Q_{b}$ corresponding to the
given continuous symmetry: 
\begin{equation}
Q_{b}\defeq\sum_{\alpha}\frac{\partial L}{\partial\dot{q}_{\alpha}}g_{q_{\alpha},b}-f_{b},\quad\frac{dQ}{dt}=0.\label{eq:NoethersTheorem}
\end{equation}

Two important examples merit immediate discussion.
\begin{itemize}
\item If the Lagrangian $L\parens{q,\dot{q},t}$ of the system is invariant
under constant translations along the coordinates, 
\begin{equation}
q_{\alpha}\mapsto q_{\alpha}^{\prime}\defeq q_{\alpha}+\epsilon_{\alpha},\label{eq:InfinitesimalDefTranslTransf}
\end{equation}
 so that 
\begin{equation}
\delta_{\epsilon}q_{\alpha}=\epsilon_{\alpha}=\sum_{\beta}g_{q_{\alpha},\beta}\epsilon_{\beta},\quad g_{q_{\alpha},\beta}=\delta_{\alpha\beta},\label{eq:InfinitesimalTranslTransfDOF}
\end{equation}
 with 
\begin{equation}
\delta_{\epsilon}L=0,\label{eq:InfinitesimalTranslTransfLagrangianZero}
\end{equation}
 then the functions in \eqref{eq:InfinitesimalSymmetryChangeLagrangian}
vanish, $f_{\beta}=0$, and the conserved quantities \eqref{eq:NoethersTheorem}
are just the canonical momenta \eqref{eq:DefCanonicalMomenta}: 
\begin{align}
Q_{\beta} & =\sum_{\alpha}\frac{\partial L}{\partial\dot{q}_{\alpha}}g_{q_{\alpha},\beta}\nonumber \\
 & =p_{\beta}.\label{eq:NoetherCanonicalMomenta}
\end{align}
\item Consider the time translation $t\mapsto t^{\prime}\defeq t+\epsilon$
in which we shift $t$ by a small constant $\epsilon$. We require
that the values $q_{\alpha}^{\prime}\parens{t^{\prime}}$ of the system's
transformed degrees of freedom at the new time $t^{\prime}\defeq t+\epsilon$
agree with their original values $q_{\alpha}\parens t$ at the time
$t$, so that 
\begin{equation}
q_{\alpha}\parens t\mapsto q_{\alpha}^{\prime}\parens{t^{\prime}}=q_{\alpha}\parens t.\label{eq:TimeTranslDOFRule}
\end{equation}
 Equivalently, the values $q_{\alpha}^{\prime}\parens t$ of the system's
transformed degrees of freedom at the original time $t$ agree with
their values $q_{\alpha}\parens{t-\epsilon}$ at the earlier time
$t-\epsilon$, 
\begin{equation}
q_{\alpha}\parens t\mapsto q_{\alpha}^{\prime}\parens t=q_{\alpha}\parens{t-\epsilon}.\label{eq:TimeTranslDOFRuleAlt}
\end{equation}
 Then, by the chain rule, the system's degrees of freedom $q_{\alpha}$
and the Lagrangian $L$ both transform by total time derivatives:
\begin{equation}
\delta_{\epsilon}q_{\alpha}=-\dot{q}_{\alpha}\epsilon,\quad g_{q_{\alpha}}=-\dot{q}_{\alpha},\label{eq:InfinitesimalTimeTranslTransfDOF}
\end{equation}
\begin{equation}
\delta_{\epsilon}L=-\frac{dL}{dt}\epsilon.\label{eq:InfinitesimalTimeTranslTransfLagrangian}
\end{equation}
 If the Lagrangian $L\parens{q,\dot{q}}$ has no explicit dependence
on the time $t$, meaning no dependence on $t$ outside of the degrees
of freedom $q_{\alpha}$ and their rates of change $\dot{q}_{\alpha}$,
then 
\[
\frac{\partial L}{\partial\epsilon}\defeq\frac{\partial L}{\partial t}=0,
\]
 so all the conditions of Noether's theorem are satisfied with $f\defeq-L$,
and the associated conserved quantity is just the system's Hamiltonian
\eqref{eq:DefHamiltonian}, up to an overall minus sign: 
\begin{align}
Q & =\sum_{\alpha}\frac{\partial L}{\partial\dot{q}_{\alpha}}g_{q_{\alpha}}-f=-\sum_{\alpha}p_{\alpha}\dot{q}_{\alpha}+L\nonumber \\
 & =-H.\label{eq:NoetherHamiltonian}
\end{align}
\end{itemize}
Noether's theorem \eqref{eq:NoethersTheorem} generalizes naturally
to the manifestly covariant Lagrangian framework described in \citep{Barandes:2019mcl},
with the time $t$ replaced by a more general smooth, strictly monotonic
parameter $\lambda$, and with the Lagrangian $L$ replaced by the
manifestly covariant Lagrangian $\mathscr{L}=\parens{dt/d\lambda}L$,
as in \eqref{eq:ReparamInvLagrangian}.

\subsection{Energy-Momentum Tensors for Classical Field Theories}

Noether's theorem \eqref{eq:NoethersTheorem} is a powerful tool
for studying the possible conservation laws for various classical
systems, including classical field theories.

Given a classical field theory with local field degrees of freedom
$\varphi_{\alpha}\parens x$ and an action functional \eqref{eq:ClassicalFieldActionFromLagrangianDensity},
\[
S\bracks{\varphi}=\int dt\int d^{3}x\,\mathcal{L}\parens{\varphi,\partial\varphi,x},
\]
 we will start by considering the infinitesimal transformation $x^{\mu}\mapsto x^{\prime\mu}\defeq x^{\mu}+\epsilon^{\mu}$
in which we translate the spacetime coordinates $x^{\mu}$ by a small
constant four-vector $\epsilon^{\mu}$. We will then require that
the transformed values $\varphi_{\alpha}^{\prime}\parens{x^{\prime}}$
of the field degrees of freedom at the new spacetime point $x^{\prime\mu}=x^{\mu}+\epsilon^{\mu}$
are equal to their values $\varphi_{\alpha}\parens x$ at the original
spacetime point $x^{\mu}$: 
\begin{equation}
\varphi_{\alpha}^{\prime}\parens{x^{\prime}}=\varphi_{\alpha}\parens x.\label{eq:SpacetimeTranslationFieldTransfRule}
\end{equation}
 Replacing $x^{\prime\mu}$ with $x^{\mu}$ and replacing $x^{\mu}$
with $x^{\mu}-\epsilon^{\mu}$, and using the chain rule, we obtain
the following infinitesimal transformation rule for the field degrees
of freedom: 
\begin{align}
\varphi_{\alpha}\parens x & \mapsto\varphi_{\alpha}^{\prime}\parens x\defeq\varphi_{\alpha}\parens{x-\epsilon}\nonumber \\
 & \qquad=\varphi_{\alpha}\parens x-\partial_{\mu}\varphi_{\alpha}\parens x\epsilon^{\mu}.\label{eq:SpacetimeTranslationForNoetherField}
\end{align}
 That is, the infinitesimal changes in the field degrees of freedom
are given by 
\begin{equation}
\delta_{\epsilon}\varphi_{\alpha}=-\partial_{\mu}\varphi_{\alpha}\,\epsilon^{\mu},\quad g_{\varphi_{\alpha},\mu}=-\partial_{\mu}\varphi_{\alpha}.\label{eq:InfinitesimalSpacetimeTranslTransfFieldDOF}
\end{equation}

If the Lagrangian density $\mathcal{L}\parens{\varphi,\partial\varphi}$
has no explicit dependence on the spacetime coordinates $x^{\mu}$,
meaning no dependence on $x^{\mu}$ apart from any dependence arising
through $\varphi_{\alpha}$ and $\partial_{\mu}\varphi_{\alpha}$,
then all the conditions of Noether's theorem will be satisfied if
we can determine the corresponding functions $f_{\mu}$ appearing
in \eqref{eq:InfinitesimalSymmetryChangeLagrangian}. Assuming that
the fields go to zero sufficiently rapidly at spatial infinity so
that we can neglect boundary terms, and remembering from \eqref{eq:Def4DSpacetimeDerivative}
that $\partial_{t}\defeq\parens{1/c}\partial/\partial t$, we have
from the chain rule that 
\begin{align*}
\delta_{\epsilon}L & =\int d^{3}x\,\parens{-\partial_{\mu}\mathcal{L}\,\epsilon^{\mu}}=-\int d^{3}x\,\partial_{t}\mathcal{L}\,\epsilon^{t}\\
 & =-\frac{1}{c}\frac{d}{dt}\int d^{3}x\,\mathcal{L}\,\epsilon^{t}\\
 & =\frac{df_{\nu}}{dt}\epsilon^{\nu},
\end{align*}
 for 
\begin{equation}
f_{\nu}=-\int d^{3}x\,\frac{1}{c}\delta_{\nu}^{t}\mathcal{L}.\label{eq:InfinitesimalSpacetimeTranslFieldLagrangianTransfFuncs}
\end{equation}

From Noether's theorem \eqref{eq:NoethersTheorem}, suitably generalized
to the context of a classical field theory, we therefore obtain the
following collection of conserved quantities: 
\begin{align}
Q_{\nu} & =\int d^{3}x\,\biggparens{\sum_{\alpha}\frac{\partial\mathcal{L}}{\partial\parens{\partial\varphi_{\alpha}/\partial t}}g_{\varphi_{\alpha},\nu}}-f_{\nu}\nonumber \\
 & =\frac{1}{c}\int d^{3}x\,\biggparens{-\sum_{\alpha}\frac{\partial\mathcal{L}}{\partial\parens{\partial_{t}\varphi_{\alpha}}}\partial_{\nu}\varphi_{\alpha}+\delta_{\nu}^{t}\mathcal{L}}.\label{eq:NoetherFieldFourMomIntermed}
\end{align}
 Introducing a unit timelike four-vector $n_{\mu}\defeq\parens{-1,\vec 0}_{\mu}$
that is orthogonal to the three-dimensional spatial hypersurface of
integration, where the $-1$ in the temporal component of $n_{\mu}$
is merely due to its lowered Lorentz index, we can write the conserved
quantities \eqref{eq:NoetherFieldFourMomIntermed} more covariantly
as 
\begin{equation}
Q_{\nu}=\frac{1}{c}\int d^{3}x\,\parens{-n_{\mu}}\biggparens{-\sum_{\alpha}\frac{\partial\mathcal{L}}{\partial\parens{\partial_{\mu}\varphi_{\alpha}}}\partial_{\nu}\varphi_{\alpha}+\delta_{\nu}^{\mu}\mathcal{L}}.\label{eq:NoetherFieldFourMomCovariant}
\end{equation}

The conservation law $dQ_{\nu}/dt=0$ then corresponds to the vanishing
of the difference between three-dimensional integrations \eqref{eq:NoetherFieldFourMomCovariant}
on two adjacent spatial hypersurfaces separated by an infinitesimal
amount of time $dt$. Hence, by the four-dimensional divergence theorem,
and under the assumption that the fields go to zero sufficiently rapidly
at spatial infinity, the equation $dQ_{\nu}/dt=0$ implies that the
quantity in parentheses in \eqref{eq:NoetherFieldFourMomCovariant}
has vanishing spacetime divergence: 
\begin{equation}
\partial_{\mu}\biggparens{-\sum_{\alpha}\frac{\partial\mathcal{L}}{\partial\parens{\partial_{\mu}\varphi_{\alpha}}}\partial_{\nu}\varphi_{\alpha}+\delta_{\nu}^{\mu}\mathcal{L}}=0.\label{eq:NoetherSpacetimeTranslFieldConservationLaw}
\end{equation}

Raising the $\nu$ index on the quantity in parentheses using the
Minkowski metric tensor, we define the system's canonical energy-momentum tensor
to be the result: 
\begin{equation}
T_{\textrm{can}}^{\mu\nu}\defeq-\sum_{\alpha}\frac{\partial\mathcal{L}}{\partial\parens{\partial_{\mu}\varphi_{\alpha}}}\partial^{\nu}\varphi_{\alpha}+\eta^{\mu\nu}\mathcal{L}.\label{eq:DefCanonicalEnergyMomentumTensor}
\end{equation}
 From \eqref{eq:NoetherSpacetimeTranslFieldConservationLaw}, this
tensor then satisfies the local conservation law 
\begin{equation}
\partial_{\mu}T_{\textrm{can}}^{\mu\nu}=0,\label{eq:LocalConservationCanonicalEnergyMomentumTensor}
\end{equation}
 and naturally generalizes the Hamiltonian \eqref{eq:NoetherHamiltonian}
to a local, Lorentz-covariant density of energy and momentum.

Notice that Noether's theorem does not determine $T_{\textrm{can}}^{\mu\nu}$
uniquely, because we are free to add terms to the definition \eqref{eq:DefCanonicalEnergyMomentumTensor}
that have vanishing spacetime divergence without affecting the local
energy-momentum conservation law \eqref{eq:LocalConservationCanonicalEnergyMomentumTensor}:
\begin{equation}
T^{\mu\nu}\defeq T_{\textrm{can}}^{\mu\nu}+\parens{\cdots}^{\mu\nu},\qquad\partial_{\mu}\parens{\cdots}^{\mu\nu}=0.\label{eq:DefPhysicalEnergyMomTensorFromCanonicalWithAdditiveTerm}
\end{equation}
That is, this redefined energy-momentum tensor $T^{\mu\nu}$ would
continue to satisfy the equation 
\begin{equation}
\partial_{\mu}T^{\mu\nu}=0.\label{eq:LocalConservationEnergyMomentumTensor}
\end{equation}

The addition of terms as in \eqref{eq:DefPhysicalEnergyMomTensorFromCanonicalWithAdditiveTerm}
may be necessary to ensure that the energy-momentum tensors $T^{\mu\nu}$
for certain field theories have particular properties, like gauge
invariance. However, even when such a redefinition \eqref{eq:DefPhysicalEnergyMomTensorFromCanonicalWithAdditiveTerm}
provides a better description of a system's underlying physics, the
canonical energy-momentum tensor $T_{\textrm{can}}^{\mu\nu}$ may
still be more convenient for certain calculations, as we will see
in our work ahead. 

The first index $\mu$ on $T^{\mu\nu}$ determines whether we are
referring to a volume density or to a flux density, the latter representing
a rate of flow per unit time, per unit cross-sectional area, so we
will refer to $\mu$ as the flux index of $T^{\mu\nu}$. The second
index $\nu$ tells us whether the physical quantity in question is
energy or momentum, so we will refer to $\nu$ as the four-momentum index
of $T^{\mu\nu}$. In analogy with \eqref{eq:InterpretationCurrentDensity}
for the charge-current density $j^{\mu}$, we therefore have the schematic
formula 
\begin{equation}
T^{\mu\nu}=\begin{cases}
\textrm{density of }\parens{\textrm{momentum}}^{\nu} & \textrm{for }\mu=t,\\
\textrm{flux density of }\parens{\textrm{momentum}}^{\nu} & \textrm{for }\mu=x,y,z.
\end{cases}\label{eq:EnergyMomentumTensorInterpretation}
\end{equation}
More concretely, the individual components of $T^{\mu\nu}$ have the
following physical interpretations.
\begin{itemize}
\item The three-dimensional scalar 
\begin{equation}
u=T^{tt}\label{eq:DefFieldEnergyDensity}
\end{equation}
 represents the volume density of the field's mass-energy.
\item The three-dimensional vector 
\begin{equation}
\vec S=c\,\parens{T^{xt},T^{yt},T^{zt}},\label{eq:DefFieldEnergyFluxDensity}
\end{equation}
 not to be confused with our notation for a spin three-vector, represents
the flux density of the field's energy, meaning the rate of energy
flow per unit time, per unit cross-sectional area.
\item The three-dimensional vector 
\begin{equation}
\vec g=\frac{1}{c}\parens{T^{tx},T^{ty},T^{tz}}\label{eq:DefFieldMomentumDensity}
\end{equation}
 represents the field's momentum density.
\item The three-dimensional tensor 
\begin{equation}
\mathbb{T}_{ij}=-T^{ij},\label{eq:DefFieldStressTensor}
\end{equation}
 called the field's stress tensor, represents the field's momentum
flux densities, with the $\parens{i,j}$ component representing the
flux density of the $j$th component of momentum in the $i$th direction.
The diagonal components $\mathbb{T}_{xx},\mathbb{T}_{yy},\mathbb{T}_{zz}$
encode the pressures in each of the three Cartesian directions, and
the off-diagonal components $\mathbb{T}_{xy},\mathbb{T}_{xz},\mathbb{T}_{yx},\mathbb{T}_{yz},\mathbb{T}_{zx},\mathbb{T}_{zy}$
encode shearing effects.
\end{itemize}
If we introduce terms into the action functional \eqref{eq:ClassicalFieldActionFromLagrangianDensity}
that describe interactions between the field and source systems, such
as mechanical particles, then these source systems will generically
exchange energy and momentum with the field in the form of work and
forces. Because these flows of energy and momentum imply that the
field can gain or lose energy and momentum, they appear as violations
of the local conservation equation \eqref{eq:LocalConservationEnergyMomentumTensor},
$\partial_{\mu}T^{\mu\nu}=0$, that would have otherwise held for
the field alone.

Specifically, any energy entering or leaving the field corresponds
to violations of the $\nu=t$ component of \eqref{eq:LocalConservationEnergyMomentumTensor}
that describe the rate at which work is done by the field on sources.
Any momentum entering or leaving the field corresponds to violations
of the $\nu=x,y,z$ components of \eqref{eq:LocalConservationEnergyMomentumTensor}
that describe forces due to the field on sources.

We can capture all these violations in terms of a new four-vector
$f^{\nu}$ that is related to the spacetime divergence of the field's
energy-momentum tensor $T^{\mu\nu}$ according to the local four-force law
\begin{equation}
f^{\nu}=-\partial_{\mu}T^{\mu\nu}.\label{eq:4DLocalEnergyMomentumConservationLawWithPowerForceDensity}
\end{equation}
 Letting $\partial w/\partial t$ denote the power density on sources,
meaning the rate at which the field does work on sources per unit
volume, and letting $\vec f=\parens{f_{x},f_{y},f_{z}}$ denote the
field's force density on sources, the preceding analysis implies that
\begin{equation}
f^{\nu}\defeq\biggparens{\frac{1}{c}\frac{\partial w}{\partial t},f_{x},f_{y},f_{z}}^{\nu},\label{eq:4DPowerForceDensity}
\end{equation}
and so we naturally refer to $f^{\nu}$ as the field's four-force density.
(Four-forces are also called Minkowski forces.)

Given a knowledge of a field's energy-momentum tensor, the local four-force
equation \eqref{eq:4DLocalEnergyMomentumConservationLawWithPowerForceDensity}
provides a very general way to derive force laws on source particles.
In particular, we will see in the example of the electromagnetic field
that \eqref{eq:4DLocalEnergyMomentumConservationLawWithPowerForceDensity}
will end up yielding the Lorentz force law in the more general form
\eqref{eq:LorentzForceLawNonRelApproxIntro} that includes forces
on elementary dipoles.

\subsection{Angular-Momentum Flux Tensors for Classical Field Theories}

We can also use Noether's theorem to determine the local conservation
law corresponding to Lorentz invariance. Under Lorentz transformations,
the spacetime coordinates $x^{\mu}$ transform as in \eqref{eq:SpacetimeCoordsLorentzTransf}:
\begin{equation}
x^{\mu}\mapsto x^{\prime\mu}\defeq\tud{\Lambda}{\mu}{\nu}x^{\nu}.\label{eq:LorentzTransfSpacetimeCoords}
\end{equation}
 We require that the new values $\varphi_{\alpha}^{\prime}\parens{x^{\prime}}$
of the field degrees of freedom at $x^{\prime\mu}$ are related to
their values $\varphi_{\alpha}\parens x$ at $x^{\mu}$ according
to a general rule of the form 
\begin{equation}
\varphi_{\alpha}\parens x\mapsto\varphi_{\alpha}^{\prime}\parens{x^{\prime}}\defeq\parens{F\parens{\Lambda}\varphi}_{\alpha}\parens x,\label{eq:LorentzTransfFieldDOF}
\end{equation}
 where $F\parens{\Lambda}$ captures the possibility that the field
index $\alpha$ has a nontrivial behavior under Lorentz transformations.
Equivalently, replacing $x^{\prime\mu}=\parens{\Lambda x}^{\mu}$
with $x^{\mu}$ and replacing $x^{\mu}$ with $\parens{\Lambda^{-1}x}^{\mu}$,
we have 
\begin{equation}
\varphi_{\alpha}^{\prime}\parens x=\parens{F\parens{\Lambda}\varphi}_{\alpha}\parens{\Lambda^{-1}x}.\label{eq:LorentzTransfFieldDOFAlt}
\end{equation}

Specializing now to an infinitesimal Lorentz transformation \eqref{eq:InfinitesimalLorentzTransfFromGenerators},
chosen to be an active transformation by replacing $-d\theta^{\mu\nu}\mapsto+\epsilon^{\mu\nu}$,
we have 
\begin{equation}
\Lambda_{\textrm{inf}}=1+\frac{i}{2}\epsilon^{\mu\nu}\sigma_{\mu\nu},\label{eq:InfinitesimalLorentzTransfFromGeneratorsForNoether}
\end{equation}
 with $\Lambda_{\textrm{inf}}^{-1}$ acting as a passive transformation,
\begin{equation}
\Lambda_{\textrm{inf}}^{-1}=1-\frac{i}{2}\epsilon^{\mu\nu}\sigma_{\mu\nu}.\label{eq:InfinitesimalInverseLorentzTransfFromGeneratorsForNoether}
\end{equation}
 The field degrees of freedom then transform as 
\begin{align}
 & \varphi_{\alpha}^{\prime}\parens x=\parens{F\parens{1+\parens{i/2}\epsilon^{\mu\nu}\sigma_{\mu\nu}}\varphi}_{\alpha}\parens{\parens{1-\parens{i/2}\epsilon^{\rho\sigma}\sigma_{\rho\sigma}}x}\nonumber \\
 & =\varphi_{\alpha}\parens x-\partial_{\mu}\varphi\parens x\frac{i}{2}\epsilon^{\rho\sigma}\tud{\bracks{\sigma_{\rho\sigma}}}{\mu}{\nu}x^{\nu}+\frac{1}{2}\parens{\Delta_{\rho\sigma}\varphi}_{\alpha}\parens x\epsilon^{\rho\sigma},\label{eq:LorentzTranslationForNoetherField}
\end{align}
 where the final term represents the infinitesimal changes in the
fields at fixed spacetime coordinates $x^{\mu}$: 
\begin{equation}
\frac{1}{2}\parens{\Delta_{\rho\sigma}\varphi}_{\alpha}\parens x\epsilon^{\rho\sigma}\defeq\parens{F\parens{1+\parens{i/2}\epsilon^{\mu\nu}\sigma_{\mu\nu}}\varphi}_{\alpha}\parens x-\varphi_{\alpha}\parens x.\label{eq:DefInfinitesimalFieldTransfFixedArgument}
\end{equation}
 Dropping factors of $1/2$ to avoid double-counting independent variables,
we can therefore identify 
\begin{equation}
g_{\varphi_{\alpha},\rho\sigma}=-\partial_{\mu}\varphi\,i\tud{\bracks{\sigma_{\rho\sigma}}}{\mu}{\nu}x^{\nu}+\parens{\Delta_{\rho\sigma}\varphi}_{\alpha}.\label{eq:InfintesimalLorentzTransfFieldNoetherFunc}
\end{equation}

If the field theory's Lagrangian density is Lorentz invariant, then
all we have left to do is determine the functions $f_{\rho\sigma}$
appearing in \eqref{eq:InfinitesimalSymmetryChangeLagrangian}. We
find 
\begin{align*}
\delta_{\epsilon}L & =\int d^{3}x\,\parens{\partial_{\mu}\mathcal{L}}\biggparens{-\frac{i}{2}\epsilon^{\rho\sigma}\tud{\bracks{\sigma_{\rho\sigma}}}{\mu}{\nu}x^{\nu}}\\
 & =\int d^{3}x\,\partial_{\mu}\left(\mathcal{L}\biggparens{-\frac{i}{2}\epsilon^{\rho\sigma}\tud{\bracks{\sigma_{\rho\sigma}}}{\mu}{\nu}x^{\nu}}\right)\\
 & \qquad+\frac{i}{2}\epsilon^{\rho\sigma}\int d^{3}x\,\mathcal{L}\,\tud{\bracks{\sigma_{\rho\sigma}}}{\mu}{\nu}\partial_{\mu}x^{\nu}\\
 & =-\frac{1}{c}\frac{d}{dt}\int d^{3}x\,\mathcal{L}\,\biggparens{-\frac{i}{2}\epsilon^{\rho\sigma}\tud{\bracks{\sigma_{\rho\sigma}}}t{\nu}x^{\nu}}\\
 & \qquad+\frac{i}{2}\epsilon^{\rho\sigma}\int d^{3}x\,\mathcal{L}\,\tud{\bracks{\sigma_{\rho\sigma}}}{\mu}{\nu}\delta_{\mu}^{\nu}.\\
\end{align*}
The antisymmetry of the Lorentz generators \eqref{eq:LorentzGeneratorsMixedIndices}
implies that $\tud{\bracks{\sigma_{\rho\sigma}}}{\mu}{\nu}\delta_{\mu}^{\nu}=0$,
so the second term vanishes, and we have 
\[
\delta_{\epsilon}L=\frac{1}{2}\frac{df_{\rho\sigma}}{dt}\epsilon^{\rho\sigma},
\]
 with 
\begin{equation}
f_{\rho\sigma}=-\int d^{3}x\,\frac{1}{c}\delta_{\mu}^{t}\mathcal{L}\,i\tud{\bracks{\sigma_{\rho\sigma}}}{\mu}{\nu}x^{\nu}.\label{eq:InfinitesimalLorentzTransfFieldLagrangianTransfFuncs}
\end{equation}

Thus, according to Noether's theorem \eqref{eq:NoethersTheorem},
we end up with the following conserved quantities: 
\begin{align}
Q_{\nu\rho} & =\frac{1}{c}\int d^{3}x\,\parens{-n_{\mu}}\nonumber \\
 & \qquad\quad\times\biggparens{-\sum_{\alpha}\frac{\partial\mathcal{L}}{\partial\parens{\partial_{\mu}\varphi_{\alpha}}}\partial_{\sigma}\varphi_{\alpha}+\delta_{\sigma}^{\mu}\mathcal{L}}i\tud{\bracks{\sigma_{\nu\rho}}}{\sigma}{\lambda}x^{\lambda}\nonumber \\
 & \quad+\frac{1}{c}\int d^{3}x\,\parens{-n_{\mu}}\biggparens{\sum_{\alpha}\frac{\partial\mathcal{L}}{\partial\parens{\partial_{\mu}\varphi_{\alpha}}}}\parens{\Delta_{\nu\rho}\varphi}_{\alpha},\label{eq:NoetherFieldTotAngMomCovariant}
\end{align}
 where, again, $n_{\mu}\defeq\parens{-1,\vec 0}_{\mu}$ is a unit
timelike four-vector that is orthogonal to the three-dimensional spatial
hypersurface of integration. Raising the $\nu$ and $\rho$ indices,
and recalling the definition \eqref{eq:DefCanonicalEnergyMomentumTensor}
of the field's canonical energy-momentum tensor $T_{\textrm{can}}^{\mu\nu}$
together with the formula \eqref{eq:LorentzGeneratorsMixedIndices}
for the Lorentz generators $\tud{\bracks{\sigma_{\mu\nu}}}{\alpha}{\beta}$,
we can write these conserved quantities as 
\begin{equation}
Q^{\nu\rho}=-\int d^{3}x\,\parens{-n_{\mu}}\mathcal{J}_{\textrm{can}}^{\mu\nu\rho},\label{eq:NoetherFieldTotAngMomCovariantFromAngMomFluxTensor}
\end{equation}
 where 
\begin{equation}
\mathcal{J}_{\textrm{can}}^{\mu\nu\rho}\defeq\mathcal{L}^{\mu\nu\rho}+\mathcal{S}^{\mu\nu\rho}=-\mathcal{J}_{\textrm{can}}^{\mu\rho\nu},\label{eq:DefFieldAngMomFluxTensor}
\end{equation}
\begin{equation}
\mathcal{L}^{\mu\nu\rho}\defeq x^{\nu}\frac{1}{c}T_{\textrm{can}}^{\mu\rho}-x^{\rho}\frac{1}{c}T_{\textrm{can}}^{\mu\nu}=-\mathcal{L}^{\mu\rho\nu},\label{eq:DefFieldOrbAngMomFluxTensor}
\end{equation}
 and 
\begin{equation}
\mathcal{S}^{\mu\nu\rho}\defeq-\frac{1}{c}\sum_{\alpha}\frac{\partial\mathcal{L}}{\partial\parens{\partial_{\mu}\varphi_{\alpha}}}\parens{\Delta^{\nu\rho}\varphi}_{\alpha}=-\mathcal{S}^{\mu\rho\nu}\label{eq:DefFieldSpinFluxTensor}
\end{equation}
 are all antisymmetric on their final two indices, and where $\mathcal{J}_{\textrm{can}}^{\mu\nu\rho}$
is locally conserved: 
\begin{equation}
\partial_{\mu}\mathcal{J}_{\textrm{can}}^{\mu\nu\rho}=0.\label{eq:LocalConservationFieldTotAngMomCanonical}
\end{equation}

The tensor $\mathcal{L}^{\mu\nu\rho}$ generalizes the mechanical
definition $\vec L=\vec X\crossprod\vec p$ of orbital angular momentum
for particles, whereas the tensor $\mathcal{S}^{\mu\nu\rho}$ represents
intrinsic spin angular momentum in the field itself, so $\mathcal{J}_{\textrm{can}}^{\mu\nu\rho}$
is called the canonical total angular-momentum flux tensor. 

The local conservation laws \eqref{eq:LocalConservationCanonicalEnergyMomentumTensor}
for $T_{\textrm{can}}^{\mu\nu}$ and \eqref{eq:LocalConservationFieldTotAngMomCanonical}
for $\mathcal{J}_{\textrm{can}}^{\mu\nu\rho}$ together imply that
the spacetime divergence of the field's spin flux tensor $\mathcal{S}^{\mu\nu\rho}$
characterizes the lack of symmetry in the two indices of the field's
canonical energy-momentum tensor $T_{\textrm{can}}^{\mu\nu}$: 
\begin{equation}
T_{\textrm{can}}^{\nu\rho}-T_{\textrm{can}}^{\rho\nu}=-c\,\partial_{\mu}\mathcal{S}^{\mu\nu\rho}.\label{eq:AsymmetryCanEnergyMomTensorFromDivSpin}
\end{equation}
 As reviewed in \citep{DiFrancescoMathieuSenechal:1996cft}, we can
use this relation to construct a symmetric energy-momentum tensor
and simplify the formula \eqref{eq:DefFieldAngMomFluxTensor} for
the canonical total angular-momentum flux tensor. We start by defining
the Belinfante-Rosenfeld tensor, 
\begin{equation}
\mathcal{B}^{\mu\rho\nu}\defeq\frac{c}{2}\parens{\mathcal{S}^{\mu\nu\rho}+\mathcal{S}^{\nu\mu\rho}+\mathcal{S}^{\rho\mu\nu}},\label{eq:DefBelinfanteRosenfeldCorrection}
\end{equation}
 which is antisymmetric on its first two indices, 
\begin{equation}
\mathcal{B}^{\mu\rho\nu}=-\mathcal{B}^{\rho\mu\nu},\label{eq:BelinfanteRosenfeldCorrectionAntisymm}
\end{equation}
 is \emph{asymmetric} (rather than \emph{antisymmetric}) on its first
and last indices according to 
\begin{equation}
\mathcal{B}^{\mu\rho\nu}=\mathcal{B}^{\nu\rho\mu}+c\mathcal{S}^{\rho\mu\nu},\label{eq:BelinfanteRosenfeldCorrectionAsymmetric}
\end{equation}
and has the property that its spacetime divergence $\partial_{\rho}\mathcal{B}^{\mu\rho\nu}$
on its second index is automatically locally conserved: 
\begin{equation}
\partial_{\mu}\parens{\partial_{\rho}\mathcal{B}^{\mu\rho\nu}}=0.\label{eq:BelinfanteRosenfeldCorrectionLocallyConserved}
\end{equation}
 The redefined energy-momentum tensor 
\begin{equation}
T^{\mu\nu}\defeq T_{\textrm{can}}^{\mu\nu}+\partial_{\rho}\mathcal{B}^{\mu\rho\nu}\label{eq:DefBelinfanteRosenfeldEnegyMomTensor}
\end{equation}
 then continues to satisfy the local conservation equation \eqref{eq:LocalConservationCanonicalEnergyMomentumTensor},
\[
\partial_{\mu}T^{\mu\nu}=0,
\]
 is symmetric on its two indices, 
\begin{equation}
T^{\mu\nu}=T^{\nu\mu},\label{eq:BelinfanteRosenfeldEnergyMomTensorSymmetric}
\end{equation}
 and, assuming that the fields go to zero sufficiently rapidly at
spatial infinity, $T^{t\nu}$ has the same integrated value over all
of three-dimensional space as $T_{\textrm{can}}^{t\nu}$: 
\begin{equation}
\int d^{3}x\,T^{t\nu}=\int d^{3}x\,T_{\textrm{can}}^{t\nu}.\label{eq:BelinfanteRosenfeldEnergyMomTensorIntegrated}
\end{equation}
 Moreover, the new total angular-momentum flux tensor defined by
\begin{equation}
\mathcal{J}^{\mu\nu\rho}\defeq x^{\nu}\frac{1}{c}T^{\mu\rho}-x^{\rho}\frac{1}{c}T^{\mu\nu}\label{eq:DefAngMomentumFluxTensor}
\end{equation}
 differs from the canonical total angular-momentum flux tensor $\mathcal{J}_{\textrm{can}}^{\mu\nu\rho}$
by a term that is antisymmetric on its final two indices and has vanishing
spacetime divergence, so $\mathcal{J}^{\mu\nu\rho}$ is still locally
conserved: 
\begin{equation}
\partial_{\mu}\mathcal{J}^{\mu\nu\rho}=0.\label{eq:4DLocalConservationAngMom}
\end{equation}
The tensor $\mathcal{J}^{\mu\nu\rho}$ also has the same integrated
value over all of three-dimensional space as $\mathcal{J}_{\textrm{can}}^{\mu\nu\rho}$,
so we are free to use $\mathcal{J}^{\mu\nu\rho}$ instead of $\mathcal{J}_{\textrm{can}}^{\mu\nu\rho}$
to describe the field's total angular momentum.  

If we include terms in the field's action functional \eqref{eq:ClassicalFieldActionFromLagrangianDensity}
that describe interactions with source systems, then the spacetime
divergence $\partial_{\mu}\mathcal{J}^{\mu\nu\rho}$ characterizes
the degree to which the angular momentum of the field is locally conserved,
and satisfies the equation 
\begin{equation}
-c\partial_{\mu}\mathcal{J}^{\mu\nu\rho}=x^{\nu}f^{\rho}-x^{\rho}f^{\nu},\label{eq:4DLocalTorqueLaw}
\end{equation}
 where $f^{\nu}=-\partial_{\mu}T^{\mu\nu}$ is the four-force density
from \eqref{eq:4DLocalEnergyMomentumConservationLawWithPowerForceDensity}.
 The terms $x^{\nu}f^{\rho}-x^{\rho}f^{\nu}$, which generalize the
mechanical definition $\vecgreek{\vec{\tau}}=\vec X\crossprod\vec F$
of torque, describe the density of torques exerted by the field on
the source system. If this torque density vanishes, then we get back
the local conservation law \eqref{eq:4DLocalConservationAngMom},
\[
\partial_{\mu}\mathcal{J}^{\mu\nu\rho}=0,
\]
 thereby implying that the field's angular momentum is locally conserved.

As an aside, notice the formal resemblance between the decomposition
\eqref{eq:DefBelinfanteRosenfeldEnegyMomTensor} of the redefined
energy-momentum tensor, 
\[
T^{\mu\nu}\defeq T_{\textrm{can}}^{\mu\nu}+\partial_{\rho}\mathcal{B}^{\mu\rho\nu},
\]
 and the first two terms of the series expansion \eqref{eq:4DCurrentDensityDerivativeExpansionExplicit}
for the current density $j^{\nu}$, 
\[
j^{\nu}=j_{\textrm{e}}^{\nu}+\partial_{\mu}M^{\mu\nu}+\dotsb.
\]
 We see that the spacetime-divergence term in $T^{\mu\nu}$ representing
the intrinsic spin of the classical field is analogous to the spacetime-divergence
term in $j^{\nu}$ representing the contribution from electric and
magnetic dipoles.

Observe also that if we use the energy-momentum tensor \eqref{eq:DefBelinfanteRosenfeldEnegyMomTensor},
which is symmetric on its two indices, $T^{\mu\nu}=T^{\nu\mu}$, then
\[
\parens{T^{xt},T^{yt},T^{zt}}=\parens{T^{tx},T^{ty},T^{tz}},
\]
 so we obtain the following simple relationship between the field's
energy flux density \eqref{eq:DefFieldEnergyFluxDensity} and the
field's momentum density \eqref{eq:DefFieldMomentumDensity}: 
\begin{equation}
\vec S=\vec gc^{2}.\label{eq:SymmetricEnergyMomentumTensorEqualEnergyFluxMomDensity}
\end{equation}
 If we consider spatially compact distributions of the field propagating
at an overall velocity $\vec v$, then integrating this formula over
three-dimensional space yields a relationship between the total field
energy $E$ and the total field momentum $\vec p$, 
\[
\vec vE=\vec pc^{2},
\]
 or, equivalently, 
\begin{equation}
\vec v=\frac{\vec pc^{2}}{E},\label{eq:RelativisticVelocityMomentumEnergyRelationshipForFields}
\end{equation}
 which we first saw in our formula \eqref{eq:RelativisticVelocityMomentumEnergyRelationship}
for relativistic particles.

Note that the existence of the relationships \eqref{eq:SymmetricEnergyMomentumTensorEqualEnergyFluxMomDensity}
and \eqref{eq:RelativisticVelocityMomentumEnergyRelationshipForFields}
does not fundamentally depend on our use of $T^{\mu\nu}$ rather than
$T_{\textrm{can}}^{\mu\nu}$. Using $T^{\mu\nu}$ merely makes them
easier to derive.

\subsection{Local Conservation of Energy and Momentum for the Free Electromagnetic
Field}

For the electromagnetic field in the absence of charges and currents,
meaning that $j^{\mu}=\parens{\rho c,\vec J}^{\mu}=0$, the action
functional is \eqref{eq:MaxwellActionInVacuum}: 
\begin{align*}
S_{\textrm{field}}\bracks A & \defeq\int dt\int d^{3}x\,\mathcal{L}_{\textrm{field}}\\
 & =\int dt\int d^{3}x\,\biggparens{-\frac{1}{4\mu_{0}}F^{\mu\nu}F_{\mu\nu}}.
\end{align*}
 Thus, the definition \eqref{eq:DefCanonicalEnergyMomentumTensor}
of the electromagnetic field's canonical energy-momentum tensor yields
\begin{align}
T_{\textrm{can}}^{\mu\nu} & \defeq-\frac{\partial\mathcal{L}_{\textrm{field}}}{\partial\parens{\partial_{\mu}A_{\rho}}}\partial^{\nu}A_{\rho}+\eta^{\mu\nu}\mathcal{L}_{\textrm{field}}\nonumber \\
 & =\frac{1}{\mu_{0}}F^{\mu\rho}\partial^{\nu}A_{\rho}-\eta^{\mu\nu}\frac{1}{4\mu_{0}}F^{\rho\sigma}F_{\rho\sigma}.\label{eq:Canonical4DElectromagneticEnergyMomentumTensor}
\end{align}
 As a consequence of the invariance of the dynamics under constant
translations in time and space, Noether's theorem guarantees that
this canonical energy-momentum tensor satisfies the local conservation
law \eqref{eq:LocalConservationCanonicalEnergyMomentumTensor}: 
\begin{equation}
\partial_{\mu}T_{\textrm{can}}^{\mu\nu}=0.\label{eq:Canonical4DElectromagneticEnergyMomentumTensorConserved}
\end{equation}

However, $T_{\textrm{can}}^{\mu\nu}$ is not invariant under gauge
transformations \eqref{eq:4DGaugeTransformation}, due to the explicit
appearance of the gauge potential $A_{\rho}$ in its first term, 
\begin{equation}
\frac{1}{\mu_{0}}F^{\mu\rho}\partial^{\nu}A_{\rho}.\label{eq:Canonical4DElectromagneticEnergyMomentumTensorNonGaugeInvTerm}
\end{equation}
 Notice that we could remedy this issue by adding on a new term 
\begin{equation}
T_{\textrm{add}}^{\mu\nu}\defeq-\frac{1}{\mu_{0}}F^{\mu\rho}\partial_{\rho}A^{\nu},\label{eq:AddOn4DElectromagneticEnergyMomentumTensor}
\end{equation}
 which would have the effect of converting the non-gauge-invariant
term \eqref{eq:Canonical4DElectromagneticEnergyMomentumTensorNonGaugeInvTerm}
into the manifestly gauge-invariant combination 
\begin{equation}
\frac{1}{\mu_{0}}F^{\mu\rho}\parens{\partial^{\nu}A_{\rho}-\partial_{\rho}A^{\nu}}=\frac{1}{\mu_{0}}F^{\mu\rho}\tud F{\nu}{\rho}.\label{eq:4DElectromagnetcEnergyMomentumTensorNowGaugeInvTerm}
\end{equation}
 Invoking the inhomogeneous Maxwell equation \eqref{eq:4DInhomogMaxwellEq}
in the absence of sources, $\partial_{\rho}F^{\mu\rho}=0$, we can
write $T_{\textrm{add}}^{\mu\nu}$ alternatively as a total spacetime
divergence: 
\begin{equation}
T_{\textrm{add}}^{\mu\nu}=\partial_{\rho}\biggparens{-\frac{1}{\mu_{0}}F^{\mu\rho}A^{\nu}}.\label{eq:AddOn4DElectromagneticEnergyMomentumTensorAsDivergence}
\end{equation}
 It follows immediately from the antisymmetry of the indices $\mu$
and $\rho$ on $F^{\mu\rho}$ that this proposed new term has vanishing
spacetime divergence, 
\begin{equation}
\partial_{\mu}T_{\textrm{add}}^{\mu\nu}=\partial_{\mu}\partial_{\rho}\biggparens{-\frac{1}{\mu_{0}}F^{\mu\rho}A^{\nu}}=0,\label{eq:AddOn4DElectromagneticEnergyMomentumTensorZeroDivergence}
\end{equation}
 so adding it to the canonical energy-momentum tensor $T_{\textrm{can}}^{\mu\nu}$
would have no effect on the local conservation equation \eqref{eq:Canonical4DElectromagneticEnergyMomentumTensorConserved}.
Furthermore, if we integrate the energy-momentum volume density $T_{\textrm{add}}^{t\nu}$
over three-dimensional space, then because $F^{tt}=0$, we end up
with the integral of a total three-dimensional divergence that vanishes
under the assumption that our fields go to zero sufficiently rapidly
at spatial infinity: 
\begin{align*}
\int d^{3}x\,T_{\textrm{add}}^{t\nu} & =\int d^{3}x\,\partial_{\rho}\biggparens{-\frac{1}{\mu_{0}}F^{t\rho}A^{\nu}}\\
 & =\int d^{3}x\,\nabla\dotprod\parens{\cdots}=0.
\end{align*}
 Hence, adding $T_{\textrm{add}}^{\mu\nu}$ to $T_{\textrm{can}}^{\mu\nu}$
does not alter the field's total energy and momentum.

The sum $T_{\textrm{can}}^{\mu\nu}+T_{\textrm{add}}^{\mu\nu}$ gives
us the physical (and gauge-invariant) electromagnetic energy-momentum tensor:
\begin{align}
T^{\mu\nu} & =T_{\textrm{can}}^{\mu\nu}-\frac{1}{\mu_{0}}F^{\mu\rho}\partial_{\rho}A^{\nu}\nonumber \\
 & =\frac{1}{\mu_{0}}F^{\mu\rho}\tud F{\nu}{\rho}-\eta^{\mu\nu}\frac{1}{4\mu_{0}}F^{\rho\sigma}F_{\rho\sigma}.\label{eq:4DElectromagneticEnergyMomentumTensor}
\end{align}
 By construction, in the absence of charged sources, this energy-momentum
continues to satisfy the local conservation law \eqref{eq:LocalConservationCanonicalEnergyMomentumTensor},
\begin{equation}
\partial_{\mu}T^{\mu\nu}=0,\label{eq:4DEnergyMomentumTensorLocalConservation}
\end{equation}
 and its individual components describe the density and flux of electromagnetic
energy and momentum throughout three-dimensional space.
\begin{itemize}
\item The electromagnetic energy density is 
\begin{equation}
u=T^{tt}=\frac{1}{2}\biggparens{\epsilon_{0}\vec E^{2}+\frac{1}{\mu_{0}}\vec B^{2}}.\label{eq:ElectromagneticEnergyDensity}
\end{equation}
\item The electromagnetic energy flux density is 
\begin{equation}
\vec S=c\parens{T^{xt},T^{yt},T^{zt}}=\frac{1}{\mu_{0}}\vec E\crossprod\vec B,\label{eq:PoyntingVector}
\end{equation}
 which is also known as the Poynting vector.
\item The electromagnetic momentum density is 
\begin{equation}
\vec g=\frac{1}{c}\parens{T^{tx},T^{ty},T^{tz}}=\epsilon_{0}\vec E\crossprod\vec B.\label{eq:ElectromagneticMomentumDensity}
\end{equation}
\item The electromagnetic momentum flux density is given by the Maxwell stress tensor,
\begin{align}
\mathbb{T} & =-\begin{pmatrix}T^{xx} & T^{xy} & T^{xz}\\
T^{yx} & T^{yy} & T^{yz}\\
T^{zx} & T^{zy} & T^{zz}
\end{pmatrix}\nonumber \\
 & =\epsilon_{0}\vec E\vec E+\frac{1}{\mu_{0}}\vec B\vec B-\mathbb{I}\frac{1}{2}\biggparens{\epsilon_{0}\vec E^{2}+\frac{1}{\mu_{0}}\vec B^{2}},\label{eq:MaxwellStressTensor}
\end{align}
 where $\mathbb{I}$ is the identity tensor.
\end{itemize}

\subsection{Local Conservation of Energy and Momentum for the Electromagnetic
Field Coupled to an Elementary Dipole}

When we couple the electromagnetic field to a charged particle with
elementary dipole moments, the energy and momentum of the field become
mixed together with those of the particle. As a result, in order to
study local conservation of energy and momentum for the overall system,
we will need to look again at the full action functional \eqref{eq:MaxwellActionWithParticleDipoleIntsAfterIBP},
which we can use \eqref{eq:IntegratedElectricMonopoleTerm} and \eqref{eq:IntegratedDipoleInteractionTerm}
to write as 
\begin{align}
S\bracks{X,\Lambda,A} & =\int dt\int d^{3}x\,\mathcal{L}\nonumber \\
 & =\int d\lambda\,\biggparens{p_{\nu}\dot{X}^{\nu}+\frac{1}{2}\Trace\bracks{S\dot{\Lambda}\Lambda^{-1}}}\nonumber \\
 & \qquad+\int dt\int d^{3}x\,\biggparens{-\frac{1}{4\mu_{0}}F^{\mu\nu}F_{\mu\nu}}\nonumber \\
 & \qquad+\int d\lambda\,q\dot{X}^{\nu}A_{\nu}\nonumber \\
 & \qquad-\frac{1}{2c}\int d\lambda\,\sqrt{-\dot{X}^{2}}m^{\mu\nu}F_{\mu\nu}.\label{eq:MaxwellActionWithParticleDipoleIntsAfterReducingInts}
\end{align}
  Our plan will be to use the symmetry of the dynamics under constant
translations in spacetime together with Noether's theorem \eqref{eq:NoethersTheorem}
to determine the canonical energy-momentum tensor for the overall
system.

To begin, we consider infinitesimal spacetime translations, for which
the particle's degrees of freedom $X^{\mu}\parens{\lambda}$ and $\tud{\Lambda}{\mu}{\nu}\parens{\lambda}$
transform according to 
\begin{equation}
\left.\begin{aligned}X^{\mu}\parens{\lambda} & \mapsto X^{\prime\mu}\parens{\lambda}\defeq X^{\mu}\parens{\lambda}+\epsilon^{\mu},\\
\tud{\Lambda}{\mu}{\nu}\parens{\lambda} & \mapsto\tud{\Lambda}{\prime\mu}{\nu}\parens{\lambda}\defeq\tud{\Lambda}{\mu}{\nu}\parens{\lambda},
\end{aligned}
\quad\right\} \label{eq:SpacetimeTranslationForNoetherParticleDOF}
\end{equation}
 where $\epsilon^{\mu}$ is a four-vector consisting of small, constant
components. In order for this transformation to be a symmetry of
the action functional, we will need the gauge field $A_{\mu}\parens x$
to transform in such a way that its new value $A_{\mu}^{\prime}\parens{x^{\prime}}$
at the new spacetime point $x^{\prime\mu}=x^{\mu}+\epsilon^{\mu}$
is equal to its original value $A_{\mu}\parens x$ at the original
spacetime point $x^{\mu}$: 
\begin{equation}
A_{\mu}^{\prime}\parens{x^{\prime}}=A_{\mu}\parens x.\label{eq:SpacetimeTranslationGaugeFieldCondition}
\end{equation}
 Replacing $x^{\prime\mu}\defeq x^{\mu}+\epsilon^{\mu}$ with $x^{\mu}$
and replacing $x^{\mu}$ with $x^{\mu}-\epsilon^{\mu}$, we obtain
the following infinitesimal transformation rule for the gauge field:
\begin{align}
A_{\mu}\parens x & \mapsto A_{\mu}^{\prime}\parens x\defeq A_{\mu}\parens{x-\epsilon}\nonumber \\
 & \qquad=A_{\mu}\parens x-\partial_{\nu}A_{\mu}\parens x\epsilon^{\nu}.\label{eq:SpacetimeTranslationForNoetherGaugeField}
\end{align}
 We therefore identify 
\begin{align}
\delta X^{\mu} & =\epsilon^{\mu}=\delta_{\nu}^{\mu}\epsilon^{\nu} &  & \implies g_{X^{\mu},\nu}=\delta_{\nu}^{\mu},\label{eq:SpacetimeTranslInfChangeCoordsForEMDipole}\\
\delta A_{\mu} & =-\partial_{\nu}A_{\mu}\epsilon^{\nu} &  & \implies g_{A_{\mu},\nu}=-\partial_{\nu}A_{\mu}.\label{eq:SpacetimeTranslInfChangeGaugeFieldForEMDipole}
\end{align}

We can write the system's action functional \eqref{eq:MaxwellActionWithParticleDipoleIntsAfterReducingInts}
as 
\[
S\bracks{X,\Lambda,A}=\int dt\,L,
\]
 with a standard, non-manifestly-covariant Lagrangian 
\begin{align}
L & =p_{\nu}\frac{dX^{\nu}}{dt}+\frac{1}{2}\Trace\biggbracks{S\frac{d\Lambda}{dt}\Lambda^{-1}}\nonumber \\
 & \qquad+\int d^{3}x\,\biggparens{-\frac{1}{4\mu_{0}}F^{\mu\nu}F_{\mu\nu}}\nonumber \\
 & \qquad+q\frac{dX^{\nu}}{dt}A_{\nu}\nonumber \\
 & \qquad-\frac{1}{2c}\sqrt{-\parens{dX/dt}^{2}}m^{\mu\nu}F_{\mu\nu}.\label{eq:MaxwellLagrangianWithParticleNonCovariant}
\end{align}
  Before we can employ Noether's theorem, it will be crucial to determine
the correct functions $f_{\nu}$ that appear on the right-hand side
of \eqref{eq:InfinitesimalSymmetryChangeLagrangian}: 
\[
\delta_{\epsilon}L=\frac{df_{\nu}}{dt}\epsilon^{\nu}.
\]
 Only the second line in \eqref{eq:MaxwellLagrangianWithParticleNonCovariant}
gives a nonzero contribution, and we find 
\begin{align}
f_{\nu} & =\int d^{3}x\,\frac{1}{c}\delta_{\nu}^{t}\biggparens{\frac{1}{4\mu_{0}}F^{\rho\sigma}F_{\rho\sigma}}\nonumber \\
 & =\int d^{3}x\,\frac{1}{c}\parens{-n_{\mu}}\delta_{\nu}^{\mu}\biggparens{\frac{1}{4\mu_{0}}F^{\rho\sigma}F_{\rho\sigma}},\label{eq:NoetherTotalChangeLagrangianFunctions}
\end{align}
 where, as before, $n_{\mu}\defeq\parens{-1,\vec 0}_{\mu}$ is a unit
timelike four-vector.  

Putting everything together, and recalling our expression \eqref{eq:ParticleLagrangianWithDipoleMoments}
for the particle's manifestly covariant Lagrangian $\mathscr{L}\defeq\mathscr{L}_{\textrm{particle+int}}$
together with our formula \eqref{eq:MaxwellActionWithParticleDipoleIntsAfterReducingInts}
for the overall system's Lagrangian density $\mathcal{L}$, Noether's
theorem \eqref{eq:NoethersTheorem} then tells us that the conserved
canonical four-momentum of the overall system is 
\begin{align}
P_{\nu} & =\frac{\partial\mathscr{L}}{\partial\dot{X}^{\rho}}g_{X^{\rho},\nu}+\int d^{3}x\,\parens{-n_{\mu}}\frac{\partial\mathcal{L}}{\partial\parens{c\partial_{\mu}A_{\rho}}}g_{A_{\rho},\nu}-f_{\nu}\nonumber \\
 & =p_{\nu}+qA_{\nu}+\frac{1}{2c^{2}}u_{\nu}m^{\sigma\tau}F_{\sigma\tau}\nonumber \\
 & \quad+\frac{1}{c}\int d^{3}x\,\parens{-n_{\mu}}\biggparens{H^{\mu\rho}\partial_{\nu}A_{\rho}-\delta_{\nu}^{\mu}\biggparens{\frac{1}{4\mu_{0}}F^{\rho\sigma}F_{\rho\sigma}}}\nonumber \\
 & =\frac{1}{c}\int d^{3}x\,\parens{-n_{\mu}}T_{\textrm{can},\nu}^{\mu},\label{eq:TotalCanonicalMomentumFromEnergyMomTensor}
\end{align}
 where we have identified the overall system's canonical energy-momentum
tensor as 
\begin{align}
T_{\textrm{can}}^{\mu\nu} & =u^{\mu}p^{\nu}\frac{1}{\gamma}\delta^{3}\parens{\vec x-\vec X}\nonumber \\
 & \qquad+H^{\mu\rho}\partial^{\nu}A_{\rho}+j_{\textrm{e}}^{\mu}A^{\nu}-\eta^{\mu\nu}\frac{1}{4\mu_{0}}F^{\rho\sigma}F_{\rho\sigma}\nonumber \\
 & \qquad+\frac{1}{2c^{2}}u^{\mu}u^{\nu}m^{\rho\sigma}F_{\rho\sigma}\frac{1}{\gamma}\delta^{3}\parens{\vec x-\vec X}.\label{eq:TotalCanonicalEnergyMomentumTensor}
\end{align}
 Here we have invoked the definition \eqref{eq:Def4DAuxiliaryFaradayTensor}
of the auxiliary Faraday tensor $H^{\mu\nu}$, specialized to the
case $Q^{\mu\nu}=M^{\mu\nu}$ in which quadrupole moments and higher
multipole moments are absent, 
\begin{align}
H^{\mu\nu} & =\frac{1}{\mu_{0}}F^{\mu\nu}+M^{\mu\nu}\nonumber \\
 & =\frac{1}{\mu_{0}}F^{\mu\nu}+m^{\mu\nu}\frac{1}{\gamma}\delta^{3}\parens{\vec x-\vec X},\label{eq:4DAuxiliaryFaradayTensorForDipoleParticle}
\end{align}
 and $j_{\textrm{e}}^{\mu}$ is the particle's electric-monopole current
density \eqref{eq:4DElectricMonopoleCurrentDensity}: 
\[
j_{\textrm{e}}^{\mu}=qu^{\mu}\frac{1}{\gamma}\delta^{3}\parens{\vec x-\vec X}.
\]

The terms $H^{\mu\rho}\partial^{\nu}A_{\rho}+j_{\textrm{e}}^{\mu}A^{\nu}$
in the canonical energy-momentum tensor \eqref{eq:TotalCanonicalEnergyMomentumTensor}
do not look gauge invariant. However, we can use the auxiliary inhomogeneous
Maxwell equation \eqref{eq:4DInhomogMaxwellEqFromAuxiliaryTensor}
to write the interaction term $j_{\textrm{e}}^{\mu}A^{\nu}$ as 
\begin{equation}
j_{\textrm{e}}^{\mu}A^{\nu}=-H^{\mu\rho}\partial_{\rho}A^{\nu}+\partial_{\rho}\parens{H^{\mu\rho}A^{\nu}},\label{eq:IntTermFromAuxiliarTensor}
\end{equation}
so when the equations of motion hold, the canonical energy-momentum
tensor \eqref{eq:TotalCanonicalEnergyMomentumTensor} is equivalent
to 
\begin{align}
T_{\textrm{can}}^{\mu\nu} & =u^{\mu}p^{\nu}\frac{1}{\gamma}\delta^{3}\parens{\vec x-\vec X}\nonumber \\
 & \qquad+H^{\mu\rho}\tud F{\nu}{\rho}-\eta^{\mu\nu}\frac{1}{4\mu_{0}}F^{2}\nonumber \\
 & \qquad+\frac{1}{2c^{2}}u^{\mu}u^{\nu}m^{\rho\sigma}F_{\rho\sigma}\frac{1}{\gamma}\delta^{3}\parens{\vec x-\vec X}\nonumber \\
 & \qquad+\partial_{\rho}\parens{H^{\mu\rho}A^{\nu}},\label{eq:TotalCanonicalEnergyMomentumTensorGaugeInv}
\end{align}
 which is a new result. The last term in \eqref{eq:TotalCanonicalEnergyMomentumTensorGaugeInv}
is a total spacetime divergence, and taking its spacetime divergence
on its $\mu$ index yields zero: 
\[
\partial_{\mu}\partial_{\rho}\parens{H^{\mu\rho}A^{\nu}}=0.
\]
  Moreover, the integral of its $\mu=t$ component over three-dimensional
space gives a boundary term that vanishes if we assume that our fields
go to zero sufficiently rapidly at spatial infinity: 
\[
\int d^{3}x\,\partial_{\rho}\parens{H^{t\rho}A^{\nu}}=\int d^{3}x\,\nabla\dotprod\parens{\cdots}=0.
\]
 We can therefore ignore this term in our calculations ahead.

Notice the crucial role played here by the interaction term $j_{\textrm{e}}^{\mu}A^{\nu}$,
which gave us the correction $-H^{\mu\rho}\partial_{\rho}A^{\nu}$
that we needed to yield a gauge-invariant combination $H^{\mu\rho}\tud F{\nu}{\rho}$
in the canonical energy-momentum tensor \eqref{eq:TotalCanonicalEnergyMomentumTensorGaugeInv}.
Despite the fact that it arises from the interaction term $j_{\textrm{e}}^{\mu}A^{\nu}$,
it is natural to regard the correction $-H^{\mu\rho}\partial_{\rho}A^{\nu}$
as part of the electromagnetic field's internal energy, even when
dipoles are absent and $H^{\mu\rho}$ reduces to $\parens{1/\mu_{0}}F^{\mu\rho}$.

We can divide up $T_{\textrm{can}}^{\mu\nu}$ into the canonical energy-momentum
tensor for the particle alone, 
\begin{equation}
T_{\textrm{can},\textrm{particle}}^{\mu\nu}\defeq u^{\mu}p^{\nu}\frac{1}{\gamma}\delta^{3}\parens{\vec x-\vec X},\label{eq:CanonicalEnergyMomentumTensorParticle}
\end{equation}
 and the canonical energy-momentum tensor for the field, 
\begin{align}
T_{\textrm{can},\textrm{field}}^{\mu\nu} & \defeq H^{\mu\rho}\tud F{\nu}{\rho}-\eta^{\mu\nu}\frac{1}{4\mu_{0}}F^{2}\nonumber \\
 & \qquad+\frac{1}{2c^{2}}u^{\mu}u^{\nu}m^{\rho\sigma}F_{\rho\sigma}\frac{1}{\gamma}\delta^{3}\parens{\vec x-\vec X}\nonumber \\
 & \qquad+\partial_{\rho}\parens{H^{\mu\rho}A^{\nu}},\label{eq:CanonicalEnergyMomentumTensorEMField}
\end{align}
 which we can equivalently write as 
\begin{align}
T_{\textrm{can},\textrm{field}}^{\mu\nu} & =H^{\mu\rho}\tud F{\nu}{\rho}-\eta^{\mu\nu}\frac{1}{4}\parens{H^{\rho\sigma}+M^{\rho\sigma}}F_{\rho\sigma}\nonumber \\
 & \qquad+\frac{1}{2}\biggparens{\eta^{\mu\nu}+\frac{u^{\mu}u^{\nu}}{c^{2}}}M^{\rho\sigma}F_{\rho\sigma}\nonumber \\
 & \qquad+\partial_{\rho}\parens{H^{\mu\rho}A^{\nu}}.\label{eq:CanonicalEnergyMomentumTensorEMFieldAlt}
\end{align}
We have 
\begin{align}
\int d^{3}x\,T_{\textrm{can},\textrm{particle}}^{t\nu} & =p^{\nu}c,\label{eq:4MomParticleFromIntegralEnergyMomTensor}
\end{align}
 as expected, and\footnote{Note that the authors of \citep{GrallaHarteWald:2009rdesf} break
up the total energy-momentum tensor differently by including the interaction
terms with the energy-momentum tensor for the \emph{particle}. This
approach obscures the work being done by the electromagnetic field
on the particle, and, indeed, the authors end up concluding that magnetic
forces are incapable of doing work on elementary magnetic dipole moments.} 
\begin{align}
\int d^{3}x\,T_{\textrm{can},\textrm{field}}^{t\nu} & =\int d^{3}x\,\biggparens{H^{t\rho}\tud F{\nu}{\rho}-\eta^{t\nu}\frac{1}{4\mu_{0}}F^{2}}\nonumber \\
 & \qquad+\frac{1}{2c}u^{\nu}m^{\rho\sigma}F_{\rho\sigma}.\label{eq:4MomFieldFromIntegralEnergyMomTensor}
\end{align}

In close analogy with the construction \eqref{eq:2DOFDerivExampleEnergyTimeDerivGivesOtherRate}
from the example of our $xy$ system, we can integrate the local conservation
law \eqref{eq:LocalConservationCanonicalEnergyMomentumTensor}, $\partial_{\mu}T_{\textrm{can}}^{\mu\nu}=0$,
over three-dimensional space to compute the time derivative of the
particle's four-momentum $p^{\nu}$: 
\begin{align*}
\frac{dp^{\nu}}{dt} & =\frac{1}{c}\frac{d}{dt}\int d^{3}x\,T_{\textrm{can},\textrm{particle}}^{t\nu}\\
 & =-\frac{1}{c}\frac{d}{dt}\int d^{3}x\,T_{\textrm{can},\textrm{field}}^{t\nu}\\
 & =\int d^{3}x\,\biggparens{-\partial_{\mu}\biggparens{H^{\mu\rho}\tud F{\nu}{\rho}-\eta^{\mu\nu}\frac{1}{4\mu_{0}}F^{2}}}\\
 & \qquad-\frac{1}{2c^{2}}\frac{d}{dt}\parens{u^{\nu}m^{\rho\sigma}F_{\rho\sigma}}.
\end{align*}
 By a straightforward calculation, we have 
\begin{align*}
 & -\partial_{\mu}\biggparens{H^{\mu\rho}\tud F{\nu}{\rho}-\eta^{\mu\nu}\frac{1}{4\mu_{0}}F^{2}}\\
 & \qquad\qquad=-j_{\textrm{e},\rho}F^{\rho\nu}-M^{\mu\rho}\partial_{\mu}\tud F{\nu}{\rho},
\end{align*}
 and so, using $dt/d\tau=\gamma$ from \eqref{eq:TimeDilation}, we
obtain 
\[
\frac{dp^{\nu}}{d\tau}=-qu_{\mu}F^{\mu\nu}+m_{\rho\mu}\partial^{\mu}F^{\nu\rho}-\frac{1}{2c^{2}}\frac{d}{d\tau}\parens{u^{\nu}m^{\rho\sigma}F_{\rho\sigma}}.
\]
Invoking the electromagnetic Bianchi identity \eqref{eq:EMBianchiIdentity},
\[
\partial^{\mu}F^{\nu\rho}+\partial^{\rho}F^{\mu\nu}+\partial^{\nu}F^{\rho\mu}=0,
\]
 we can write the second term as 
\[
m_{\rho\mu}\partial^{\mu}F^{\nu\rho}=-\frac{1}{2}m_{\rho\sigma}\partial^{\nu}F^{\rho\sigma}.
\]
 Relabeling indices, we find
\[
\frac{dp^{\mu}}{d\tau}=-qu_{\nu}F^{\nu\mu}-\frac{1}{2}m_{\rho\sigma}\partial^{\mu}F^{\rho\sigma}-\frac{1}{2c^{2}}\frac{d}{d\tau}\parens{u^{\mu}m^{\rho\sigma}F_{\rho\sigma}},
\]
 which precisely replicates the particle's equation of motion \eqref{eq:ParticleWithDipoleMomentsCoordEOMProperTime},
and thereby gives further support for the main claim of this paper\textemdash that
magnetic forces can classically do work on particles with elementary
dipole moments.

\subsection{Local Conservation of Angular Momentum}

Observe that the overall system's canonical energy-momentum tensor
\eqref{eq:TotalCanonicalEnergyMomentumTensorGaugeInv} is not symmetric
on its two indices, $T_{\textrm{can}}^{\mu\nu}\ne T_{\textrm{can}}^{\nu\mu}$,
reflecting the fact that it does not encode the system's intrinsic
spin. To analyze local conservation of angular momentum for the overall
system comprehensively, we return once again to the full action functional
\eqref{eq:MaxwellActionWithParticleDipoleIntsAfterReducingInts}:
\begin{align*}
S\bracks{X,\Lambda,A} & =\int d\lambda\,\biggparens{p_{\nu}\dot{X}^{\nu}+\frac{1}{2}\Trace\bracks{S\dot{\Lambda}\Lambda^{-1}}}\\
 & \qquad+\int dt\int d^{3}x\,\biggparens{-\frac{1}{4\mu_{0}}F^{\mu\nu}F_{\mu\nu}}\\
 & \qquad+\int d\lambda\,q\dot{X}^{\nu}A_{\nu}\\
 & \qquad-\frac{1}{2c}\int d\lambda\,\sqrt{-\dot{X}^{2}}m^{\mu\nu}F_{\mu\nu}.
\end{align*}
 Our next goal will be to invoke the symmetry of this action functional
under Lorentz transformations, along with Noether's theorem, to compute
the system's canonical angular-momentum tensor.

We start by noting that under an active ($-d\theta^{\rho\sigma}\mapsto+\epsilon^{\rho\sigma}$)
infinitesimal Lorentz transformation \eqref{eq:InfinitesimalLorentzTransfFromGenerators},
\begin{equation}
\Lambda_{\textrm{inf}}=1+\frac{i}{2}\epsilon^{\rho\sigma}\sigma_{\rho\sigma},\label{eq:InfinitesimalLorentzTransfFromGeneratorsForNoetherElectromagnetism}
\end{equation}
the particle's degrees of freedom $X^{\mu}\parens{\lambda}$ and $\tud{\Lambda}{\mu}{\nu}\parens{\lambda}$
transform according to 
\begin{equation}
\left.\begin{aligned}X^{\mu}\parens{\lambda} & \mapsto X^{\prime\mu}\parens{\lambda}\defeq\parens{\Lambda_{\textrm{inf}}X\parens{\lambda}}^{\mu}\\
 & \qquad=X^{\mu}\parens{\lambda}+\frac{i}{2}\epsilon^{\rho\sigma}\tud{\bracks{\sigma_{\rho\sigma}}}{\mu}{\nu}X^{\nu}\parens{\lambda},\\
\tud{\Lambda}{\mu}{\nu}\parens{\lambda} & \mapsto\tud{\Lambda}{\prime\mu}{\nu}\parens{\lambda}\defeq\tud{\parens{\Lambda_{\textrm{inf}}\Lambda\parens{\lambda}}}{\mu}{\nu}\\
 & \qquad=\tud{\Lambda}{\mu}{\nu}\parens{\lambda}+\frac{i}{2}\epsilon^{\rho\sigma}\tud{\bracks{\sigma_{\rho\sigma}}}{\mu}{\lambda}\tud{\Lambda}{\lambda}{\nu}\parens{\lambda},
\end{aligned}
\quad\right\} \label{eq:LorentzTranslationForNoetherParticleDOF}
\end{equation}
 where $\epsilon^{\rho\sigma}$ is an antisymmetric tensor consisting
of small constants. Note that the lower Lorentz index on $\tud{\Lambda}{\mu}{\nu}\parens{\lambda}$
does not participate in the second transformation rule, which fundamentally
arises from the composition property $\Lambda^{\prime}\defeq\Lambda_{\textrm{inf}}\Lambda\parens{\lambda}$.
 Observe also that we can rephrase this second transformation rule
as the statement that the underlying antisymmetric array $\theta^{\mu\nu}\parens{\lambda}$
of boost and angular parameters transforms as 
\begin{equation}
\theta^{\mu\nu}\parens{\lambda}\mapsto\theta^{\prime\mu\nu}\parens{\lambda}\defeq\theta^{\mu\nu}\parens{\lambda}+\epsilon^{\mu\nu}.\label{eq:LorentzTranslationForNoetherParticleBoostAngleDOF}
\end{equation}
  Meanwhile, the gauge field $A_{\mu}\parens x$ transforms as 
\begin{align}
A_{\mu}\parens x & \mapsto A_{\mu}^{\prime}\parens x\defeq\parens{A\parens{\Lambda_{\textrm{inf}}^{-1}x}\Lambda_{\textrm{inf}}^{-1}}_{\mu}\nonumber \\
 & \defeq A_{\lambda}\parens{\parens{1-\parens{i/2}\epsilon^{\rho\sigma}\sigma_{\rho\sigma}}x}\parens{\delta_{\mu}^{\lambda}-\parens{i/2}\epsilon^{\rho\sigma}\tud{\bracks{\sigma_{\rho\sigma}}}{\lambda}{\mu}}\nonumber \\
 & =A_{\mu}\parens x-\partial_{\nu}A_{\mu}\parens x\parens{i/2}\epsilon^{\rho\sigma}\tud{\bracks{\sigma_{\rho\sigma}}}{\nu}{\lambda}x^{\lambda}\nonumber \\
 & \qquad-A_{\lambda}\parens x\parens{i/2}\epsilon^{\rho\sigma}\tud{\bracks{\sigma_{\rho\sigma}}}{\lambda}{\mu}.\label{eq:LorentzTranslationForNoetherGaugeField}
\end{align}
 Again dropping factors of $1/2$ to avoid double-counting independent
variables, we can therefore identify 
\begin{align}
\delta X^{\mu}= & \frac{i}{2}\epsilon^{\rho\sigma}\tud{\bracks{\sigma_{\rho\sigma}}}{\mu}{\nu}X^{\nu}\nonumber \\
 & \implies g_{X^{\mu},\rho\sigma}=i\tud{\bracks{\sigma_{\rho\sigma}}}{\mu}{\nu}X^{\nu},\label{eq:InfLorentzTransfChangeCoordsEMDipole}\\
\delta\theta^{\mu\nu} & =\epsilon^{\mu\nu}=\frac{1}{2}\parens{\delta_{\rho}^{\mu}\delta_{\sigma}^{\nu}-\delta_{\sigma}^{\mu}\delta_{\rho}^{\nu}}\epsilon^{\rho\sigma}\nonumber \\
 & \implies g_{\theta^{\mu\nu},\rho\sigma}=\delta_{\rho}^{\mu}\delta_{\sigma}^{\nu}-\delta_{\sigma}^{\mu}\delta_{\rho}^{\nu},\label{eq:InfLorentzTransfBoostAngleParamsEMDipole}\\
\delta A_{\mu} & =-\partial_{\nu}A_{\mu}\parens{i/2}\epsilon^{\rho\sigma}\tud{\bracks{\sigma_{\rho\sigma}}}{\nu}{\lambda}x^{\lambda}-A_{\nu}\parens{i/2}\epsilon^{\rho\sigma}\tud{\bracks{\sigma_{\rho\sigma}}}{\nu}{\mu}\nonumber \\
 & \implies g_{A_{\mu},\rho\sigma}=-\partial_{\nu}A_{\mu}i\tud{\bracks{\sigma_{\rho\sigma}}}{\nu}{\lambda}x^{\lambda}-A_{\nu}i\tud{\bracks{\sigma_{\rho\sigma}}}{\nu}{\mu}.\label{eq:InfLorentzTransfGaugeFieldEMDipole}
\end{align}
   Finally, the functions $f_{\rho\sigma}$ that appear on the right-hand
side of \eqref{eq:InfinitesimalSymmetryChangeLagrangian}, 
\[
\delta_{\epsilon}L=\frac{1}{2}\frac{df_{\rho\sigma}}{dt}\epsilon^{\rho\sigma},
\]
 are given by 
\begin{align}
f_{\rho\sigma} & =\int d^{3}x\,\frac{1}{c}\delta_{\nu}^{t}\biggparens{\frac{1}{4\mu_{0}}F^{2}}i\tud{\bracks{\sigma_{\rho\sigma}}}{\nu}{\lambda}x^{\lambda}\nonumber \\
 & =\int d^{3}x\,\frac{1}{c}\parens{-n_{\mu}}\delta_{\nu}^{\mu}\biggparens{\frac{1}{4\mu_{0}}F^{2}}i\tud{\bracks{\sigma_{\rho\sigma}}}{\nu}{\lambda}x^{\lambda},\label{eq:NoetherTotalChangeLorentzTransformationLagrangianFunctions}
\end{align}
 where, as usual, $n_{\mu}\defeq\parens{-1,\vec 0}_{\mu}$. 

We then have from Noether's theorem \eqref{eq:NoethersTheorem} that
the conserved angular-momentum tensor of the overall system is, up
to an overall minus sign, given by 
\begin{align}
 & -J_{\nu\rho}=\frac{\partial\mathscr{L}}{\partial\dot{X}^{\alpha}}g_{X^{\alpha},\nu\rho}+\frac{1}{2}\frac{\partial\mathscr{L}}{\partial\dot{\theta}^{\alpha\beta}}g_{\theta^{\alpha\beta},\nu\rho}\nonumber \\
 & \qquad+\int d^{3}x\,\parens{-n_{\mu}}\frac{\partial\mathcal{L}}{\partial\parens{c\partial_{\mu}A_{\alpha}}}g_{A_{\alpha},\nu\rho}-f_{\nu\rho}\nonumber \\
 & =-\biggparens{p_{\alpha}+qA_{\alpha}-\frac{1}{2}\parens{-u_{\alpha}/c^{2}}m^{\sigma\lambda}F_{\sigma\lambda}}\parens{X_{\nu}\delta_{\rho}^{\alpha}-X_{\rho}\delta_{\nu}^{\alpha}}\nonumber \\
 & \quad-S_{\nu\rho}\nonumber \\
 & \quad-\frac{1}{c}\int d^{3}x\,\parens{-n_{\mu}}\biggparens{H^{\mu\alpha}-\delta_{\sigma}^{\mu}\biggparens{\frac{1}{4\mu_{0}}F^{2}}}\nonumber \\
 & \qquad\qquad\qquad\times\partial_{\sigma}A_{\alpha}\parens{x_{\nu}\delta_{\rho}^{\sigma}-x_{\rho}\delta_{\nu}^{\sigma}}\nonumber \\
 & \quad-\frac{1}{c}\int d^{3}x\,\parens{-n_{\mu}}\parens{\tud H{\mu}{\nu}A_{\rho}-\tud H{\mu}{\rho}A_{\nu}}\nonumber \\
 & =-\int d^{3}x\,\parens{-n_{\mu}}\mathcal{J}_{\textrm{can},\nu\rho}^{\mu},\label{eq:TotalAngMomentumFromAngMomTensor}
\end{align}
 where the overall system's canonical angular-momentum flux tensor
is 
\begin{align}
\mathcal{J}_{\textrm{can}}^{\mu\nu\rho} & =\frac{1}{c}\parens{x^{\nu}T_{\textrm{can}}^{\mu\rho}-x^{\rho}T_{\textrm{can}}^{\mu\nu}}\nonumber \\
 & \quad+\frac{1}{c}u^{\mu}S^{\nu\rho}\frac{1}{\gamma}\delta^{3}\parens{\vec x-\vec X}+\frac{1}{c}\parens{H^{\mu\nu}A^{\rho}-H^{\mu\rho}A^{\nu}},\label{eq:TotalCanonicalAngMomFluxTensorExplicit}
\end{align}
 which is another new result. Here $T_{\textrm{can}}^{\mu\nu}$ is
the canonical energy-momentum tensor \eqref{eq:TotalCanonicalEnergyMomentumTensor}:
\begin{align*}
T_{\textrm{can}}^{\mu\nu} & =u^{\mu}p^{\nu}\frac{1}{\gamma}\delta^{3}\parens{\vec x-\vec X}\\
 & \qquad+H^{\mu\rho}\partial^{\nu}A_{\rho}+j_{\textrm{e}}^{\mu}A^{\nu}-\eta^{\mu\nu}\frac{1}{4\mu_{0}}F^{\rho\sigma}F_{\rho\sigma}\\
 & \qquad+\frac{1}{2c^{2}}u^{\mu}u^{\nu}m^{\rho\sigma}F_{\rho\sigma}\frac{1}{\gamma}\delta^{3}\parens{\vec x-\vec X}.
\end{align*}

Observe that the canonical angular-momentum flux tensor \eqref{eq:TotalCanonicalAngMomFluxTensorExplicit}
has precisely the form \eqref{eq:DefFieldAngMomFluxTensor}, 
\[
\mathcal{J}_{\textrm{can}}^{\mu\nu\rho}=\mathcal{L}^{\mu\nu\rho}+\mathcal{S}^{\mu\nu\rho},
\]
 with $\mathcal{L}^{\mu\nu\rho}$ representing the contribution \eqref{eq:DefFieldOrbAngMomFluxTensor}
from orbital angular momentum, 
\[
\mathcal{L}^{\mu\nu\rho}\defeq x^{\nu}\frac{1}{c}T_{\textrm{can}}^{\mu\rho}-x^{\rho}\frac{1}{c}T_{\textrm{can}}^{\mu\nu},
\]
 and with $\mathcal{S}^{\mu\nu\rho}$ representing the contribution
\eqref{eq:DefFieldSpinFluxTensor} from the intrinsic spin of both
the particle and the electromagnetic field: 
\begin{equation}
\mathcal{S}^{\mu\nu\rho}=\frac{1}{c}u^{\mu}S^{\nu\rho}\frac{1}{\gamma}\delta^{3}\parens{\vec x-\vec X}+\frac{1}{c}\parens{H^{\mu\nu}A^{\rho}-H^{\mu\rho}A^{\nu}}.\label{eq:SpinFluxTensorEMAndParticle}
\end{equation}
 Specifically, the first term in \eqref{eq:SpinFluxTensorEMAndParticle}
describes the particle's intrinsic spin, 
\begin{equation}
\mathcal{S}_{\textrm{particle}}^{\mu\nu\rho}=\frac{1}{c}u^{\mu}S^{\nu\rho}\frac{1}{\gamma}\delta^{3}\parens{\vec x-\vec X},\label{eq:ParticleSpinAngMomTensor}
\end{equation}
 and the second term arises from the field's spin: 
\begin{equation}
\mathcal{S}_{\textrm{field}}^{\mu\nu\rho}=\frac{1}{c}\parens{H^{\mu\nu}A^{\rho}-H^{\mu\rho}A^{\nu}}.\label{eq:SpinFluxTensorEMAlone}
\end{equation}

Integrating the local conservation law \eqref{eq:LocalConservationFieldTotAngMomCanonical},
$\partial_{\mu}\mathcal{J}_{\textrm{can}}^{\mu\nu\rho}=0$, over three-dimensional
space, and taking advantage of the local conservation \eqref{eq:LocalConservationCanonicalEnergyMomentumTensor}
of the total canonical energy-momentum tensor $T_{\textrm{can}}^{\mu\rho}$,
we can compute the time derivative of the particle's spin tensor as
follows: 
\begin{align*}
 & \frac{dS^{\nu\rho}}{dt}=\frac{d}{dt}\int d^{3}x\,\mathcal{S}_{\textrm{particle}}^{t\nu\rho}\\
 & \quad=-\frac{d}{dt}\int d^{3}x\,\frac{1}{c}\parens{x^{\nu}T_{\textrm{can}}^{t\rho}-x^{\rho}T_{\textrm{can}}^{t\nu}+H^{t\nu}A^{\rho}-H^{t\rho}A^{\nu}}\\
 & \quad=-\int d^{3}x\,\partial_{\mu}\parens{x^{\nu}T_{\textrm{can}}^{\mu\rho}-x^{\rho}T_{\textrm{can}}^{\mu\nu}+H^{\mu\nu}A^{\rho}-H^{\mu\rho}A^{\nu}}\\
 & \quad=-\frac{1}{\gamma}\parens{u^{\nu}p^{\rho}-u^{\rho}p^{\nu}}-\frac{1}{\gamma}\parens{m^{\nu\sigma}\tud F{\rho}{\sigma}-m^{\rho\sigma}\tud F{\nu}{\sigma}}.
\end{align*}
 Using $dt/d\tau=\gamma$ from \eqref{eq:TimeDilation} and relabeling
indices, we therefore find 
\[
\frac{dS^{\mu\nu}}{d\tau}=-\parens{u^{\mu}p^{\nu}-u^{\nu}p^{\mu}}-\parens{m^{\mu\rho}\tud F{\nu}{\rho}-m^{\nu\rho}\tud F{\mu}{\rho}},
\]
 which precisely agrees with the particle's equation of motion \eqref{eq:ParticleSpinTensorEOMProperTime}
for $S^{\mu\nu}$.

Now that we have calculated the system's canonical angular-momentum
flux tensor $\mathcal{J}_{\textrm{can}}^{\mu\nu\rho}$ and identified
the spin flux tensor $\mathcal{S}^{\mu\nu\rho}$, as given by \eqref{eq:SpinFluxTensorEMAndParticle},
we can construct a symmetric, gauge-invariant energy-momentum tensor
\eqref{eq:DefBelinfanteRosenfeldEnegyMomTensor}, $T^{\mu\nu}=T_{\textrm{can}}^{\mu\nu}+\partial_{\rho}\mathcal{B}^{\mu\rho\nu}$,
from the system's Belinfante-Rosenfeld tensor \eqref{eq:DefBelinfanteRosenfeldCorrection}:
\begin{align}
 & \mathcal{B}^{\mu\rho\nu}\defeq\frac{c}{2}\parens{\mathcal{S}^{\mu\nu\rho}+\mathcal{S}^{\nu\mu\rho}+\mathcal{S}^{\rho\mu\nu}}\nonumber \\
 & =-H^{\mu\rho}A^{\nu}+\frac{1}{2}\parens{u^{\mu}S^{\nu\rho}+u^{\nu}S^{\mu\rho}+u^{\rho}S^{\mu\nu}}\frac{1}{\gamma}\delta^{3}\parens{\vec x-\vec X}.\label{eq:BelinfanteRosenfeldCorrectionEMAndParticle}
\end{align}
 We obtain\footnote{This formula differs from the corresponding result in \citep{VanDamRuijgrok:1980crepsmef},
whose energy-momentum tensor yields the correct equations of motion
for the particle only after an unjustified four-dimensional integration
by parts.} 
\begin{align}
T^{\mu\nu} & =\frac{1}{2}\parens{u^{\mu}p^{\nu}+u^{\nu}p^{\mu}}\frac{1}{\gamma}\delta^{3}\parens{\vec x-\vec X}\nonumber \\
 & \qquad+\frac{1}{2}H^{\mu\rho}\tud F{\nu}{\rho}+\frac{1}{2}H^{\nu\rho}\tud F{\mu}{\rho}-\eta^{\mu\nu}\frac{1}{4\mu_{0}}F^{\rho\sigma}F_{\rho\sigma}\nonumber \\
 & \qquad+\frac{1}{2c^{2}}u^{\mu}u^{\nu}m^{\rho\sigma}F_{\rho\sigma}\frac{1}{\gamma}\delta^{3}\parens{\vec x-\vec X}\nonumber \\
 & \qquad+\frac{1}{2}\partial_{\rho}\parens{\mathcal{S}_{\textrm{particle}}^{\mu\nu\rho}+\mathcal{S}_{\textrm{particle}}^{\nu\mu\rho}}.\label{eq:TotalBelinfanteRosenfeldEnergyMomentumTensor}
\end{align}
 In the free-field limit\textemdash meaning in the absence of the
particle\textemdash this energy-momentum tensor reduces to \eqref{eq:4DElectromagneticEnergyMomentumTensor},
as expected: 
\[
T^{\mu\nu}=\frac{1}{\mu_{0}}F^{\mu\rho}\tud F{\nu}{\rho}-\eta^{\mu\nu}\frac{1}{4\mu_{0}}F^{\rho\sigma}F_{\rho\sigma}.
\]

\section{Conclusion}

In this paper, we have employed the Lagrangian formulation of classical
physics to show that a massive particle with four-momentum $p^{\mu}$,
spin tensor $S^{\mu\nu}$, electric charge $q$, and elementary dipole
tensor $m^{\mu\nu}$ in an external electromagnetic field $F_{\mu\nu}$
obeys the relativistic equations of motion \eqref{eq:ParticleWithDipoleMomentsCoordEOMProperTime}
and \eqref{eq:ParticleSpinTensorEOMProperTime}: 
\begin{align*}
\frac{dp}{d\tau}^{\mu} & =-qu_{\nu}F^{\nu\mu}-\frac{1}{2}m^{\rho\sigma}\partial^{\mu}F_{\rho\sigma}-\frac{1}{2c^{2}}\frac{d}{d\tau}\parens{u^{\mu}m^{\rho\sigma}F_{\rho\sigma}},\\
\frac{dS^{\mu\nu}}{d\tau} & =-\parens{u^{\mu}p^{\nu}-u^{\nu}p^{\mu}}-\parens{m^{\mu\rho}\tud F{\nu}{\rho}-m^{\nu\rho}\tud F{\mu}{\rho}}.
\end{align*}
 To verify that these equations of motion are compatible with local
conservation of energy, momentum, and angular momentum, we have effectively
divided up the locally conserved, canonical energy-momentum tensor
$T_{\textrm{can}}^{\mu\nu}=T_{\textrm{can},\textrm{particle}}^{\mu\nu}+T_{\textrm{can},\textrm{field}}^{\mu\nu}$
of the overall system by defining the canonical energy-momentum tensor
for the particle to be \eqref{eq:CanonicalEnergyMomentumTensorParticle},
\[
T_{\textrm{can},\textrm{particle}}^{\mu\nu}\defeq u^{\mu}p^{\nu}\frac{1}{\gamma}\delta^{3}\parens{\vec x-\vec X},
\]
 and the canonical energy-momentum tensor for the electromagnetic
field to be \eqref{eq:CanonicalEnergyMomentumTensorEMField}, 
\begin{align*}
T_{\textrm{can},\textrm{field}}^{\mu\nu} & \defeq H^{\mu\rho}\tud F{\nu}{\rho}-\eta^{\mu\nu}\frac{1}{4\mu_{0}}F^{2}\\
 & \qquad+\frac{1}{2c^{2}}u^{\mu}u^{\nu}m^{\rho\sigma}F_{\rho\sigma}\frac{1}{\gamma}\delta^{3}\parens{\vec x-\vec X}\\
 & \qquad+\partial_{\rho}\parens{H^{\mu\rho}A^{\nu}}.
\end{align*}
 The local conservation equation $\partial_{\mu}T_{\textrm{can}}^{\mu\nu}=0$
then translates into the relativistic equation of motion \eqref{eq:ParticleWithDipoleMomentsCoordEOMProperTime}
for the particle's four-momentum, and the local conservation law $\partial_{\mu}\mathcal{J}_{\textrm{can}}^{\mu\nu\rho}=0$
satisfied by the overall system's canonical angular-momentum flux
tensor $\mathcal{J}_{\textrm{can}}^{\mu\nu\rho}$, as defined in \eqref{eq:TotalCanonicalAngMomFluxTensorExplicit},
yields the equation of motion \eqref{eq:ParticleSpinTensorEOMProperTime}
for the particle's spin tensor.

In the non-relativistic limit with time-independent external fields,
the equation of motion \eqref{eq:ParticleWithDipoleMomentsCoordEOMProperTime}
generalizes the Lorentz force law to \eqref{eq:LorentzForceLawWithDipoles},
\[
\vec F=q\vec E_{\textrm{ext}}+q\vec v\crossprod\vec B_{\textrm{ext}}+\nabla\parens{\vecgreek{\pi}\dotprod\vec E_{\textrm{ext}}}+\nabla\parens{\vecgreek{\mu}\dotprod\vec B_{\textrm{ext}}},
\]
 and gives the power law \eqref{eq:LorentzPowerLawWithDipoles}, 
\begin{align*}
\frac{dW}{dt} & =\frac{d}{dt}\parens{-q\Phi_{\textrm{ext}}+\vecgreek{\pi}\dotprod\vec E_{\textrm{ext}}+\vecgreek{\mu}\dotprod\vec B_{\textrm{ext}}}\\
 & =\vec v\dotprod\vec F.
\end{align*}
 These formulas are consistent with the fact that magnetic forces
cannot do work on \emph{electric monopoles}, but also make clear that
magnetic forces are fully capable of doing work on \emph{elementary
magnetic dipoles}, in accordance with the basic definition \eqref{eq:DefWork}
of what it means for a force to do mechanical work on an object in
moving the object from a location $A$ to another location $B$, as
we showed explicitly in \eqref{eq:EMWorkDoneOnParticle}: 
\begin{align*}
W & \defeq\int_{A}^{B}d\vec X\dotprod\vec F\\
 & =\int_{A}^{B}d\vec X\dotprod q\vec E_{\textrm{ext}}+\Delta\parens{\vecgreek{\pi}\dotprod\vec E_{\textrm{ext}}}+\Delta\parens{\vecgreek{\mu}\dotprod\vec B_{\textrm{ext}}}.
\end{align*}

As an interesting application, these results provide a loophole in
the Bohr-van Leeuwen theorem, which Niels Bohr first proved in his
1911 doctoral thesis \citep{BohrRosenfeldNielsen:1972ddtt} and which
was later independently proved by Hendrika Johanna van Leeuwen in
her own doctoral thesis in 1919 \citep{VanLeeuwen:1921crepsmef}.
The Bohr-van Leeuwen theorem asserts on the basis of the original
Lorentz force law (that is, without contributions from elementary
dipoles) that a non-rotating system of particles, when treated \emph{classically},
always has a vanishing average magnetization in thermal equilibrium.
A key implication of the Bohr-van Leeuwen theorem is that phenomena
like diamagnetism cannot arise without quantum mechanics. Our results
in this paper provide a theoretical exception to this corollary.\footnote{We thank Sebastiano Covone for suggesting the consideration of the
Bohr-van Leeuwen theorem in the context of our results.}

Returning to our equations describing forces and work done on a classical
particle with elementary dipole moments, it is important to note that
we do not require any external, ad hoc sources of energy and momentum
to ensure the validity of these equations. The energy and momentum
that flow into the particle are fully accounted for in the energy
and momentum that arise from the overall classical action functional
describing the coupling of the particle to the electromagnetic field,
regardless of whether, at the level of interpretation, we attribute
all that energy and momentum to the electromagnetic field alone or
to the interactions between the electromagnetic field and the particle.

Magnetic forces can do work. In this paper, we have shown how.

\begin{acknowledgments}
J.\,A.\,B. has benefited tremendously from personal communications
with Gary Feldman, Howard Georgi, Andrew Strominger, Bill Phillips,
David Griffiths, David Kagan, David Morin, Logan McCarty, Monica Pate,
Alex Lupsasca, and Sebastiano Covone.
\end{acknowledgments}

\bibliographystyle{1_home_jacob_Documents_Work_My_Papers_Magnetic_Forces_Can_Do_Work__2020__custom-abbrvunsrturl}
\bibliography{0_home_jacob_Documents_Work_My_Papers_Bibliography_Global-Bibliography}

\end{document}